%% file: frame.tex
\documentclass[12pt,twoside]{article}
\makeatletter \c@tocdepth \@ne
\makeatother
\usepackage{times}
\usepackage{latexsym}
\usepackage{mathptm}
\usepackage{ling1}
\usepackage{nd}
\usepackage{restrained}
\usepackage{fullpage}
\usepackage{epsfig}
\usepackage{doublespace}
\newlength{\localjunklength}

\newcommand{\op}[1]{\ensuremath{[\mbox{\textsc{#1}}]}}

\newcommand{\qed}{\mbox{\rule[.5ex]{1ex}{1ex}}}

\newcommand{\bimp}{\ensuremath{>}}

\newcommand{\access}{{\ensuremath{{\cal R}\,}}}
\newcommand{\domain}{{\ensuremath{\cal D}}}
\newcommand{\cdomain}{{\ensuremath{\cal C}}}
\newcommand{\types}{{\ensuremath{\cal A}}}
\newcommand{\inclusions}{{\ensuremath{\cal N}}}
\newcommand{\dconstraint}{{\ensuremath{\cal Q}}}
\newcommand{\valuation}{{\ensuremath{\cal J}}}

\newcommand{\pipie}{\mbox{\vrule{\raisebox{0in}[1ex][0in]{~}}}}
\newcommand{\forces}{\ensuremath{\mathrel{\pipie\!\pipie\!\!\!-}}}

\newcommand{\verax}{\textsc{(ver)}}
\newcommand{\piax}{\textsc{(pi)}}
\newcommand{\incax}{\textsc{(inc)}}
   
\newcommand{\dialup}{{\sc dialup}}
\newtheorem{definition}{Definition}
\newtheorem{prop}{Proposition}
\newtheorem{lemma}{Lemma}   
\newtheorem{thrm}[prop]{Theorem}

\newcommand{\worlds}{\mbox{\bf w}}

\begin{document}   
\bibliographystyle{apalike}   

\markboth{\hfill{\sc disjunction and modular proof
   search}\hfill}{\hfill{\sc disjunction and modular proof
   search}\hfill}
   
\thispagestyle{empty}
\sloppy
\begin{center}
\begin{singlespace}
{\bf Disjunction and Modular Goal-directed Proof Search}\footnote{
	Thanks to three anonymous reviewers, Mark Steedman, Rich
   Thomason, L. Thorne McCarty and Michael Fourman for extensive
   comments.  This work was supported by an NSF graduate fellowship,
   an IRCS graduate fellowship, and a postdoctoral fellowship from
   RUCCS, as well as NSF grant IRI95-04372, ARPA grant N6601-94-C6043,
   and ARO grant DAAH05-94-G0426.  \today.
} \\
Matthew Stone \\
Department of Computer Science and Center for Cognitive Science \\
Rutgers University \\
110 Frelinghuysen Road, Piscataway NJ 08854-8019 \\
{\tt mdstone@cs.rutgers.edu} \\[1em]
\end{singlespace}
\vspace{\fill}

{\bf Summary}
\end{center}
\begin{singlespace}
\begin{quote}
\small
\vspace*{-.5in}
	This paper explores goal-directed proof search in first-order
   multi-modal logic.  The key issue is to design a proof system that
   respects the modularity and locality of assumptions of many modal
   logics.  By forcing ambiguities to be considered independently,
   modular disjunctions in particular can be used to construct
   efficiently executable specifications in reasoning tasks involving
   partial information that otherwise might require prohibitive
   search.  To achieve this behavior requires prior proof-theoretic
   justifications of logic programming to be extended, strengthened,
   and combined with proof-theoretic analyses of modal deduction in a
   novel way.
\tableofcontents
\end{quote}
\end{singlespace}
   
\pagebreak

\input{nljp}

\bibliography{/fac/u/mdstone/info/biblio}

\end{document}

%% file: nljp.tex
\section{Introduction}
\label{intro-sec}

	This paper explores the proof-theoretic interaction between
   the goal-directed application of logical inferences and
   \emph{information-flow}---that is, the possible connections between
   assumptions and conclusions in proofs.  My own starting-point for
   this exploration was the result of \cite{permute-paper}, that
   intuitionistic sequent calculi can be formulated so as exhibit the
   characteristically intuitionistic \emph{modular} information-flow
   (as underlying the correspondence between proofs and programs
   \cite{howard:iso}, for example) while nevertheless allowing logical
   inferences to be applied in any order whatsoever.  This raises the
   question whether it is possible to enforce this kind of modularity
   \emph{incrementally} during goal-directed proof search.  Of course,
   the well-known flexibility of deduction in nonclassical logics
   \cite{fitting:tableau,fitting:proof,wallen:nonclassical} is ample
   motivation for the question.

\subsection{Problem Statement}
\label{problem-subsec}

   	I begin by delineating the focus of the paper more precisely.
   I will work with a family of first-order multi-modal logics in this
   paper.  The generalization from intuitionistic logic reflects the
   utility of more general ways of structuring logical specifications
   \cite{baldoni/giordano/martelli:modal,baldoni:jlc98}, as well as
   the broader importance of expressive modality in practical
   knowledge representation \cite{mccarthy:notes,mccarthy:no}.
   Qualitatively, what distinguishes the logics I consider (for which
   formal definitions are provided in Section~\ref{logic-sec}) is that
   they permit rules of modal inference to be formulated in two
   equivalent ways
   \cite{fitting:tableau,fitting:proof,wallen:nonclassical}.  I
   illustrate the alternatives for the case of S4, a logic that we can
   perhaps regard as the pure modal logic of local and global modular
   assumptions \cite{giordano/martelli:structuring}.

\subsubsection{Structural scope and modularity}
\label{structure-subsubsec}

	The first formulation of modal inference is illustrated by the
   sequent inference figure below:
\[
\derivi{$(\proveslabel\Box)$}
       {\Gamma^* \proveslabel G, \Delta^*}
       {\Gamma \proveslabel \Box G, \Delta}
\]
	Such inferences set up a discipline of structural scope in
   proofs.  Read upward, as a description of proof search, the figure
   describes how to accomplish generic reasoning about a modal
   context, such as the conclusion $\Box G$ here.  We have to
   transform the sequent we are considering, by restricting our
   attention just to the generic modal statements in the sequent.
   Specifically, $\Gamma^*$ contains the formula occurrences of the
   form $\Box A$ in $\Gamma$, and $\Delta^*$ contains the formula
   occurrences of the form $\Diamond A$ in $\Delta$.  The effect of
   the transformation is that we move from our current scope into a
   new, nested scope in which just generic information is available.
   Figure~\ref{structure-scope-fig} illustrates all the
   structurally-scoped S4 sequent inferences that I will draw on in
   this motivating discussion; I refer the reader to
   \cite{fitting:proof,wallen:nonclassical} for more
   details on structurally-scoped proof.
\begin{figure}
\[
\begin{array}{c}
\derivi{$(\proveslabel\Box)$}
       {\Gamma^* \proveslabel G, \Delta^*}
       {\Gamma \proveslabel \Box G, \Delta}
\\
\derivi{$(\Box\proveslabel)$}
       {\Gamma, G \proveslabel \Delta}
       {\Gamma, \Box G \proveslabel \Delta}
\\
\derivi{$(\proveslabel\imp)$}
       {\Gamma, P \proveslabel G, \Delta}
       {\Gamma \proveslabel P \imp G, \Delta}
\\
\derivii{$(\imp\proveslabel)$}
       {\Gamma \proveslabel G, \Delta}
       {\Gamma, P \proveslabel \Delta}
       {\Gamma, G \imp P \proveslabel \Delta}
\\
\derivii{$(\vee\proveslabel)$}
       {\Gamma, A \proveslabel \Delta}
       {\Gamma, B \proveslabel \Delta}
       {\Gamma, A \vee B \proveslabel \Delta}
\\
\derivii{$(\proveslabel\wedge)$}
       {\Gamma \proveslabel A, \Delta}
       {\Gamma \proveslabel B, \Delta}
       {\Gamma \proveslabel A \wedge B, \Delta}
\\
\Gamma, A \proveslabel A, \Delta \;\;\; (\m{Axiom})
\end{array}
\]
\caption{Six inference figures and the axiom for structurally-scoped
   S4.  After \cite{fitting:proof,wallen:nonclassical}.  Sequents are
   multisets of modal formulas; this formulation (though not others
   that we will consider later) requires a structural rule of
   contraction.  See Section~\ref{logic-sec}.}
\label{structure-scope-fig}
\end{figure}

	The ability to define structural scope is intimately connected
   with the ability to describe modular and local reasoning.  In
   specifying reasoning, we can think of antecedent formulas in
   sequents as program statements and succedent formulas in
   sequents as goals.  In modal logics with structural scope, a
   necessary goal $\Box G$ can be seen as a {\it modular} goal
   because, as enforced by the structurally-scoped inference figure,
   only program statements of the form $\Box P$ can contribute to its
   proof.  In other words, we cannot use the entire program to prove
   $G$; rather, we must use a designated \emph{part} of the program:
   formulas of the form $\Box P$.  This is the \emph{module} we use to
   prove $G$.  Multi-modal logic allows us to name modules in a
   general way \cite{baldoni/giordano/martelli:modal,baldoni:jlc98}.

	In fragments of logic without the operator $\Diamond$,
   including S4 translations of intuitionistic formulas in particular,
   modularity brings \emph{locality}.  A goal $\Box(P \imp G)$
   introduces a {\it local} assumption $P$.  The assumption is local
   in the sense that it can only contribute to the proof of $G$, and
   cannot contribute to any other goal.  We can motivate this locality
   in logical terms by examining the sequent inferences for
   $(\proveslabel\Box)$ and $(\proveslabel\imp)$ in combination:
\[
\derivi{$(\proveslabel\Box)$}
       {\derivi{$(\proveslabel\imp)$}
	       {\Gamma^*, P \proves G}
	       {\Gamma^* \proves P \imp G}}
       {\Gamma \proves \Box (P \imp G), \Delta}
\]
	Observe that this logical fragment is constructed so that the
   succedent context $\Delta^*$ above $(\proveslabel\Box)$ is empty,
   and so we introduce $P$ into a subproof where $G$ is the only goal.

	Logical modularity and locality underlie the use of the proof
   theory of modal logic as a declarative framework for structuring
   specifications, and thereby facilitating their design and reuse
   \cite{dale:modules,giordano/martelli:structuring,baldoni/giordano/martelli:modal,bgm:framework,baldoni:jlc98}.%
\footnote{
   	The model theory of modal logic can also be used to structure
   specifications \cite{sakakibara:iclp87}.
}
	Concretely, a goal that specifies the part of the program to
   be used in its proof will give rise to the same operational
   behavior when other parts of the program change.  In this paper, I
   further emphasize that logical modularity and locality provide
   declarative tools for constraining the complexity of proof search
   itself.  My motivating example is the proof in
   Figure~\ref{structure-module-fig}, which establishes that the
   conclusion
\[
	(\Box A) \wedge (\Box C)
\]
	follows from the assumptions
\[
	\Box(A \vee B), \Box(C\vee D), \Box(B \imp A), \Box(D\imp C)
\]	
\begin{figure}
\small
\begin{center}
\makebox[0in]{
$\derivii{$(\proveslabel\wedge)$}
        {\derivi{$(\proveslabel\Box)$}
	        {\derivi{$(\Box\proveslabel)$}
		        {\derivii{$(\vee\proveslabel)$}
   			         {A, \ldots \proves A}
				 {\derivi{$(\Box\proveslabel)$} 
					 {\derivii{$(\imp\proveslabel)$}
						  {B, \ldots \proves B, A}
						  {A, \ldots \proves A}
						  {B, B \imp A, \ldots \proves A}}
					 {B, \Box (B \imp A), \ldots \proves A}}
				 {A \vee B, \Box (B \imp A), \ldots \proves A}}
   			{\Box (A \vee B), \Box (B \imp A), \ldots \proves A}}
	        {\Box (A \vee B), \Box (B \imp A), \ldots \proves \Box A}}
        {\derivi{$(\proveslabel\Box)$}
	        {\derivi{$(\Box\proveslabel)$}
		        {\derivii{$(\vee\proveslabel)$}
   			         {C, \ldots \proves C}
				 {\derivi{$(\Box\proveslabel)$} 
					 {\derivii{$(\imp\proveslabel)$}
						  {D, \ldots \proves D, C}
						  {C, \ldots \proves C}
						  {D, D \imp C, \ldots \proves C}}
					 {D, \Box (D \imp C),
   \ldots \proves C}}
				 {C \vee D, \Box (D \imp C),
   \ldots \proves C}}
   			{\Box (C \vee D), \Box (D \imp C), \ldots \proves C}}
	        {\Box (C \vee D), \Box (D \imp C), \ldots \proves \Box C}}
        {\Box (A \vee B), \Box (C \vee D), \Box (B \imp A), \Box (D
   \imp C) \proves (\Box A) \wedge (\Box C)}
$}
\end{center}
\caption{This structurally-scoped S4 proof shows how the locality of
   modular assumptions limits the possible interactions in proof.
   Ellipses mark points in sequents where I have suppressed additional
   formula occurrences that no longer contribute to the inference.}
\label{structure-module-fig}
\end{figure}

	The assumptions in this proof---the program
   statements---specify two ambiguities.  Either $A$ or $B$ holds, and
   either $C$ or $D$ holds.  As part of the specification, we use
   modal operators to say how to reason with these ambiguities: we
   have $\Box (A\vee B)$ and $\Box(C \vee D)$.  This means that the
   ambiguities themselves are \emph{generic}; we can use them to
   perform case analysis at any time.  However, when we reason about
   any particular case, we make \emph{local} assumptions---we will
   assume $A$ rather than $\Box A$ for example.
	
	This specification limits the way case analyses in the proof
   interact.  Consider our goal here: $(\Box A) \wedge (\Box C)$.  We
   must prove each conjunct separately, \emph{using generic
   information}; that is, each conjunct is proved in its own new
   nested scope.  Thus, in the proof of
   Figure~\ref{structure-module-fig}, we perform case analysis from
   $\Box(A \vee B)$ within the nested scope for $\Box A$, and perform
   case analysis from $\Box(C \vee D)$ within the nested scope for
   $\Box C$.  Observe that the logic dictates the choice for us.  For
   instance, performing case analysis from $\Box(A \vee B)$ within the
   nested scope for $\Box C$ is useless---the assumption of $A$ and
   $B$ cannot help here.  Importantly, performing case analysis from
   $\Box(A \vee B)$ at the initial, outermost scope is also useless.
   Whatever assumptions we make will have to be discarded when we try
   to prove the conjuncts, and consider only generic information.
   This specification therefore cordons off the two ambiguities from
   one another in this proof problem.  We have to consider the
   ambiguities separately.

	Effectively, it is part of the \emph{meaning} of the
   specification of Figure~\ref{structure-module-fig} that proofs must
   be short.  A proof in this specification must be a combined record
   of independent steps, not an interacting record with combined
   resolutions of ambiguities.  To my knowledge, the possibility for
   this kind of declarative search control in disjunctive modal
   specifications has not received comment previously.  But it seems
   to me to be one of the most exciting and unique uses for modal
   logic in representation and problem-solving.

\subsubsection{Explicit scope and goal-directed search}
\label{explicit-subsubsec}

	The second formulation of modal reasoning is illustrated by
   the sequent figure below:
\[
\derivi{$(\proveslabel\Box)$}
	{\Gamma \proves G^{\mu\alpha}, \Delta}
	{\Gamma \proves \Box G^{\mu}, \Delta}
\]
	Such inferences institute an explicitly-scoped sequent
   calculus; each formula is tagged with a label indicating the modal
   context which it describes.  These labels are sequences of terms,
   each of which corresponds to an inference that changes scope.
   Superscripts are my notation for labels; above, $\mu$ labels the
   scope of the succedent formula $\Box G$.  To reason about a generic
   modal formula, we again introduce a new, nested scope in which just
   generic information is available; we now label the formula with its
   new scope.  Thus above $G$ is labeled $\mu\alpha$; and $\alpha$ is
   subject to an eigenvariable condition---it cannot occur elsewhere
   in the sequent---and so represents a generic possibility.  At
   axioms, the scopes of premises and conclusions must match;
   therefore modal inferences can dispense with destructive
   transformation of sequents.  

	Figure~\ref{explicit-scope-fig} illustrates the other
   explicitly-scoped S4 sequent inferences that I will draw on in this
   motivating discussion.  
\begin{figure}
\[
\begin{array}{c}
\derivi{$(\proveslabel\Box)$}
       {\Gamma \proveslabel G^{\mu\alpha}, \Delta}
       {\Gamma \proveslabel \Box G^{\mu}, \Delta}
\\
\derivi{$(\Box\proveslabel)$}
       {\Gamma, G^{\mu\nu} \proveslabel \Delta}
       {\Gamma, \Box G^{\mu} \proveslabel \Delta}
\\
\derivi{$(\proveslabel\imp)$}
       {\Gamma, P^{\mu} \proveslabel G^{\mu}, \Delta}
       {\Gamma \proveslabel P \imp G^{\mu}, \Delta}
\\
\derivii{$(\imp\proveslabel)$}
       {\Gamma \proveslabel G^{\mu}, \Delta}
       {\Gamma, P^{\mu} \proveslabel \Delta}
       {\Gamma, G \imp P^{\mu} \proveslabel \Delta}
\\
\derivii{$(\vee\proveslabel)$}
       {\Gamma, A^{\mu} \proveslabel \Delta}
       {\Gamma, B^{\mu} \proveslabel \Delta}
       {\Gamma, A \vee B^{\mu} \proveslabel \Delta}
\\
\derivii{$(\proveslabel\wedge)$}
       {\Gamma \proveslabel A^{\mu}, \Delta}
       {\Gamma \proveslabel B^{\mu}, \Delta}
       {\Gamma \proveslabel A \wedge B^{\mu}, \Delta}
\\
\Gamma, A^{\mu} \proveslabel A^{\mu}, \Delta \;\;\; (\m{Axiom})
\end{array}
\]
\caption{Six inference figures and the axiom for explicitly-scoped S4.
   See \cite{fitting:proof,wallen:nonclassical,permute-paper}.  The
   $(\proveslabel\Box)$ inference is subject to an eigenvariable
   condition that $\alpha$ is new.  In the $(\Box\proveslabel)$
   inference, $\mu\nu$ refers to any sequence of terms that extends
   $\mu$ by a suffix $\nu$.}
\label{explicit-scope-fig}
\end{figure}
	Explicitly-scoped proof systems have a long history as
   \emph{prefixed tableaus}; see
   \cite{fitting:proof,wallen:nonclassical} and references therein.
   Each label sequence can be viewed as representing a possible world
   in possible-worlds semantics, so for example the
   $(\proveslabel\Box)$ inference figure represents a transition from
   the world named by $\mu$ to a new world $\mu\alpha$ that represents
   a generic possibility accessible from $\mu$.  The more general
   study of such systems has put them in a new proof-theoretic
   perspective recently.  They are closely related to semantics-based
   translation systems \cite{ohlbach:translation,nonnengart:ijcai} and
   \emph{labelled deductive systems}
   \cite{gabbay:lds,basin/matthews/vigano:jlli98}.  I use the term
   \emph{explicitly-scoped} from \cite{permute-paper} because I
   continue to emphasize the extent to which the two formulations of
   reasoning represent the same inferences, just in different ways.

	The ability to define explicit scope is intimately connected
   with the ability to carry out goal-directed proof.  I adopt the
   perspective due to \cite{mnps:uniform} that goal-directed proof
   simply amounts to a specific strategy for constructing sequent
   calculus deductions.  The strategy is first to apply inferences
   that decompose goals to atoms and then to apply inferences that use
   a specific program statement to match a specific goal.  Proofs that
   respect this strategy are called \emph{uniform}.  On this strategy,
   logical connectives amount to explicit instructions for search, and
   this is in fact what lets us view a logical formula concretely as a
   \emph{program} \cite{mnps:uniform}.

	Unlike other, more procedural characterizations of algorithmic
   proof, such as \cite{gabbay:htcs92}, this view largely abstracts
   away from the exact state of computations during search.  The key
   questions are purely proof-theoretic.  In particular, goal-directed
   proof is possible in a logic if and only if any theorem has a
   uniform proof.  In systems of structural scope, this is not
   possible, and we must instead restrict our to inference in specific
   logical fragments, as described for the intuitionistic case in,
   e.g., \cite{mnps:uniform,harland:jlc93,harland:iccl00}.%
\footnote{
	A further case of structural control of inference that has
   attracted particular interest is linear logic, where linear
   disjunction must be understood to specify synchronization between
   concurrent processes rather than proof by case analysis; see, e.g.,
   \cite{andreoli:focusing,hodas/miller:lolli,pym/harland:jlc94,miller:tcs96,kobayashi:tcs99}.
   The investigation of fragments of linear logic remains essential,
   as linear logic has no analogue of an explicitly-scoped proof
   system, and so---unlike intuitionistic logic and modal logic---must
   be understood as a refinement of classical logic rather than an
   extension to it \cite{girard:lu}.
}

	By contrast, systems of explicit scope can be lifted by a
   suitable analogue to the Herbrand-Skolem-G\"odel theorem for
   classical logic so that \emph{any} pair of unrelated inferences can
   be interchanged
   \cite{kleene:permute,wallen:nonclassical,lincoln/shankar:search,permute-paper}.
   Thus, unlike systems of structural scope, systems of explicit scope
   permit \emph{general} goal-directed reasoning.  If we adopt
   Miller's characterization of uniform proof for sequent calculi with
   multiple conclusions \cite{dale:forum,miller:tcs96}, then any modal
   theorem has a uniform proof in a lifted, explicitly-scoped
   inference system.  Put another way, explicitly-scoped inference
   assimilates modal proof to classical proof, and we know that
   uniformity is not really a restriction on classical proof
   \cite{harland:cl97,nadathur:uniform/classical}.  This is why my
   investigation emphasizes questions of information-flow, such as
   modularity and locality, rather than questions of goal-directed
   proof \emph{per se}.

   	I will refer to the proof of Figure~\ref{explicit-search-fig}
   to illustrate some of the properties of information-flow in
   goal-directed search.
\begin{figure}
\small
\begin{center}
\begin{tabular}{rl}
\fbox{3} &
\derivii{$(\imp\proveslabel)$}
	{\derivii{$(\proveslabel\wedge)$}
	         {\underline{B}, \ldots \proves \ldots, \underline{B}}
		 {\underline{D}, \ldots \proves \ldots, \underline{D}}
		 {B, D, \ldots \proves \underline{B \wedge D}}}
	{\underline{F}, \ldots \proves \ldots, \underline{F}}
	{B, D, \underline{(B\wedge D) \imp F}, \ldots \proves \ldots,
   \underline{F}}
\\[2em]
\fbox{2} &
{\derivii{$(\imp\proveslabel)$}
	 {\derivii{$(\vee\proveslabel)$}
	          {\underline{C}, \ldots \proves \ldots, \underline{C}}
		  {\fbox{3}}
		  {B, \underline{C\vee D}, (B\wedge D) \imp F, \ldots
   \proves \ldots, \underline{C}, F}}
	 {\underline{F}, \ldots \proves \ldots, \underline{F}}
	 {B, C\vee D, \underline{C \imp F}, (B\wedge D) \imp F \proves
   \ldots, \underline{F}}}
\\[2em]
\fbox{1} &
\derivii{$(\imp\proveslabel)$}
	{\derivii{$(\vee\proveslabel)$}
		 {\underline{A}, \ldots \proves \underline{A}, F}
		 {\fbox{2}}
		 {\underline{A \vee B}, C\vee D, C \imp F, (B\wedge D) \imp F \proves \underline{A}, F}}
	{\underline{F}, \ldots \proves \underline{F}}
	{A \vee B, C\vee D, \underline{A \imp F}, C \imp F, (B\wedge D)
   \imp F \proves \underline{F}}
\end{tabular}
\end{center}
\caption{A goal-directed proof in which multiple cases are
   considered.  Each case is displayed in a separate block.}
\label{explicit-search-fig}
\end{figure}
   	The proof establishes the conclusion
\[
	F
\]
	from assumptions
\[
	A \vee B, C \vee D, A \imp F, C \imp F, (B \wedge D) \imp F
\]

	First we must get clear on the reasoning
   Figure~\ref{explicit-search-fig} represents.  The assumptions in
   this proof again specify two ambiguities, $A\vee B$ and $C\vee D$.
   In modal terms, these are local ambiguities that introduce local
   assumptions; but actually, Figure~\ref{explicit-search-fig} uses
   only classical connectives, and this classical reasoning suffices
   for my discussion here.  In Figure~\ref{explicit-search-fig}, the
   two ambiguities interact to require inference for three separate
   cases: one case where $A$ is true, one case where $C$ is true, and
   a final case where $B$ and $D$ are true together.  (These cases are
   laid out separately in Figure~\ref{explicit-search-fig}.)  In
   goal-directed inference, we discover these cases by backward
   chaining from the main goal $F$ through a series of implications:
   $A \imp F$, $C \imp F$, and $(B \wedge D) \imp F$.

   	Now we can describe the structure of the proof of
   Figure~\ref{explicit-search-fig} more precisely.  The inference is
   segmented out into three chunks, one for each case.  The chunks are
   indexed to indicate how they should be assembled into a single
   proof-tree; the chunk indexed \fbox{3} should appear as a subtree
   where the index \fbox{3} is used in chunk \fbox{2}, and that chunk
   should in turn appear as a subtree where the index \fbox{2} is used
   in chunk \fbox{1}.  We could imagine writing out that tree in
   full---on an ample blackboard!  However, the chunks are actually
   natural units of the proof of Figure~\ref{explicit-search-fig};
   they are what Loveland calls \emph{blocks}
   \cite{loveland:nhp,nadathur/loveland:disj}.  In general, a
   \emph{block} of a derivation is a maximal tree of contiguous
   inferences in which the right subtree of any $(\vee\proveslabel)$
   inference in the block is omitted.  (Check this in
   Figure~\ref{explicit-search-fig}.)  Each block presents reasoning
   that describes a single case from the specification.

	Within blocks, we can trace the progress of goal-directed
   reasoning, as follows.  At each step, our attention is directed to
   a distinguished goal formula---the current goal---and at most one
   distinguished program formula---the selected statement.  For
   illustration, these distinguished formulas are underlined in each
   sequent in Figure~\ref{explicit-search-fig}.  Logical operations
   apply only to distinguished formulas; we first decompose the goal
   down to atomic formulas, then select a program formula and reason
   from it to establish the current goal.

	There are two things to notice about this derivation.  First,
   we use a \emph{restart} discipline when handling disjunctive case
   analysis across blocks \cite{loveland:nhp,nadathur/loveland:disj}.
   In each new block, the current goal is reset to the original goal
   $F$ to restart proof search.  It is easy to see that it does not
   suffice, in general, to keep the current goal across blocks; in
   Figure~\ref{explicit-search-fig}, for example, keeping the current
   goal would mean continuing to try to prove $A$ after we turn from
   the case of $A$ to the case of $B$.  The more general restart rule
   is however complete; in fact, the restart rule is a powerful way to
   extend a goal-directed proof system to logics where a single proof
   must sometimes analyze the same goal formula in qualitatively
   different ways
   \cite{gabbay/reyle:n-prolog,gabbay:n-prolog,gabbay:htcs92}.

	Second, note when and how newly-assumed disjuncts are used in
   new blocks.  For example, $B$ is assumed in block \fbox{2} but it
   is not used until block \fbox{3}.  By contrast, $D$ is assumed in
   block \fbox{3} and used immediately there.  Following
   \cite{loveland:nhp}, I will refer to any use of a disjunctive
   premise in the first block of case analysis where it is assumed as
   a \emph{cancellation}; I will also say that the inference that
   introduces the disjunct, and the new block it creates, are
   \emph{canceled}.  The proof of Figure~\ref{explicit-search-fig}
   cannot be recast in terms of canceled inferences using the sequent
   rules of Figure~\ref{explicit-scope-fig}.  Whichever case of $A\vee
   B$ or $C\vee D$ is treated first cannot be canceled; the second
   disjunct of the one must wait to be used until the second disjunct
   of the other is introduced.  This is a gap between Loveland's
   original Near-Horn Prolog interpreter \cite{loveland:nhp}, which
   requires cancellations, and the generalized reformulation in terms
   of sequent calculi given in
   \cite{nadathur/loveland:disj,nadathur:uniform/classical} and
   suggested in Figure~\ref{explicit-search-fig}.  Loveland suggests
   that cancellation is just an optimization, but we shall see that
   modal logic establishes an important proof-theoretic link between
   cancellation and modularity.

\subsubsection{On modular goal-directed proof search}
\label{intro-example-sec}

	As befits alternative proof methods for the same logic,
   structurally-scoped systems and explicitly-scoped systems are very
   close.  In fact, in the case of intuitionistic logic, they define
   not just the same theorems but the same proofs \cite{permute-paper}.
   This correspondence suggests that we use insights about
   information-flow in structurally-scoped proofs---including the
   modularity and locality exhibited by
   Figure~\ref{structure-module-fig}---to restrict goal-directed
   proof-search in explicitly-scoped systems.

	In fact, we know from \cite{permute-paper} that we can
   sometimes enforce a straightforward requirement of locality in
   explicitly-scoped inferences, as follows.  Assume that we have an
   explicitly-scoped proof system for a logic with modularity and
   locality, with an eigenvariable condition on $(\proveslabel\Box)$,
   and we work in a fragment of logic without negation (this again
   includes the S4 translation of intuitionistic formulas).  Then when
   we consider a sequent of this form in proof-search:
\[
	\Gamma \proves \Delta
\]
	we apply inferences to a formula $P$ in $\Gamma$ only when $P$
   is labeled with a prefix of a label of a formula in $\Delta$.  That
   is, we can consider inferences on $P^{\mu} \in \Gamma$ only when
   there is some $G^{\mu\nu} \in \Delta$.  The prefix relationship is
   required for $P$ to eventually contribute to the proof of any
   $\Delta$ formula.  For example here:
\[
	A, B^{\beta}, C^{\gamma}, D^{\delta} \proves  E^{\beta}, F^{\gamma}
\]
	We can consider $A$, $B^{\beta}$ or $C^{\gamma}$, but not
   $D^{\delta}$.

	This invariant is weaker than one might want or expect for
   certain kinds of goal-directed search.  Specifically, we have seen
   that when we use goal-directed search as a model for logic
   programming, we understand the interpreter to be working on at most
   one goal and one program statement at a time.  In this setting, we
   would like to require that the program statement must be able to
   contribute to the current goal.  However, we must understand the
   result of \cite{permute-paper} to require only that the current
   program statement must contribute to \emph{some} goal, not necessarily
   the \emph{current} one.  In addition, we must take into account
   other, inactive goals, such as the goals that we may potentially
   restart later in proof search.  In the preceding example, even if
   $E^{\beta}$ were the current goal, we might have to consider
   reasoning with $C^{\gamma}$ because of the possibility of a restart
   with $F^{\gamma}$.

	For most inferences, we can rule out their contribution to
   inactive goals, on independent grounds.  Most inferences from
   assumptions in goal-directed proofs must contribute to the current
   goal or not at all.  But there is one difficult case, which happens
   also the most meaningful one.  This is the case of disjunction
   itself, where only one disjunct contributes to the current goal.
   The other disjunct may contribute to some other goal; we will set
   up a new proof problem by assuming this disjunct and making some
   inactive goal active.  Modularity and locality suggest that we
   should be able now to select a goal that our newly-assumed disjunct
   could contribute to.  In other words, if the new disjunct is
   $P^{\mu}$ the next goal should take the form $G^{\mu\nu}$ with
   $\mu$ a prefix of $\mu\nu$.  Call this a modular restart.  The
   alternative is that there is no relationship of scope between the
   new disjunct and the next goal.

	Modular restarts would be quite powerful.  For example, they
   would allow us to capture the declarative search control
   illustrated in Figure~\ref{structure-module-fig}.  In an
   explicitly-scoped goal-directed proof corresponding to
   Figure~\ref{structure-module-fig}, the case analysis for
   $\Box(A\vee B)$ will look like this:
\[
\derivi{$\Box\proveslabel$}
{\derivii{$\vee\proveslabel$}
	 {\underline{A}^{\alpha}, \ldots \proves \underline{A}^{\alpha}, \ldots}
	 {B^{\alpha}, \ldots \proves A^{\alpha}, \ldots}
	 {\underline{A\vee B}^{\alpha}, \ldots \proves \underline{A}^{\alpha}, \ldots}}
{\underline{\Box(A \vee B)}, \ldots \proves \underline{A}^{\alpha}, \ldots}
\]
	With modular restarts, we know we must continue to take
   $A^{\alpha}$ as the current goal in the right top ($B^{\alpha}$)
   subproof.  In effect, we know to build a short proof in which
   ambiguities are considered independently.  We can cut down the
   space for proof search accordingly---for example, there will be no
   question of introducing the other ambiguity from $\Box(C\vee D)$ in
   the new modular block.  On the other hand, without modular
   restarts, we are free to reconsider the initial goal $(\Box A)
   \wedge (\Box C)$ at this stage; in subsequent search we will
   reconsider both $\Box(A \vee B)$ and $\Box(C \vee D)$.  Thus, even
   though the logic guarantees that ambiguities do not interact in a
   proof, we still wind up considering interacting ambiguities in
   proof search.

	The main result of this paper is to provide an
   explicitly-scoped goal-directed proof system in which modular
   restarts are complete.  The proof system has modular restarts
   because, in the new proof system, any proof can be presented in
   such a way that all disjunctions are canceled.  Each new disjunct
   $P^{\mu}$ therefore contributes to the proof of the restart goal in
   the current block, and so we know to choose a restart goal
   $G^{\mu\nu}$ that the new disjunct could contribute to.

	It turns out that modular restarts are not automatic; you need
   to design the policy for disjunctive inference to respect it.
   Figure~\ref{explicit-search-fig} already makes the problem clear.
   How can we enforce cancellations here?  The sequent rules seem not
   to allow it.  The new idea is simple actually---to allow a new
   inference figure for disjunction that considers disjuncts out of
   their textual order:
\[
\derivii{$\vee\proveslabel*$}
	{\Gamma, D^{\mu} \proves \Delta}
	{\Gamma, C^{\mu} \proves \Delta}
	{\Gamma, C\vee D^{\mu} \proves \Delta}
\]
	This is the direct analogue of the Near-Horn Prolog inference
   scheme, which can proceed by matching any of the heads of a
   disjunctive clause at any time \cite{loveland:nhp}.  The new
   sequent rule will allow us to reanalyze the constitution of higher
   blocks so that, wherever we use the new disjunct in the original
   proof, we can always reanalyze it as part of the current block.
   Figure~\ref{explicit-modular-fig} demonstrates this reanalysis for
   Figure~\ref{explicit-search-fig}.
\begin{figure}
\small
\begin{center}
\begin{tabular}{rl}
\fbox{$2'$} &
{\derivii{$(\imp\proveslabel)$}
	 {\underline{C}, \ldots \proves \ldots, \underline{C}}
	 {\underline{F}, \ldots \proves \ldots, \underline{F}}
	 {B, C, \underline{C \imp F}, \ldots \proves
   \ldots, \underline{F}}}
\\[2em]
\fbox{$3'$} &
\derivii{$(\imp\proveslabel)$}
	{\derivii{$(\proveslabel\wedge)$}
	         {\underline{B}, \ldots \proves \ldots, \underline{B}}
		 {\derivii{$(\vee\proveslabel*)$}
			  {\underline{D}, \ldots \proves \ldots,
   \underline{D}}
			  {\fbox{$2'$}}
			  {B, \underline{C\vee D}, C\imp F, \ldots \proves \ldots, \underline{D}}}
		 {B, C\vee D, C \imp F, \ldots \proves \underline{B \wedge D}}}
	{\underline{F}, \ldots \proves \ldots, \underline{F}}
	{B, C\vee D, C\imp F, \underline{(B\wedge D) \imp F}, \ldots \proves \ldots,
   \underline{F}}
\\[2em]
\fbox{1} &
\derivii{$(\imp\proveslabel)$}
	{\derivii{$(\vee\proveslabel)$}
		 {\underline{A}, \ldots \proves \underline{A}, F}
		 {\fbox{$3'$}}
		 {\underline{A \vee B}, C\vee D, C \imp F, (B\wedge D) \imp F \proves \underline{A}, F}}
	{\underline{F}, \ldots \proves \underline{F}}
	{A \vee B, C\vee D, \underline{A \imp F}, C \imp F, (B\wedge D)
   \imp F \proves \underline{F}}
\end{tabular}
\end{center}
\caption{A reanalysis of the proof of Figure~\ref{explicit-search-fig}
   to enforce cancellations.  We rewrite block \fbox{$3'$}, in which
   $B$ is canceled, to use the new disjunctive inference figure; block
   \fbox{$3'$} thus becomes the second block after \fbox{1}.  At the
   same time, we introduce a simplified block \fbox{$2'$} which uses
   the assumption $C$, without disjunction at all.}
\label{explicit-modular-fig}
\end{figure}
	In fact, demonstrating the generality of such reanalysis will
   prove to be quite involved.  Explicitly-scoped inferences with an
   eigenvariable condition give blocks in modal proofs an inherently
   hierarchical structure, because of the different modal scopes that
   are introduced and the local assumptions that are made.  Loveland's
   construction for cancellations, by contrast, assumes that the
   structure of blocks is flat.  Instead, we must use the natural
   tools of the sequent calculus to develop suitable constructions for
   reanalyzed inferences.

\subsection{The results and their context}
\label{tech-overview-subsec}

	The problem sketched in Section~\ref{problem-subsec} is a pure
   problem of modal proof.  Accordingly, all the proof systems I
   consider will describe sound and complete inference under the usual
   Kripke semantics for modal logic \cite{kripke:modal,fitting:proof}.
   I will not consider interactions of disjunction with
   negation-by-failure and other operational features of of logic
   programming proof-search systems.  For such issues in disjunctive
   logic programming, see for example
   \cite{lobo/minker/raj:disjunction}.  Nor will I attempt to describe
   a minimal model or fixed-point construction in which exactly the
   consequences of a modal program hold, as in
   \cite{orgun/wadge:jlp92}.

	Moreover, my interest is in specific fragments of specific
   modal logics in particular.  \emph{Modularity} and \emph{locality}
   allow consideration of the logics T, K and K4 in addition to S4,
   but are not compatible with such logics as S5, temporal logics with
   symmetric past and future operators \cite{gabbay:87}, the logic of
   context of \cite{mccarthy/buvac:context} or the modal logic of
   named addresses of \cite{kobayashi:tcs99}.  For example, in S5, if
   $\Box A$ is true at any world, then $\Box A$ is true at all worlds;
   thus the logic prohibits making such an assumption locally.  To see
   the problem, observe, for example, that $\Box(\Box A \imp B) \vee
   \Box A$ is a theorem of S5.  Modal proofs in such cases require
   global restarts \cite{gabbay/olivetti:sl98}.  \emph{Locality}
   further rules out logical fragments with possibility or negation.
   Such fragments can be used to pose goals about that access
   otherwise local assumptions, as in the theorem $\Box(A \imp B) \vee
   \Diamond A$ of all normal modal logics.  (Goal-directed proof of
   this theorem also involves a global restart.)  Moreover, such
   fragments make it more difficult to enforce modularity as well,
   since they do not permit an eigenvariable condition at
   $(\proveslabel\Box)$ inferences in goal-directed proofs.  My
   investigation therefore sticks closely to the treatments of logical
   modularity and locality originally explored in
   \cite{dale:modules,giordano/martelli:structuring}.  Indeed, I
   continue to restrict implications and universal quantifiers in
   goals to strict statements of the form $\Box(P\imp G)$ and $\Box
   \forall x G$.

	The basic strategy that I adopt is to start with a relatively
   straightforward proof system, and gradually to narrow the
   formulation of its inference rules---preserving soundness and
   completeness with respect to the underlying semantics---until we
   have a proof system, SCLP, with the desired characteristics, namely
   goal-directed search and modular restarts.  I have been
   particularly influenced by Lincoln and Shankar's presentation of
   proof-theoretic results in terms of simple transformations among
   successive proof systems \cite{lincoln/shankar:search}; and by
   Andreoli's construction of focusing sequent calculi that embody the
   discipline of goal-directed proof directly in the form of inference
   figures \cite{andreoli:focusing}.

	However, the correct design of the final proof system requires
   a variety of proof-theoretic ideas about logic programming to be
   extended, strengthened, and combined with proof-theoretic results
   about modal logic in a novel way.  To describe logic programming,
   we start with the idea of uniform proof search described in
   \cite{mnps:uniform} and extended to multiple-conclusion calculi in
   \cite{dale:forum}.  To derive a uniform proof system in the
   presence of indefinite information in assumptions, however, we can
   no longer use the familiar quantifier rules used in previous logic
   programming research, which simply introduce fresh parameters; we
   must apply a generalization of Herbrand's Theorem
   \cite{lincoln/shankar:search} and work with quantifier rules that
   introduce structured terms.  The calculus of Herbrand terms, SCL,
   lifts the explicitly-scoped proof systems considered in
   Section~\ref{explicit-subsubsec} and
   \cite{fitting:proof,wallen:nonclassical}.  The key property of SCL
   is that inferences can be freely interchanged.  This allows
   arbitrary proofs to be transformed easily into uniform proofs.

   	The \emph{modular} behavior of this uniform system depends on
   the further proof-theoretic analyses of path-based sequent calculi
   adapted, in part, from \cite{permute-paper}.  These analyses
   establish that path representations enforce modularity and locality
   in the uses of formulas in proofs, even with otherwise classical
   reasoning.  Hence, although path-based calculi obscure the natural
   modularity of modal inference, they do not eliminate it.  I finish
   with a streamlined uniform proof system that takes advantage of
   these results; as a consequence, proof search in this calculus can
   dynamically exploit the local use of modular assumptions.  

   	The justification of this new proof system makes much of a
   strategy originally due to \cite{kleene:permute}, in which the
   inferences in a proof are reordered so as to satisfy a global
   invariant.  The strategy achieves termination despite generous
   copying and deepening of inferences by a judicious choice of
   transformations within a double induction.  In our cases, these
   transformations are guided by the constraints of uniform proof, and
   by the cancellations of disjunctive assumptions that we know we
   must maintain in proofs, to achieve modularity.  This provides an
   analogue of Loveland's transformations on restart proofs
   \cite{loveland:nhp} in the sequent calculus setting.

   	Of course, modal logic is not just a modular logic.  Modal
   logic provides a general, declarative formalism for specifying
   change over time, the knowledge of agents, and other
   special-purpose
   domains~\cite{prior:tense,hintikka:scope,schild:terminology}.
   Goal-directed systems for modal proof are often motivated by such
   specifications
   \cite{cerro:molog,debart:modal/lp,baldoni/giordano/martelli:modal,bgm:framework}.
   In generalizing goal-directed modal proof to indefinite
   specifications, SCLP can play an important role in applying modal
   formalisms to planning, information-gathering and communication
   \cite{aaai98,jlac00}.  Even when content, not modularity, is
   primary, the modular treatment of disjunction limits the size of
   proofs and the kinds of interactions that must be considered in
   proof search.  Such constraints are crucial to the use of logical
   techniques in applications that require automatic assessment of
   incomplete information, such as planning and natural language
   generation.  The interest of these more general applications helps
   explain why I pursue this investigation in the full first-order
   language.

\subsection{Outline}
\label{outline-subsec}

   	The structure of the rest of this paper is as follows.  I
   begin by presenting first-order multi-modal logic in
   Section~\ref{logic-sec}.  I consider syntax
   (Section~\ref{syntax-subsec}), semantics
   (Section~\ref{semantics-subsec}), and finally proof
   (Section~\ref{proof-subsec}); I describe the explicitly-scoped
   Herbrand proof system for modal logic that is my starting point.
   Section~\ref{uniform-proof-subsec} shows that this calculus offers
   a suitable framework for goal-directed proof because uniform proof
   search in this calculus is complete.

   	Section~\ref{module-sec} describes and justifies a modular
   goal-directed proof system, as advertised in
   Section~\ref{problem-subsec}.  I introduce the calculus itself in
   Section~\ref{final-subsec}, along with key definitions and
   examples.  Then in
   Sections~\ref{segment-subsec}--\ref{modularity-subsec} I outline
   how this sequent calculus is derived in stages from the calculus of
   Section~\ref{logic-sec}.  Full details are provided in an appendix.

	Finally, Section~\ref{assessment-sec} offers a broader
   assessment of these results.  I consider some further optimizations
   that the new sequent calculus invites in
   Section~\ref{implementation-subsec}, and briefly conclude in
   Section~\ref{use-subsec} with some applications of
   first-order multi-modal inference that the new sequent calculus
   suggests.

\section{First-order multi-modal deduction}
\label{logic-sec}

   	I begin by supporting the informal presentation of first-order
   multi-modal logic from Section~\ref{intro-sec} more explicitly.  I
   will adopt a number of techniques that are individually quite
   familiar.  I allow an arbitrary number of modal operators and a
   flexible regime for relating different modal operators to one
   another, following many applied investigations
   \cite{debart:modal/lp,baldoni/giordano/martelli:modal,bgm:framework,baldoni/giordano/martelli:ld98}.
   I use prefix terms for worlds and sequent calculus inference,
   following the comprehensive treatment of the first-order modal
   logic using prefix terms and analytic tableaux (or, seen
   upside-down, in the cut-free sequent calculus) of
   \cite{fitting/mendelsohn}.  I factor out reasoning about
   accessibility into side conditions on inference rules, similar to
   the proof-theoretic view of \cite{basin/matthews/vigano:jlli98}, in
   which reasoning about accessibility and boolean reasoning are
   clearly distinguished.  And I use Herbrand terms to reason
   correctly about parameterized instances of formulas, avoiding the
   usual eigenvariable condition on quantifier (and modal) rules, as
   in \cite{lincoln/shankar:search}.

	Though the techniques are routine, the combination is still
   somewhat unusual.  Research in modal logic---whether the
   investigation is more mathematical
   \cite{gore:diss,massacci:1step,massacci:tableau,gore:tableau} or
   primarily concerns algorithms for proof search
   \cite{otten/kreitz,beckert/gore:97,schmidt:uni}---is dominated by
   the study of the propositional logic of a single modal operator (or
   accessibility relation).  Moreover, researchers who have
   investigated modal logic in a first-order setting have tended to
   jump directly into a discussion of particular theorem-proving
   strategies, particularly resolution
   \cite{jackson:ijcai,wallen:nonclassical,catach:tableau,frisch/scherl:constraint,auffray:equations,nonnengart:ijcai,ohlbach:multimodal}.

\subsection{Syntax}
\label{syntax-subsec}

   	Our language depends on a \emph{signature} including a
   suitable set of atomic constants $C$ (and suitable predicate
   symbols and modalities).  We then consider program statements of
   the syntactic category $D(C)$ and goals of the category $G(C)$
   defined recursively as in \sref{dialup.syntax}; we refer to the
   union of these two languages as $L(C)$.  \sref{dialup.syntax} makes
   explicit the conditions observed in
   Section~\ref{tech-overview-subsec}; there is no possibility or
   negation, and universal and hypothetical goals must be modal.
\sentence{dialup.syntax}{
$\begin{array}[t]{l@{\;::=\;}l}
G & A \; | \; \op{m}G \; | \; G \wedge G \; | \; G \vee G \; |
   \; \op{m}(\forall x G) \; | \; \exists x G \; | \;
   \op{m}(D\imp G) \\
D & A \; | \; \op{m}D \; | \; D \wedge D \; | \; D \vee D
   \; | \; \forall x D \; | \; \exists x D \; | \; G \imp
   D
\end{array}$
}
	In \sref{dialup.syntax}, $A$ schematizes an atomic formula;
   atomic formulas take the form $p_i(a_1,\ldots,a_k)$ where $p_i$ is
   a predicate symbol of arity $k$ and each $a_i$ is either a variable
   or an atomic constant in the set $C$.  We assume some initial
   non-empty set of constants $CONST$.  But it will be convenient to
   consider languages in which a countably infinite number of
   \emph{parameters} are included in the language to supplement the
   symbols in CONST.

	In \sref{dialup.syntax}, \op{m} schematizes a modal operator
   of necessity; intuitively, such modal operators allow a
   specification to manipulate constrained sources of information.
   That is, a program statement $\op{m}D$ explicitly indicates that
   $D$ holds in the constrained source of information associated with
   the operator $\op{m}$.  Conversely, a goal $\op{m}G$ can be
   satisfied only when $G$ is established by using information
   from the constrained source associated with $\op{m}$.  

	We will work in a multi-modal logic, in which any finite
   number $m$ of distinct necessity operators or \emph{modalities} may
   be admitted.  (Necessity operators will also be written as $\Box$
   or $\Box_i$.)  In addition to ordinary program statements, a
   specification may contain any of the following axiom schemes
   describing the modalities to be used in program statements and
   goals:
\sentence{dialup.axioms}{
$\begin{array}[t]{ll}
   \Box_i p \imp p \;\; &
	\mbox{\emph{veridicality}} \; \verax \\
   \Box_i p \imp \Box_i \Box_i p \;\; &
	\mbox{\emph{positive introspection}} \; \piax \\
   \Box_i p \imp \Box_j p \;\; &
	\mbox{\emph{inclusion}} \; \incax \\
\end{array}$
}
	These axioms describe the nature of the information that an
   operator provides, and spell out relationships among the different
   sources of information in a specification.  \verax\ is needed for
   information that correctly reflects the world; \piax, for
   information that provides a complete picture of how things might
   be; and \incax, for a source of information, \m j, that elaborates
   on information from another source, \m i.  Because we use this
   explicit axiomatization, we can take the names of the modal
   operators as arbitrary.

   	We appeal to the usual notions of \emph{free} and \emph{bound}
   occurrences of variables in formulas; we likewise invoke the
   \emph{depth} of a formula---the largest number of nested logical
   connectives in it.

\subsection{Semantics}
\label{semantics-subsec}

	As is standard, we describe the models for the modal language
   in two steps.  The first step is to set up \emph{frames} that
   describe the structure of any model; a full model can then be
   obtained by combining a frame with a way of assigning
   interpretations to formulas in a frame.

\begin{definition}[Frame]
\label{frame-def}
	A \emph{frame} consists of a tuple $\langle \worlds, \access,
   \domain\rangle$ where: \worlds\ is a non-empty set of
   \emph{possible worlds}; $\access$ names a family of $m$ binary
   \emph{accessibility} relations on \worlds, a relation $\access_i$
   for each modality $i$; and \domain\ is a \emph{domain function}
   mapping members of \worlds\ to non-empty sets.
\end{definition}

	Within the frame ${\cal F}$, the function \domain\ induces a
   set $\domain({\cal F})$, called the \emph{domain of the frame}, as
   $\cup \{ \domain(w) \; | \; w \in \worlds \}$.  In order to
   simplify the treatment of constant symbols, it is also convenient
   to define a set of objects that all the domains of the different
   possible worlds have in common, the \emph{common domain of the
   frame} ${\cal F}$: $\cdomain({\cal F}) = \cap \{ \domain(w) \; | \;
   w \in \worlds \}$.  We effectively insist that $\cdomain({\cal F})$
   be non-empty as well, since CONST is non-empty and each symbol in
   CONST must be interpreted by an element of $\cdomain({\cal F})$.

	The intermediate level of frames is useful in characterizing
   the meanings of modal operators and modal quantification.  In
   particular, simply by putting constraints on $\access_i$ or on
   \domain\ at the level of frames, we can obtain representative
   classes of models in which certain general patterns of inference
   are validated.  The constraints we will avail ourselves of are
   introduced in Definition~\ref{r-prop-def}.

\begin{definition} 
\label{r-prop-def}
	Let $\langle \worlds, \access, \domain\rangle$ be a
   \emph{frame}.  We say the frame is:
\begin{itemize}
\item
	\emph{reflexive} at $i$ if $w \access_i w'$ for every $w \in
   \worlds$;
\item
	\emph{transitive} at $i$ if, for any $w,w'' \in \worlds$, $w
   \access_i w''$ whenever there is a $w' \in \worlds$ such that $w
   \access_i w'$ and $w' \access_i w''$;
\item
	\emph{narrowing} from $i$ to $j$ if $w \access_j w'$ implies
   $w \access_i w'$ for all $w, w' \in \worlds$;
\item
	\emph{increasing domain} if for all $w, w' \in \worlds$,
   $\domain(w) \subseteq \domain(w')$ whenever there is some
   accessibility relationship $w \access_i w'$.
\end{itemize}
\end{definition}

   	Our scheme for using the constraints of
   Definition~\ref{r-prop-def} depends on establishing a regime for
   the $m$ modalities in the language, describing the inferences that
   should relate them.  The regime is defined as follows.
\begin{definition}[Regime]
	A \emph{regime} is a tuple $\langle \types, \inclusions,
   \dconstraint\rangle$, where: \types\ is a function mapping each
   modality into one of the symbols K, K4, T and S4; $\inclusions$ is
   a (strict) partial order on the modalities; and $\dconstraint$ is
   the symbol \emph{increasing}.
\end{definition}
	The reader will recognize the symbols in the image of \types\
   as the classic names for modal logics of a single modality.  $S4$
   is for modalities that are subject to \piax\ and \verax.  $T$ is
   for modalities that are subject just to \verax.  $K4$ is for
   modalities that are subject just to \piax.  $K$ is modalities
   subject to neither axiom.  The interactions specified by \incax\
   are determined by the partial order on modalities: $j \leq i$ when
   $\Box_i p \imp \Box_j p$.  This meaning for these symbols can be
   enforced by considering only frames that \emph{respect} the given
   \emph{regime}.
\begin{definition}[Respect]
	Let ${\cal F} = \langle\worlds, \access, \domain\rangle$ be a
   frame, and let ${\cal S} = \langle \types, \inclusions,
   \dconstraint\rangle$ be a regime.  We say ${\cal F}$
   \emph{respects} ${\cal S}$ whenever the following conditions are
   met for all modalities $i$ and $j$:
\begin{itemize}
\item
	If $\types(i)$ is T or S4 then $\access_i$ is
   reflexive.
\item
	If $\types(i)$ is K4 or S4 then
   $\access_i$ is transitive.
\item
	If $j \leq i$ according to \inclusions\ then ${\cal F}$ is
   narrowing from $i$ to $j$.
\item
	If \dconstraint\ is \emph{increasing}, then ${\cal
   F}$ is increasing domain.
\end{itemize}
\end{definition}

   	From now on, we assume that some regime ${\cal S} =
   \langle\types, \inclusions,\dconstraint\rangle$ is fixed, and
   restrict our attention to frames that respect ${\cal S}$.
   Informally, now, a model consists of a frame together with an
   interpretation. 
\begin{definition}[Interpretation]
	\valuation\ is an \emph{interpretation} in a frame $\langle
   \worlds, \access, \domain \rangle$ if \valuation\ satisfies these
   two conditions:
\begin{enumerate}
\item
	\valuation\ assigns to each $n$-place relation symbol $p_i$ and
   each possible world $w \in \worlds$ some $n$-ary relation on the
   domain of the frame $\domain({\cal F})$.  
\item
	\valuation\ assigns to each constant symbol $c$ some element
   of the common domain of the frame $\cdomain({\cal F})$.
\end{enumerate}
\end{definition}
	Thus we can define a \emph{model} over ${\cal S}$ thus:

\begin{definition}[Model]
	A first-order $k$-\emph{modal model} over a regime ${\cal S}$
   is a tuple $\langle \worlds, \access, \domain, \valuation \rangle$
   where $\langle\worlds,\access, \domain \rangle$ is a frame that
   respects ${\cal S}$ and $\valuation$ is an interpretation in
   $\langle\worlds,\access,\domain\rangle$.
\end{definition}

   	To define truth in a model, we need the usual notion of
   assignments and variants:
\begin{definition}[Assignment]
	Let ${\cal M} = \langle\worlds, \access, \domain,
   \valuation\rangle$ be a model (that respects the regime ${\cal
   S}$).  An \emph{assignment} in ${\cal M}$ is a mapping $g$ that
   assigns to each variable $x$ some member $g(x)$ of the domain of
   the frame of the model ${\cal
   D}(\langle\worlds,\access,\domain\rangle)$.
\end{definition}
	In proofs, we interpret formulas not just in the ordinary
   language $L(C)$ with a given set of modalities, relations,
   constants and variables, but in an expanded language \mbox{$L(C\cup
   P)$} which also includes a set $P$ of first-order
   \emph{parameters}; we will want to use the same models for this
   interpretation.  Over $L(C\cup P)$, we suppose that an assignment
   in ${\cal M}$ also assigns some member $g(p)$ of the domain of the
   frame of ${\cal M}$ to each parameter $p$ in $P$.
\begin{definition}[Variants]
	Let $g$ and $g'$ be two assignments in a model ${\cal M} =
   \langle\worlds,\access,\domain,\valuation\rangle$; $g'$ is an
   \emph{$x$-variant of $g$ at a world $w\in\worlds$} if $g$ and $g'$
   agree on all variables except possibly for $x$ and $g'(x) \in
   \domain(w)$.
\end{definition}

\begin{definition}[Truth in a model]
	Let ${\cal M} =
   \langle\worlds,\access,\domain,\valuation\rangle$ be a model.  Then
   the formula $A$ is \emph{true at world $w$ of model ${\cal M}$ on
   assignment $g$}---written ${\cal M}, w \forces_g A$---just in case
   the clause below selected by syntactic structure of $A$ is
   satisfied:
\begin{itemize}
\item
	$A$ is $p_i(t_1,\ldots,t_n)$: Then ${\cal M}, w \forces_g A$
   just in case $\langle e_1,\ldots e_n \rangle \in
   \valuation(p_i,w)$, where for each $t_i$, $e_i$ is
   $\valuation(t_i)$ if $t_i$ is a constant and $g(t_i)$ otherwise.
\item	
	$A$ is $B_1 \wedge B_2$: Then ${\cal M}, w \forces_g A$ just
   in case both ${\cal M}, w \forces_g B_1$ and ${\cal M}, w \forces_g
   B_2$.
\item	
	$A$ is $B_1 \vee B_2$: Then ${\cal M}, w \forces_g A$ just in
   case either ${\cal M}, w \forces_g B_1$ or ${\cal M}, w \forces_g
   B_1$.
\item
	$A$ is $\Box_i B$: Then ${\cal M}, w \forces_g A$ just in case
   for every $w' \in \worlds$, if $w \access_i w'$ then ${\cal M}, w'
   \forces_g B$.
\item
	$A$ is $\forall x B$: Then ${\cal M}, w \forces_g A$ just in
   case for every $x$-variant $g'$ of $g$ at $w$, ${\cal M}, w
   \forces_{g'} B$.
\item
	$A$ is $\exists x B$: Then ${\cal M}, w \forces_g A$ just in
   case there is some $x$-variant $g'$ of $g$ at $w$ with ${\cal
   M}, w \forces_{g'} B$.
\end{itemize}
\end{definition}

   	By a \emph{sentence} we mean a formula of $L($CONST$)$ in
   which no variables occur free.  For any sentence $A$, model ${\cal
   M}$ and world $w$ of ${\cal M}$, a simple induction on depth
   guarantees that ${\cal M}, w \forces_g A$ for some assignment $g$
   in ${\cal M}$ exactly when ${\cal M}, w \forces_g A$ for all
   assignments $g$ in ${\cal M}$.  In this case, we can write simply
   ${\cal M}, w \forces A$ and say that $A$ is \emph{true in ${\cal
   M}$ at $w$}.

\begin{definition}[Valid] \label{valid-def}
	Let $A$ be a sentence and ${\cal M} = \langle \worlds,
   \access, \domain, \valuation \rangle$ be a model.  $A$ is
   \emph{valid in ${\cal M}$} if for every world $w \in \worlds$,
   ${\cal M}, w \forces A$.  $A$ is \emph{valid} (on the regime
   $\langle\types, \inclusions,\dconstraint\rangle$) if $A$ is valid
   in any model ${\cal M}$ that respects the regime.
\end{definition}

\subsection{Proof theory}
\label{proof-subsec}

	We now present our basic deductive system---a cut-free
   path-based sequent calculus for multi-modal deduction which uses
   Herbrand terms to reason correctly about parameterized instances of
   formulas.  Since this calculus represents our basic \emph{lifted
   sequent calculus} for modal logic, we refer to it as SCL here.  Our
   starting point is Theorem~\ref{sem-thrm} that SCL provides a sound
   and complete characterization of valid formulas.

	SCL has the advantage that inferences can be freely
   interchanged, allowing arbitrary proofs to be transformed easily
   into goal-directed proofs; we show in Theorem~\ref{eager-thrm},
   presented in Section~\ref{uniform-proof-subsec}, how to obtain
   goal-directed proofs in this calculus.  The very same flexibility
   of inference, however, means that this calculus neither respects
   nor represents the potential of modal inference to give proofs an
   explicitly modular structure.

	The basic constituent in SCL is a \emph{tracked, prefixed
   formula}.  The formulas extend the basic languages $D(C)$ and
   $G(C)$ of definitions and goals defined in \sref{dialup.syntax} by
   allowing additional terms---representing arbitrary witnesses of
   first order quantifiers, and arbitrary transitions of modal
   accessibility among possible worlds---to be introduced into
   formulas for the purposes of proof.  We begin by assuming two
   countable sets of symbols: a set $H$ of \emph{first-order Herbrand
   functions} and $\Upsilon$ of \emph{modal Herbrand functions}.  We
   can now define sets $P_H$ of \emph{first-order Herbrand terms},
   $\kappa_{\Upsilon}$ of \emph{modal Herbrand terms}, and
   ${\Pi(\kappa_{\Upsilon})}$ of \emph{Herbrand prefixes} by mutual
   recursion:
\begin{definition}[Herbrand terms and prefixes]
	Assume that $t_0$ is a Herbrand prefix and let
   $t_1,\ldots,t_n$ be a sequence (possibly empty), where each $t_i$
   is either an element of $C$, a first-order Herbrand term, or a
   Herbrand prefix.  Then if $h$ is a first-order Herbrand function
   then $h(t_0,t_1,\ldots,t_n)$ is a \emph{first-order Herbrand term}.
   If $\eta$ is a modal Herbrand function then
   $\eta(t_0,t_1,\ldots,t_n)$ is a \emph{modal Herbrand term}.  A
   \emph{Herbrand prefix} is any finite sequence of modal Herbrand
   terms.
\end{definition}
	The rationale behind the use of a Herbrand term $h(X)$ at an
   existential inference $R$ goes like this.  At existential
   inferences, we need to reason about a generic individual.  We need
   to have a suitable representation for a generic individual for $R$.
   Regardless of the order in which inferences are applied in a
   sequent deduction, there will be some parameters that must occur in
   the sequent where $R$ applies.  For example, some parameters must
   appear here as a result of the instantiations that must take place
   in deriving the formula to which $R$ applies.  We must be sure that
   the individual we introduce for $R$ is different from all these
   parameters.  The terms $X$ which are supplied as an argument to the
   Herbrand term $h(X)$ identify these parameters indirectly.  The
   structure $h(X)$ therefore serves as a placeholder for a new
   parameter that could be chosen to be different from each of the
   terms in $X$.  The structure $h(X)$ thus packs all the information
   required to allow the inferences in the proof to be reordered and
   an appropriate parameter chosen so that the inference at $R$ is
   truly generic.

	In modal deduction, of course, we need generic individuals at
   modal inferences as well as existential ones.  Modal Herbrand
   inference therefore requires that we introduce Herbrand terms to
   describe transitions among possible worlds and Herbrand prefixes to
   name possible worlds, in addition to introducing first-order
   Herbrand terms to represent first-order parameters.  In this case,
   the arguments $X$ to Herbrand terms must mix first-order Herbrand
   terms and Herbrand prefixes, since logical formulas can encode
   dependencies among first-order and modal parameters.

	A \emph{prefixed formula} is now an expression of the form
   $A^{\mu}$ with $A$ a formula and $\mu$ a Herbrand prefix---we use
   $D(C\cup P_H)^{\Pi(\kappa_{\Upsilon})}$ and $G(C\cup
   P_H)^{\Pi(\kappa_{\Upsilon})}$ to refer to prefixed definitions and
   prefixed goals.  For Herbrand calculi, formulas must also be
   \emph{tracked} to indicate the sequence of instantiations that has
   taken place in the derivation of the formula.
\begin{definition}[Tracked expressions]
	If $E$ denotes the expressions of some class, then the
   \emph{tracked expressions} of that class are expressions of the
   form $e_I$ where $e$ is an expression of $E$ and $I$ is a
   finite sequence (possibly empty) of elements of $C\cup P_H \cup
   \Pi(\kappa_{\Upsilon})$.
\end{definition}
	We say that a tracked expression $e_I$ \emph{tracks} a term
   $t$ just in case $t$ occurs as a subterm of some term in $I$.

	In order to reason correctly about multiple modal operators,
   we need to keep track of the kinds of accessibility that any modal
   transition represents.  To endow the system with correct
   first-order reasoning on increasing domains, we also need to keep
   track of the worlds where first-order terms are introduced.  We
   use the following notation to record these judgments: $\mu/\nu:i$
   indicates that world $\nu$ is accessible from world $\mu$ by
   the accessibility relation for modality $i$; and $t:\mu$ indicates
   that the entity associated with term $t$ exists at world $\mu$.

	It is convenient to keep track of this information by
   anticipating the restricted reasoning required for our fragment
   $L(C)$ and exploiting the structure of Herbrand terms, as follows.
   It is clear that there are countably many first-order Herbrand
   terms, Herbrand prefixes, and formulas in \mbox{$L(C\cup P_H)$}.
   We can therefore describe a correspondence as follows.  If $A$ is a
   formula of the form $\forall x B$ or $\exists x B$ and $u$ is a
   natural number, we define a corresponding first-order Herbrand
   function $h^u_A$ so that each first-order Herbrand function is
   $h_A$ for some $A$ and no first-order Herbrand function is $h^u_A$
   and $h^v_B$ for distinct $A$ and $B$ or distinct $u$ and $v$.
   Likewise, if $A$ is a formula of the form $\Box_i B$ and $u$ is a
   natural number, we define a corresponding modal Herbrand function
   $\eta^u_A$ so that each modal Herbrand function is $\eta^u_A$ for
   some $A$ and no modal Herbrand function is $\eta^u_A$ and
   $\eta^v_B$ for distinct $A$ and $B$ or distinct $u$ and $v$.
   (Indexing Herbrand functions by natural numbers means that adapting
   a Herbrand proof to respect an eigenvariable condition can be as
   simple as reindexing its Herbrand functions.)  Now we have:
\begin{definition}[Herbrand typings]
	A \emph{Herbrand typing for the language} $L(C\cup P_H)$
   (under a correspondence as just described) is a set $\Xi$ of
   statements, each of which takes one of two forms:
\begin{enumerate}
\item
	$\mu/\mu \eta:i$ where: $\mu$ is a Herbrand prefix and $\eta$ is a
   modal Herbrand term of the form $\eta^u_A(\mu,\ldots)$ and $A$ is
   $\Box_i B$.

\item
	$t: \mu$ where $t$ is a first-order Herbrand term of the form
   $h(\mu,\ldots)$.
\end{enumerate}
	A sequence of modal and first-order Herbrand terms $X$
   determines a Herbrand typing $\Xi_X$, consisting of the
   appropriate $\mu/\mu\eta:i$ for each modal Herbrand term $\eta$
   that occurs in $X$ (possibly as a subterm) and the appropriate
   $h:\mu$ for each first-order Herbrand term $h$ that occurs in $X$
   (possibly as a subterm).  
\end{definition}
\begin{definition}[Typings] \label{transition-def}
	Suppose that $\Xi$ is a Herbrand typing over a language $L(C
   \cup P)^{\Pi(\kappa)}$, and that ${\cal S} = \langle \types,
   \inclusions, \m{increasing} \rangle$ is a modal regime.  We define
   the relation that $E$ is a \emph{derived typing} from $\Xi$ with
   respect to ${\cal S}$, written ${\cal S}, \Xi \triangleright E$, as
   the smallest relation satisfying the following conditions:
\begin{itemize}
\item
	$(K)$.  ${\cal S}, \Xi \triangleright \mu/\nu:i$ if
   $\mu/\nu:i \in \Xi$.
\item
	$(T)$.  ${\cal S}, \Xi \triangleright \mu/\mu:i$ if
   $\types(i)$ is T or S4, and $\mu$ occurs in $\Xi$.
\item
	$(4)$.  ${\cal S}, \Xi \triangleright \mu/\nu:i$ if
   $\mu/\mu':i \in \Xi$, ${\cal S}, \Xi \triangleright
   \mu'/\nu:i$, and $\types(i)$ is K4 or S4.
\item
	$(\m{Inc})$.  ${\cal S}, \Xi \triangleright \mu/\nu:i$ if
   ${\cal S}, \Xi \triangleright \mu/\nu:j$ and $j \leq i$
   according to \inclusions.
\item
	$(V)$.  ${\cal S}, \Xi \triangleright t:\mu$ if $t:\mu \in
   \Xi$.
\item
	$(I)$.  ${\cal S}, \Xi \triangleright t:\nu$ if ${\cal S},
   \Xi \triangleright \mu/\nu:i$ for some $i$ and ${\cal S}, \Xi
   \triangleright t:\mu$.
\end{itemize}
\end{definition}
	Inspection of these rules shows that ${\cal S},\Xi
   \triangleright \mu/\nu:i$ only if $\nu$ and $\mu$ occur in $\Xi$.
   Moreover, given these rules, an easy induction on the length of typing
   derivations gives that ${\cal S},\Xi \triangleright \mu/\nu:i$ only
   if $\nu = \mu \nu'$ for some prefix $\nu'$.  Thus, suppose that ${\cal
   S},\Xi \triangleright \mu/\nu:i$ for some Herbrand typing $\Xi$:
   each step in the derivation must concern some prefix of $\nu$ and thus
   ${\cal S},\Xi_{\nu} \triangleright \mu/\nu:i$.  These invariants permit
   some simplifications in reasoning in the fragment $L(C\cup P)$ over more
   expressive modal regimes containing other modal operators and other uses
   of connectives.  

	These definitions allow us to describe the modal Herbrand
   sequent calculus precisely.  This calculus, SCL, is given in
   Definition~\ref{seq-calc-def-1}.  Note that for this fragment of
   modal logic, it suffices to consider sequents of the form $\Delta
   \proves \Gamma$, where $\Delta$ is a multiset of prefixed
   definitions (from $D(C\cup P_H)^{\Pi(\kappa_{\Upsilon})}$), and
   $\Gamma$ is a multiset of prefixed goals (from $G(C\cup
   P_H)^{\Pi(\kappa_{\Upsilon})}$).  Note also that ${\cal S},\Xi
   \triangleright \mu/\nu:i$ only if $\nu$ is of the form $\mu\nu'$.
\begin{definition}[Herbrand sequent calculus] \label{seq-calc-def-1}
	For basic first-order multi-modal Herbrand deductions in our
   fragment over a regime ${\cal S}$, we will use the sequent rules
   defined here, which comprise the system SCL.  The rules consist of
   an axiom rule and recursive rules---each recursive rule relates a
   \emph{base} sequent below to one or more \emph{spur} sequents
   above; it applies to the base in virtue of an occurrence of a
   distinguished tracked, prefixed formula in the sequent; we refer to
   this as the \emph{principal expression} or simply the
   \emph{principal} of the inference.  Similarly, each of the sequent
   rules introduces new expressions onto each spur, which we refer to
   as the \emph{side expressions} of the rule.  We will also refer to
   the two named expression occurrences at axioms as the
   \emph{principal expressions} or \emph{principals} of the axiom.
   Now we have:
\begin{enumerate}
\item
	axiom---$A$ atomic:
\[
       {\Delta,  A^{\mu}_X \proves \Gamma, A^{\mu}_Y}
\]
\item
	conjunctive:   
\[
\derivi{$(\wedge\proveslabel)$}
       {\Delta,  A\wedge B^{\mu}_X,
		     A^{\mu}_X, B^{\mu}_X \proves \Gamma}
       {\Delta, A\wedge B^{\mu}_X \proves \Gamma}
\]
\[
\derivi{$(\proveslabel\vee)$}
       {\Delta \proves \Gamma, A\vee B^{\mu}_X,
		      A^{\mu}_X, B^{\mu}_X}
       {\Delta \proves \Gamma, A\vee B^{\mu}_X}
\]
\[
\derivi{$(\proveslabel\imp)$}
       {\Delta, A^{\mu}_X \proves \Gamma, A\imp B^{\mu}_X,
		      B^{\mu}_X}
       {\Delta \proves \Gamma, A\imp B^{\mu}_X}
\]
\item
	disjunctive:
\[
\derivii{$(\proveslabel\wedge)$}
       {\Delta \proves \Gamma,  A\wedge B^{\mu}_X,
			    A^{\mu}_X}
       {\Delta \proves \Gamma,  A\wedge B^{\mu}_X, 
			   B^{\mu}_X}
       {\Delta \proves \Gamma, A\wedge B^{\mu}_X}
\]
\[
\derivii{$(\vee\proveslabel)$}
       {\Delta, A\vee B^{\mu}_X,
			   A^{\mu}_X \proves \Gamma}
       {\Delta, A\vee B^{\mu}_X, 
			   B^{\mu}_X \proves \Gamma}
       {\Delta, A\vee B^{\mu}_X \proves \Gamma}
\]
\[
\derivii{$(\imp\proveslabel)$}
       {\Delta, A\imp B^{\mu}_X \proves 
			   A^{\mu}_X, \Gamma}
       {\Delta, A\imp B^{\mu}_X, 
			   B^{\mu}_X \proves \Gamma}
       {\Delta, A\imp B^{\mu}_X \proves \Gamma}
\]
\item
	possibility---where $\eta$ is $\eta^u_{\Box_i A}(\mu,X)$ for
   some $u$:
\[
\derivi{$(\proveslabel\Box)$}
       {\Delta \proves \Gamma, \Box_i A^{\mu}_X, A^{\mu\eta}_{X,\mu\eta}}
       {\Delta \proves \Gamma, \Box_i A^{\mu}_X}
\]
\item
	necessity---subject to the side condition ${\cal S}, \Xi_{\nu}
   \triangleright \mu/\mu\nu:i$:
\[
\derivi{$(\Box\proveslabel)$}
       {\Delta, \Box_i A^{\mu}_X,
		 A^{\mu\nu}_{X,\mu\nu} \proves \Gamma}
       {\Delta, \Box_i A^{\mu}_X \proves \Gamma}
\]
\item
	existential---subject to the side condition that $h$ is
   $h^u_B(\mu,X)$ for $B^{\mu}_X$ the principal of the rule (either
   $\exists x A$ or $\forall x A$) and some $u$:
\[
\derivi{$(\exists\proveslabel)$}
	{\Delta, 
	  \exists x A^{\mu}, A[h/x]^{\mu}_{X,h} \proves \Gamma}
	{\Delta, \exists x A^{\mu}_X \proves \Gamma}
\;\;\;\;\;\;\;\;\;\;\;\;\;\;
\derivi{$(\proveslabel\forall)$}
	{\Delta \proves \Gamma, 
	  \forall x A^{\mu}_X, A[h/x]^{\mu}_{X,h}}
	{\Delta \proves \Gamma, \forall x A^{\mu}_X}
\]
\item
	universal---subject to the side condition ${\cal S},
   \Xi_{t,\mu} \triangleright t:\mu$:
\[
\derivi{$(\forall\proveslabel)$}
	{\Delta, \forall x A^{\mu}_X, A[t/x]^{\mu}_{X,t}
	 \proves \Gamma}
	{\Delta, \forall x A^{\mu}_X \proves \Gamma}
\;\;\;\;\;\;\;\;\;\;\;\;\;\;
\derivi{$(\proveslabel\exists)$}
	{\Delta \proves \Gamma, 
	  \exists x A^{\mu}_X, A[t/x]^{\mu}_{X,t}}
	{\Delta \proves \Gamma, \exists x A^{\mu}_{X}}
\]
\end{enumerate}
\end{definition}
	A ${\cal S}$-proof or ${\cal S}$-derivation for a sequent $\Delta
   \proves \Gamma$ is a tree built by application of these inference
   figures (in such a way that any side conditions are met for regime
   ${\cal S}$), with instances of the axiom as leaves and with the sequent
   $\Delta \proves \Gamma$ at the root.  A tree similarly constructed
   except for containing some arbitrary sequent \m S as a leaf is a
   \emph{derivation from} \m S.  

	I state the correctness theorem for this proof theory in a way
   that highlights the continuity with previous work on modal logic,
   particularly \cite{fitting:proof}.
\begin{thrm}[Soundness and Completeness]\label{sem-thrm}
	Suppose there is an ${\cal S}$-proof for a sequent $\proves
   A$.  Then $A$ is valid.  Conversely, if there is no ${\cal
   S}$-proof for the sequent $\proves A$ then there is a model ${\cal
   M}$ (that respects ${\cal S}$) and world $w$ such that ${\cal M}, w
   \not \forces A$.
\end{thrm}
	I merely sketch a proof here, which involves simply applying
   the standard techniques of
   \cite{fitting:proof,lincoln/shankar:search}.  It is convenient to
   prove an intermediate result, using slightly modified sequent
   calculus SCE which imposes an eigenvariable condition on the
   possibility and existential rules---$u$ must be new.  We can show
   the soundness of SCE by adapting the arguments presented in
   \cite[2.3]{fitting:proof} and \cite[5.3]{fitting/mendelsohn}.
   Meanwhile, we can follow \cite{fitting:proof} in developing the
   completeness argument in terms of \emph{analytic consistency
   properties} for the modal language.  This argument can be seen as a
   formalization of the motivation for sequent calculi in the
   systematic search for models.  Now, modal formulas may be satisfied
   only in infinite models, so the completeness theorem effectively
   requires us to consider infinite sequences of applications of
   sequent rules.  In moving to infinite sets in this way, we must
   formally move from deductions, viewed as syntactic objects, to a
   more abstract, algebraic characterization of sets of modal
   formulas.

	We can now establish the correctness of SCL by syntactic
   methods, which relate SCL proofs to SCE proofs.  Suppose $\Gamma$
   and $\Delta$ contains sentences of $L($CONST$)$ (labeled with the
   empty prefix).  Completeness is immediate: if there is an SCE proof
   for $\Gamma \proves \Delta$, that very proof is also an SCL proof
   of $\Gamma \proves \Delta$.  Conversely, the soundness theorem says
   that if there is an SCL proof of $\Gamma \proves \Delta$, then
   there is an SCE proof for $\Gamma \proves \Delta$.  We establish
   this by simply adapting the general Herbrand theorem of
   \cite{lincoln/shankar:search} to SCE.  The idea behind the
   soundness result is that the structure of Herbrand terms provides
   enough information to reconfigure an SCL proof (by an inductive
   process of interchanges of inference, like that considered next in
   Section~\ref{uniform-proof-subsec}) so that equivalents of the
   eigenvariable conditions are enforced.  The SCL proof may then be
   reindexed to respect SCE's eigenvariable requirements.  $\qed$

\subsection{Permutability of inference and uniform proofs}
\label{uniform-proof-subsec}

	Our syntactic methods for reasoning about derivations exploit
   \emph{permutability of inference}---the general ability to
   transform derivations so that inferences are interchanged
   \cite{kleene:permute}.  To develop the notion of permutability of
   inference, we need to make some observations about the SCL sequent
   rules.  First, the reasoning that is performed in subderivations is
   reasoning about subformulas (and vice versa).  That is, in any spur
   sequent, the occurrence of the principal expression and the side
   expression all correspond to---or as we shall say, \emph{are based
   in}---the occurrence of the principal expression in the base
   sequent.  Likewise, each of the remaining expressions in the spur
   \emph{are based in} an occurrence of an identical expression in the
   base.  Here, as in \cite{kleene:permute}, we are assuming an
   \emph{analysis} of each inference to specify this correspondence in
   the case where the same expression has multiple occurrences in the
   base or in a spur.  Thus, our proof techniques, where they involve
   copying derivations, sometimes involve (implicit) reanalyses of
   inferences.
   
	Now, in any derivation, the spur of one inference serves as the
   base for an \emph{adjacent} inference or an axiom.  We can therefore
   associate any tracked prefixed formula occurrence $E$ in any sequent in
   the derivation with the occurrence in the root (or \emph{end-sequent})
   which $E$ is based in.  A similar notion can relate inferences, as
   follows.  Suppose $O$ is the inference at the root of a (sub)derivation,
   and $L$ is another inference in the (sub)derivation.  Then $L$ \emph{is
   based in} an expression $E$ in the spur of $O$ if the principal
   expression of $L$ is based in $E$; $L$ \emph{is based in} $O$ itself if
   $E$ is a side expression of $O$.  An important special case is that of
   an axiom based in an inference $O$.  In effect, such an axiom marks a
   contribution that inference $O$ contributes to completing the deduction.

	To define interchanges of inference, we appeal to the two
   basic operations of \emph{contraction} and \emph{weakening}, which
   we cast as transformations on proofs.  (In other proof systems,
   contraction and weakening may be introduced as explicit
   \emph{structural rules}.)
\begin{lemma}[Weakening]
	Let ${\cal D}$ be an SCL proof, let $\Delta_0$ be a finite
   multiset of tracked prefixed definitions and let $\Gamma_0$ be a
   finite multiset of tracked prefixed goals (in the same language as
   ${\cal D}$).  Denote by $\Delta_0+{\cal D}+\Gamma_0$ a derivation
   exactly like ${\cal D}$, except that where any node in ${\cal D}$
   carries $\Delta \proves \Gamma$, the corresponding node in
   $\Delta_0+{\cal D}+\Gamma_0$ carries $\Delta,\Delta_0 \proves
   \Gamma, \Gamma_0$.  (When $\Delta_0$ or $\Gamma_0$ is empty, we
   drop the corresponding $+$ from the notation.)  Then
   $\Delta_0+{\cal D}+\Gamma_0$ is also an SCL proof.
\end{lemma}
\begin{lemma}[Contraction] 
	Let ${\cal D}$ be an SCL proof whose root carries $\Delta \proves
   \Gamma, E, E$.  Then we can construct an SCL proof ${\cal D}'$ whose
   root carries $\Delta \proves \Gamma, E$, whose height is at most the
   height of ${\cal D}$ and where there is a one-to-one correspondence
   (also preserving order of inferences) that takes any inference of ${\cal
   D}'$ to an inference with the same principal and side expressions in
   ${\cal D}$.  We can likewise transform an SCL proof ${\cal D}$ whose
   root carries $\Delta, E, E \proves \Gamma$ into an SCL proof ${\cal D}'$
   whose root carries $\Delta, E \proves \Gamma$.
\end{lemma}
	These lemmas follow from straightforward induction on the
   structure of derivations.  These consequences continue to hold,
   suitably adapted, for the intermediate proof systems that we will
   construct from SCL in later sections.

	Now consider two adjacent inferences in a derivation, a base
   inference $R$ and an inference $S$ (whose base is a spur of $R$).
   If $S$ is not based in $R$, we may replace the derivation rooted at
   the base of $R$ by a new derivation of the same end-sequent in
   which $S$ applies at the root, $R$ applies adjacent, and the
   remaining subderivations are copied from the original derivation
   (but possibly weakened to reflect the availability of additional
   logical premises).  Performing such a replacement constitutes an
   interchange of rules $R$ and $S$ and demonstrates the permutability
   of $R$ over $S$; see \cite{kleene:permute}.  SCL is formulated so
   that any such pair of inferences may be exchanged in this way.

	We also observe that we can correctly introduce an
   abbreviation for goal occurrences of \mbox{$\Box_i(A\imp B)$} by a
   single formula $(\m A >_i \m B)$ and the consolidation of
   corresponding inferences $(\proveslabel\Box_i)$ and
   $(\proveslabel\imp)$ into a single figure $(\proveslabel>_i)$,
   while retaining unrestricted interchange of inference.  Again when
   the inference applies to principal $A^{\mu}_X$, the figure is
   formulated using $\eta$ for $\eta^u_A(\mu,X)$ as:
\[
\derivi{$\proveslabel>_i$}
       {\Gamma, \m A^{\mu\eta}_{X,\mu\eta} \proves  
	\m B^{\mu\eta}_{X,\mu\eta}, \m A>_i\m B^{\mu}_X, \Delta}
       {\Gamma \proves \m A >_i \m B^{\mu}_X, \Delta}
\]
	We will refer to the calculus using $(\proveslabel>_i)$ in
   place of $(\proveslabel\Box_i)$ and $(\proveslabel\imp)$ as SCLI,
   and consider SCLI in the sequel.

	\cite{dale:forum,miller:tcs96} uses Definition
   \ref{uniform-def} to characterize \emph{abstract logic programming
   languages}.
\begin{definition}  \label{uniform-def}
	A cut-free sequent proof ${\cal D}$ is {\em uniform} if for for
   every subproof ${\cal D}'$ of ${\cal D}$ and for every non-atomic
   formula occurrence \m B in the right-hand side of the end-sequent of
   ${\cal D}'$ there is a proof ${\cal D}''$ that is equal to ${\cal D}'$
   up to a permutation of inferences and is such that the base inference in
   ${\cal D}''$ introduces the top-level logical connective of \m B.
\end{definition}   	
\begin{definition} \label{alp-def}
	A logic with a sequent calculus proof system is an {\em
   abstract logic programming language} if restricting to uniform
   proofs does not lose completeness.
\end{definition}
	It is easy to show that the sequent calculi SCL and SCLI are
   abstract logic programming languages in this sense.  In fact, by
   this definition \emph{every} SCL or SCLI derivation is uniform.

	To anticipate our analysis of permutability in later sections,
   let us introduce the notion of an \emph{eager} derivation in SCL
   or SCLI.  
\begin{definition} \label{delay-def}
   	Consider a derivation ${\cal D}$ containing a right inference
   \m R that applies to principal $E$.  \m R is {\em delayed} exactly
   when there is a subderivation ${\cal D}'$ of ${\cal D}$ where:
   ${\cal D}'$ contains R; ${\cal D}'$ has a left inference $L$ at the
   root; and the principal $E$ of $R$ is based in an occurrence of $E$
   in the end-sequent of ${\cal D}'$.
\end{definition}
	Consider this schematic diagram of such a subderivation ${\cal
   D}'$:
\[
\begin{derive}{
\vdots 
}
\derivitem{$R$}{
\begin{array}{c}\ldots E \ldots \\
\downarrow
\end{array}}
\derivitem{$L$}{
\ldots E \ldots
}
\end{derive}
\]
	On an intuitive conception of a sequent proof as a record of
   proof search constructed from root upwards, \m R is delayed in that
   we have waited in ${\cal D}$ to apply \m R until after consulting
   the program by applying \m L, when we might have applied \m R
   earlier.  Thus, we will also say in the circumstances of
   Definition~\ref{delay-def} that $R$ is delayed \emph{with respect
   to} $L$.
\begin{definition} \label{eager-def}
	${\cal D}$ is {\em eager} exactly when it contains no delayed
   applications of right rules.
\end{definition}
	By transforming any derivation ${\cal D}$ into an eager
   derivation ${\cal D}'$ by permutations of inferences, we make it
   clear that reasoning about goals can always precede reasoning with
   program statements, and thereby provide a starting point for
   further analysis of goal-directed proof search.

\begin{thrm} \label{eager-thrm}
	Any SCL(I) derivation ${\cal D}$ is equal to an eager derivation
   ${\cal D}'$ up to permutations of inferences.
\end{thrm}
	The {\bf proof} follows \cite[Theorem 2]{kleene:permute}.  A
   double induction transforms each derivation into an eager one; the
   inner induction rectifies the final rule of a derivation whose
   subderivations are eager by an interchange of inferences (and
   induction) \cite[Lemma 10]{kleene:permute}; the outer one rectifies
   a derivation by rectifying the furthest violation from the root
   (and induction).  See Appendix~\ref{eager-proof-sec}.  $\qed$

\section{Modular goal-directed proof search}
\label{module-sec}   

\subsection{Overview}
\label{final-subsec}

	Eager derivations do not make a satisfactory specification for
   goal-directed proof in a logic programming sense, because they do
   not embody a particularly directed search strategy.  For one thing,
   eager derivations are free to work in parallel on different
   disjuncts of a goal using different program statements; in logic
   programming we want \emph{segments} in which a single program
   statement and a single goal is in force.  Moreover, eager derivations
   can reuse work across separate case analyses; in logic programming
   we want \emph{blocks} where particular cases are investigated
   separately.  Finally, because of their classical formulation, eager
   derivations do not enforce or exploit any modularity in their
   underlying logic.  Our task is to remedy these faults of eager
   derivations.

	Our result takes the form of an alternative sequent calculus
   SCLP which is equivalent to SCL.  SCLP enforces a strictly
   goal-directed proof search through the structure of its inferences.
   First, SCLP sequents take the form 
\[
	\Gamma; U \proves V; \Delta
\]
	We understand $\Gamma$ to specify the global program and
   $\Delta$ to specify the global restart goals; both are multisets of
   tracked, prefixed formulas.  $U$ is at most one tracked, prefix
   formula, representing the current program statement; $V$ is at most
   one tracked, prefixed formula, representing the current goal.
   
	Logical rules apply only to the current program statement and
   the current goal.  In addition, if there is a current program
   statement $U$ then the current goal $V$ must be an atomic formula.
   Thus, the interpreter first breaks the goal down into its
   components.  Once an atomic goal is derived, the program is
   consulted; the selected program statement is decomposed and matched
   against the current goal by applicable logical rules.  The form of
   the $(\imp\proveslabel)$ figure ensures that the interpreter
   continues to work on at most one goal at any time; this gives SCLP
   proofs their segment structure.  Meanwhile, the form of the
   $(\vee\proveslabel)$ figures specify no current goal in its second
   case.  The new current goal can then be chosen flexibly from
   possible restart goals.  This gives SCLP proofs their block
   structure.

	The new inferences are presented in
   Definition~\ref{lp-seq-fig-a} and~\ref{lp-seq-fig-b}.
   Definition~\ref{lp-seq-fig-a} shows the rules for decomposing
   program statements; Definition~\ref{lp-seq-fig-b} shows the rules
   for decomposing goals.
\begin{definition}[Logic programming calculus---programs]  \label{lp-seq-fig-a}
	The following inference figures describe the logic programming
   sequent calculus SCLP as it applies to program statements.
\begin{enumerate}
\item
	axiom---$A$ atomic:
\[
	\Gamma ; \m A^{\nu} \proves \m A^{\nu} ; \Delta 
\]
\item
	decision (program consultation)---again $A$ atomic:
\[
\derivi{decide}
       {\Gamma, \m P^{\mu}_X ; \m P^{\mu}_X
	\proves \m A^{\nu}_Y ; \Delta}
       {\Gamma, \m P^{\mu}_X ;
	\proves \m A^{\nu}_Y ; \Delta}
\]
\item
	conjunctive:
\[
\derivi{$\wedge\proveslabel_{\s L}$}
       {\Gamma ; \m P^{\mu}_X \proves
	\m A^{\nu}_Y ; \Delta}
       {\Gamma ; \m P\wedge\m Q^{\mu}_X \proves
	\m A^{\nu}_Y ; \Delta}
\]
\[
\derivi{$\wedge\proveslabel_{\s R}$}
       {\Gamma ; \m Q^{\mu}_X \proves
	\m A^{\nu}_Y ; \Delta}
       {\Gamma ; \m P\wedge\m Q^{\mu}_X \proves
	\m A^{\nu}_Y ; \Delta}
\]
\item
	disjunctive:
\[
\derivii{$\vee\proveslabel_{\s L}$}
	{\Gamma ; \m P^{\mu}_X \proves \m A^{\nu}_Y ; \Delta}
	{\Gamma, \m Q^{\mu}_X ; \proves ; \Delta}
	{\Gamma ; \m P \vee \m Q^{\mu}_X \proves \m A^{\nu}_Y ; \Delta}
\]
\[
\derivii{$\vee\proveslabel_{\s R}$}
	{\Gamma ; \m Q^{\mu}_X \proves \m A^{\nu}_Y ; \Delta}
	{\Gamma, \m P^{\mu}_X ; \proves ; \Delta}
	{\Gamma ; \m P \vee \m Q^{\mu}_X \proves \m A^{\nu}_Y ; \Delta}
\]
\item
   	implication:
\[
\derivii{$\imp\proveslabel$}
	{\Gamma ; \proves \m Q^{\mu}_X ; \Delta }
	{\Gamma ; \m P^{\mu}_X \proves 
	 \m A^{\nu}_Y ; \Delta}
	{\Gamma ; \m Q \imp \m P^{\mu}_X \proves
	 \m A^{\nu}_Y ; \Delta}
\]
\item
	necessity---subject to the side condition that there is a
   typing derivation ${\cal S}, \Xi_{\nu} \triangleright
   \mu/\mu\nu:i$:
\[ 
\derivi{$\Box_{\s i}\proveslabel$}
       {\Gamma ; \m P^{\mu\nu}_{X,\mu\nu}
	\proves \m A^{\nu'}_Y ; \Delta}
       {\Gamma, \Box_{\s i} \m P^{\mu}_X
	\proves \m A^{\nu'}_Y ; \Delta}
\]
\item
	existential---subject to the side condition that $h$ is
   $h^u_{\exists x P}(\mu,X)$ for some $u$:
\[
\derivi{$\exists\proveslabel$}
       {\Gamma ; \m P[\m h/\m x])^{\mu}_{X,h}
	\proves \m A^{\nu}_Y ; \Delta}
       {\Gamma ; \exists \m x. \m P^{\mu}_X
	\proves \m A^{\nu}_Y ; \Delta}
\]
\item
	universal---subject to the side condition that there is a
   typing derivation ${\cal S}, \Xi_{t,\mu} \triangleright t:\mu$:
\[
\derivi{$\forall\proveslabel$}
       {\Gamma ; \m P[\m t/\m x]^{\mu}_{X,t}
	\proves \m A^{\nu}_Y ; \Delta}
       {\Gamma ; \forall \m x. \m P^{\mu}_X
	\proves \m A^{\nu}_Y ; \Delta}
\]
\end{enumerate}
\end{definition}

\begin{definition}[Logic programming calculus---goals]  \label{lp-seq-fig-b}
	The following inference figures describe the logic programming
   sequent calculus SCLP as it applies to goals.
\begin{enumerate}
\item
	restart:
\[
\derivi{restart}
       {\Gamma ; 
	\proves \m G^{\nu}_X ; \m G^{\nu}_X, \Delta}
       {\Gamma ;
	\proves ; \m G^{\nu}_X, \Delta}
\]
\item
	conjunctive goals:
\[
\derivii{$\proveslabel\wedge$}
	{\Gamma ; \proves \m F^{\mu}_X ; \Delta}
	{\Gamma ; \proves \m G^{\mu}_X ; \Delta}
	{\Gamma ; \proves \m F \wedge \m G^{\mu}_X ; \Delta}
\]
\item
	disjunctive goals:
\[
\derivi{$\proveslabel\vee_{\s L}$}
       {\Gamma ; \proves \m F^{\mu}_X ; \Delta}
       {\Gamma ; \proves \m F \vee \m G^{\mu}_X ; \Delta}
\]
\[
\derivi{$\proveslabel\vee_{\s R}$}
       {\Gamma ; \proves \m G^{\mu}_X ; \Delta}
       {\Gamma ; \proves \m F \vee \m G^{\mu}_X ; \Delta}
\]
\item
	necessary goals---where $\eta$ is $\eta^u_A(\mu,X)$ for
   $A^{\mu}_X$ the principal of the rule and for some $u$ for which
   $\eta^u_A$ does not occur in $\Delta$ or $\Gamma$:
\[
\derivi{$\proveslabel\Box_{\s i}\imp$}
       {\Gamma, \m F^{\mu\eta}_{X,\mu\eta} ; \proves  
	\m G^{\mu\eta}_{X,\mu\eta} ; \m G^{\mu\eta}_{X,\mu\eta}, \Delta}
       {\Gamma ; \proves \m F \bimp_i \m G^{\mu}_X ; \Delta}
\]
\[
\derivi{$\proveslabel\Box_{\s i}$}
       {\Gamma ; \proves 
	\m G^{\mu\eta}_{X,\eta} ; \m G^{\mu\eta}_{X,\mu\eta}, \Delta }
       {\Gamma ; \proves 
	\Box_{\s i} \m G^{\mu}_X ; \Delta} 
\]
\item
	universal goals---subject to the side condition that $h$ is
   $h^u_{\forall x G}(\mu,X)$ for some $u$:
\[
\derivi{$\proveslabel\forall$}
       {\Gamma ; \proves \m G[\m h/\m x]^{\mu}_{X,h} ; \Delta}
       {\Gamma ; \proves 
	\forall \m x. \m G^{\mu}_X ; \Delta}
\]
\item
	existential goals---subject to the side condition that there is a
   typing derivation ${\cal S}, \Xi_{t,\mu} \triangleright t:\mu$:
\[
\derivi{$\proveslabel\exists$}
       {\Gamma ; \proves 
	\m G[\m t/\m x]^{\mu}_{X,t} ; \Delta}
       {\Gamma ; \proves 
	\exists \m x. \m G^{\mu}_X ; \Delta}
\]
\end{enumerate}
\end{definition}
   
	Inspection of the figures of Definitions~\ref{lp-seq-fig-a}
   and~\ref{lp-seq-fig-b} reveals the following generalization of
   modularity and locality: in any derivation, the label of the
   current program statement must be a prefix of the label of the
   current goal.  Moreover, goal labels are always extended with novel
   symbols, because of the eigenvariable condition in the
   $(\proveslabel\Box)$ figure.  Inductively, these facts determine a
   strong invariant---consider a block beginning with a restart
   inference whose spur is
\[
	\Gamma; \proves G^{\nu}_X ; \Delta
\]
	and consider any expression $P^{\mu}_Y$ in $\Gamma$.  If $\mu$
   is not a prefix of $\nu$, then $\mu$ will not be a prefix of the
   label of any goal formula in the block.  Thus $P^{\mu}_Y$ cannot be
   used in the block.  (Compare \cite[Lemma 2]{permute-paper}.)

	This is why the (restart) rule of SCLP can be made modular, so
   that it limits the work that is reanalyzed to the scope of the
   ambiguity just introduced.  We must simply show that the new
   disjunct will contribute to its restart goal.  In particular,
   define canceled blocks as in Definition~\ref{canceled-def}.
\begin{definition}[Linked]  \label{link-def}
	An expression $E$ in a sequent in an SCLU derivation ${\cal
   D}$ is \emph{linked} if the principal formula of an axiom in the
   same block of ${\cal D}$ as that sequent is based in $E$.  An
   inference \m R is \emph{linked} in ${\cal D}$ if some side
   expression of $R$ is linked in each spur of $R$.  A derivation or
   block is \emph{linked} iff all of the inferences in it are linked.
\end{definition}
\begin{definition}[Canceled]\label{canceled-def}
   	A block is \emph{canceled} if it contains the root of ${\cal
   D}$, or if the side expression $E$ of the $(\vee\proveslabel)$
   inference whose spur is the root of the block is linked.
\end{definition}
	Thus a canceled block includes a use of any disjunctive case
   introduced in the block.  The key fact about SCLP is that it
   suffices to consider only canceled blocks in proof search.
\begin{thrm}
\label{lp-sequent-thrm}
	Let $\Gamma$ and $\Delta$ be multisets of tracked prefixed
   expression in which each formula is tracked by the empty set and
   prefixed by the empty prefix.  There is a proof of $\Gamma \proves
   \Delta$ in SCL exactly when there is a proof of $\Gamma; \proves
   ;\Delta$ in SCLP in which every block is canceled.
\end{thrm}
	The discussion of the following subsections represents an
   outline of the proof of this result.  The strategy is to transform
   eager proofs from SCL to SCLP by a series of refinements of sequent
   rules that make the logic programming strategy explicit.  We give
   force to the idea that the interpreter has a current goal and
   current program statement, in Section~\ref{segment-subsec}.  Then
   we create blocks for case analysis, in Section~\ref{block-subsec}.
   Finally, we enforce modularity, in Section~\ref{modularity-subsec}.
   See also Appendix~\ref{lp-proof-sec}.

   	Figure~\ref{sclp-fig} shows how the proof of
   Figure~\ref{explicit-modular-fig} is recast in SCLP.
\begin{figure}
\small
\begin{center}
\begin{tabular}{rl}
\fbox{$2'$} &
{\derivi{(restart)}
	{\derivi{(decide)}
	 {\derivii{$(\imp\proveslabel)$}
	  {\derivi{(decide)}
		  {\ldots; C \proves C; F}
		  {\ldots, C; \proves C; F}}
	  {\ldots; F \proves F; F}
	  { C, \ldots; C \imp F \proves F;F}}
	 { C \imp F, C \ldots; \proves F;F}}
	{A \vee B, C\vee D, A \imp F, C \imp F, (B\wedge D)
   \imp F, B, C; \proves ;F}}
\\[2em]
\fbox{$3'$} &
\derivi{(restart)}
       {\derivi{(decide)}
	{\derivii{$(\imp\proveslabel)$}
	   {\derivii{$(\proveslabel\wedge)$}
	         {\derivi{(decide)}
			 {\ldots; B \proves B; F}
			 {B, \ldots; \proves B; F}}
		 {\derivi{(decide)}
			 {\derivii{$(\vee\proveslabel_R)$}
			  { \ldots; D \proves D; F}
			  {\fbox{$2'$}}
			 { \ldots; C \vee D \proves  D;F}}
		    {C\vee D, \ldots; \proves  D;F}}
		 {C\vee D, B, \ldots; \proves  B\wedge D;F}}
	    {\ldots; F \proves F; F}
	   {C\vee D, B, \ldots; (B\wedge D) \imp F \proves F;F}}
	 {C\vee D, (B\wedge D) \imp F, B, \ldots; \proves F;F}}
	{A \vee B, C\vee D, A \imp F, C \imp F, (B\wedge D)
   \imp F, B; \proves ;F}
\\[2em]
\fbox{1} &
\derivi{(restart)}
       {\derivi{(decide)}
	       {\derivii{$(\imp\proveslabel)$}
	         {\derivi{(decide)}
		   {\derivii{$(\vee\proveslabel_L)$}
		    {\ldots; A \proves A; F}
		    {\fbox{$3'$}}
		    {\ldots; A \vee B \proves A; F}}
		 {A \vee B, \ldots; \proves  A;F}}
	       {\ldots; F \proves F; F}
	       {A \vee B, \ldots; A \imp F \proves  F;F}}
	   {A \vee B, A \imp F, \ldots; \proves  F;F}}
	{A \vee B, C\vee D, A \imp F, C \imp F, (B\wedge D)
   \imp F; \proves ;F}
\end{tabular}
\end{center}
\caption{The SCLP presentation of the proof of
   Figure~\ref{explicit-modular-fig}.}
\label{sclp-fig}
\end{figure}
   	Figure~\ref{sclp-fig} extends
   Figure~\ref{explicit-modular-fig} to make the bookkeeping of
   goal-directed proof explicit.  In Figure~\ref{sclp-fig}, the
   informal underline of Figure~\ref{explicit-modular-fig} is gone,
   and instead the current goal and the current program statement are
   displayed at distinguished positions in sequents.  New (restart)
   and (decide) inferences mark the consideration of new goals and new
   program statements.  Of course, the logical content of the two
   inferences is identical.  Applying Definition~\ref{canceled-def},
   block \fbox{1} is canceled because it contains the root; there is
   no new disjunct to discharge here.  Block \fbox{$3'$} is canceled:
   the inference whose spur is the root of block \fbox{$3'$} is the
   $(\vee\proveslabel_L)$ and its side expression is an occurrence of
   $B$, the new disjunct in the block.  This occurrence is linked in
   the block because of the leftmost axiom $\ldots;B\proves B;F$ which
   is based in it; the inference $(\vee\proveslabel_L)$ is linked in
   the block for the same reason.  Similarly block \fbox{$2'$} is
   canceled because the new disjunct $C$ (the side expression of the
   $(\vee\proveslabel_R)$ inference whose spur is the root of block
   \fbox{$2'$}) contributes to the leftmost axiom $\ldots;C\proves
   C;F$ in the block.

	Figure~\ref{sclp-fig2} shows how the proof of
   Figure~\ref{structure-module-fig} is recast in SCLP.
\begin{figure}
\small
\begin{center}
\begin{tabular}{rl}
\fbox{5} &
\derivi{(restart)}
       {\derivi{(decide)}
	       {\derivi{$(\Box\proveslabel)$}
		       {\derivii{$(\imp\proveslabel)$}
				{\derivi{(decide)}
					{\ldots; B^{\alpha} \proves
   B^{\alpha}; \ldots}
					{\ldots, B^{\alpha} ; \proves
   B^{\alpha}; \ldots}}
				{\ldots; A^{\alpha} \proves
   A^{\alpha}; \ldots}
				{B^{\alpha}, \ldots; B \imp A^{\alpha} \proves
   A^{\alpha} ; \ldots}}
		       {B^{\alpha}, \ldots; \Box (B \imp A) \proves
   A^{\alpha} ; \ldots}}
	       {\Box (B \imp A), B^{\alpha} ; \proves A^{\alpha} ; \ldots}}
       {\Box (A \vee B), \Box (B \imp A), \Box (C \vee D), \Box (D
   \imp C), B^{\alpha}; \proves ; A^{\alpha}, (\Box A) \wedge (\Box C)}
\\[2em]
\fbox{6} &
\derivi{(restart)}
       {\derivi{(decide)}
	       {\derivi{$(\Box\proveslabel)$}
		       {\derivii{$(\imp\proveslabel)$}
				{\derivi{(decide)}
					{\ldots; D^{\beta} \proves
   D^{\beta}; \ldots}
					{\ldots, D^{\beta} ; \proves
   D^{\beta}; \ldots}}
				{\ldots; C^{\beta} \proves
   C^{\beta}; \ldots}
				{D^{\beta}, \ldots; D \imp C^{\beta} \proves
   C^{\beta} ; \ldots}}
		       {D^{\beta}, \ldots; \Box (D \imp C) \proves
   C^{\beta} ; \ldots}}
	       {\Box (D \imp C), D^{\beta} ; \proves C^{\beta} ; \ldots}}
       {\Box (A \vee B), \Box (B \imp A), \Box (C \vee D), \Box (D
   \imp C), D^{\beta}; \proves ; C^{\beta}, (\Box A) \wedge (\Box C)}
\\[2em]
\fbox{4} &
\derivi{(restart)}
	{\derivii{$(\proveslabel\wedge)$}
         {\derivi{$(\proveslabel\Box)$}
	        {\derivi{(decide)}
		    {\derivi{$(\Box\proveslabel)$}
		        {\derivii{$(\vee\proveslabel_L)$}
   			         {\ldots; A^{\alpha} \proves A^{\alpha};
   \ldots}
				 {\fbox{5}}
				 {\ldots; A \vee B^{\alpha} \proves
   A^{\alpha}; \ldots}}
			{\ldots; \Box(A \vee B) \proves A^{\alpha}; \ldots}}
   			{\Box (A \vee B), \ldots ;
   \proves A^{\alpha}; \ldots}}
        {\Box (A \vee B), \ldots ; \proves \Box A; \ldots}}
         {\derivi{$(\proveslabel\Box)$}
	        {\derivi{(decide)}
		    {\derivi{$(\Box\proveslabel)$}
		        {\derivii{$(\vee\proveslabel_L)$}
   			         {\ldots; C^{\beta} \proves C^{\beta};
   \ldots}
				 {\fbox{6}}
				 {\ldots; C \vee D^{\beta} \proves
   C^{\beta}; \ldots}}
			{\ldots; \Box(C \vee D) \proves C^{\beta}; \ldots}}
   			{\Box (C \vee D), \ldots ;
   \proves C^{\beta}; \ldots}}
        {\Box (C \vee D), \ldots ; \proves \Box C; \ldots}}
        {\Box (A \vee B), \Box (C \vee D), \ldots ; \proves 
   (\Box A) \wedge (\Box C); \ldots}}
        {\Box (A \vee B), \Box (B \imp A), \Box (C \vee D), \Box (D
   \imp C) ; \proves ; (\Box A) \wedge (\Box C)}
\end{tabular}
\end{center}
\caption{The SCLP presentation of the proof of
   Figure~\ref{structure-module-fig}.  We suppress tracking of
   formulas and hide the internal structure of Herbrand terms.}
\label{sclp-fig2}
\end{figure}
   	The most dramatic change here is that the inferences of
   Figure~\ref{sclp-fig2} are segmented out into three blocks.
   Another change is the discipline of explicit scope; we introduce a
   suitable term $\alpha$ to represent the generic context in which we
   prove $\Box A$ and another suitable term $\beta$ to represent the
   generic context in which we prove $\Box C$.  Correspondingly, we
   transition to $\alpha$ in using $\Box(A\vee B)$ and transition to
   $\beta$ in using $\Box(C\vee D)$.  In the (restarts) of \fbox{5}
   and \fbox{6} the changes interact.  In \fbox{5} we pick the modular
   restart $A^{\alpha}$ in order to permit a contribution by the new
   assumption $B^{\alpha}$.  In \fbox{6} we pick the modular restart
   $C^{\beta}$ in order to permit a contribution by the new assumption
   $D^{\beta}$. 

\subsection{Segment structure}
\label{segment-subsec}

	Our first task is to formalize goal-directed search that
   directs attention to a single goal at a time.  To distinguish such
   goals, we begin with a trick that for now is purely
   formal---introducing an \emph{articulated} SCLI.  We represent
   assumptions as a pair $\Pi ; \Gamma$ with $\Pi$ encoding the global
   program and $\Gamma$ encoding local program statements; eventually
   local statements will be processed only in the current segment and
   then discarded.  (Compare the similar notation and treatment from
   \cite{girard:lu}.)  Similarly, we represent goals as a pair $\Delta
   ; \Theta$, with $\Theta$ encoding the restart goals and $\Delta$
   encoding the local goals; ultimately, we will also describe
   inference rules which will discard $\Delta$ between segments.  With
   this representation, principal formulas of logical rules are local
   formulas, in $\Gamma$ or $\Delta$; so are the side formulas---with
   these exceptions: the $(\proveslabel\Box)$ and
   $(\proveslabel\bimp)$ rules augment $\Pi$ instead of $\Gamma$ (when
   they add a new program statement) and $\Theta$ instead of $\Delta$
   (when they add new restart goals).

	New (decide) and (restart) rules keep this change general;
   they allow a global formula---a program statement or restart goal---to
   be selected and added to the local state.
\[
\begin{array}{l@{\hspace{1in}}l@{\hspace{1in}}}
\derivi{(decide)}
       {\Pi, \m A^{\mu}_X ; \Gamma, \m A^{\mu}_X
	\proves \Delta ; \Theta}
       {\Pi, \m A^{\mu}_X ; \Gamma
	\proves \Delta ; \Theta} 
&
\derivi{(restart)}
       {\Pi ; \Gamma
	\proves \Delta, \m G^{\mu}_X ; \Theta, \m G^{\mu}_X}
       {\Pi ; \Gamma
	\proves \Delta ; \Theta, \m G^{\mu}_X} 
\end{array} 
\]
\begin{lemma}[Articulation] \label{articulation-lemma}
	Every SCLI deduction can be converted into an articulated SCLI
   deduction with an end-sequent of the form $\Pi; \proves ;\Theta$ in such
   a way that if the initial derivation is eager then so is the resulting
   derivation (and vice versa).
\end{lemma}
	{\bf Proof}.  Straightforward structural induction.  $\qed$

   	The next step is to introduce an inference figure
   $(\imp\proveslabel^S)$ that imposes a \emph{segment} structure on
   derivations, thus:
\[
\derivii{$(\imp\proveslabel^S)$}
	{\Pi ; 
	 \proves \m A^{\mu}_X, \Delta ; \Theta}	
	{\Pi ; 
	 \Gamma, \m A \imp \m B^{\mu}_X, \m B^{\mu}_X
	 \proves \Delta ; \Theta}	
	{\Pi ; \Gamma, 
	 \m A \imp \m B^{\mu}_X
	 \proves \Delta ; \Theta}	
\]
\begin{definition}[Segment]
	A \emph{segment} in a derivation ${\cal D}$ is a maximal tree
   of contiguous inferences in which the left subtree of any
   $(\imp\proveslabel^S)$ inference is omitted.
\end{definition}
	The distinctive feature of the $(\imp\proveslabel^S)$ figure
   is that the local results inferred from the program are discarded
   in the subderivation where the new goal is introduced.  In an eager
   derivation, this will begin a new segment where first the new goal
   will be considered and then a new program statement will be
   selected to establish that goal.

	We will define two calculi using $(\imp\proveslabel^S)$.  The
   first, SCLS, eliminates the $(\imp\proveslabel)$ inference of the
   articulated SCLI and instead has $(\imp\proveslabel^S)$.  The
   second, SCLV, is a calculus like the articulated SCLI but also
   allows $(\imp\proveslabel^S)$; $(\imp\proveslabel)$ and
   $(\imp\proveslabel^S)$ can appear anywhere in an SCLV derivation.
   We introduce SCLV to facilitate the incremental transformation of
   articulated SCLI proofs into SCLS proofs.  

\begin{lemma} \label{segment-lemma}
	An eager articulated SCLI derivation whose end-sequent is of the
   form
\[
	\Pi ; \proveslabel \Delta ; \Theta
\]
	can be transformed to an eager SCLS derivation of the same
   end-sequent.
\end{lemma}

	\textbf{Proof.}  We proceed with an inductive construction
   that eliminates $(\imp\proveslabel)$ inferences in favor of
   $(\imp\proveslabel^S)$ inferences one at a time.  See
   Appendix~\ref{segment-proof-subsec}.  $\qed$

\subsection{Block structure}
\label{block-subsec}

	We now revise how we perform case analysis from assumptions.
   We introduce new rules where all local work is discarded in the
   subderivation written on the right.  This corresponds to a sequent
   of the form $\Pi; \proves ;\Theta$.  In addition, some
   \emph{global} work may be discarded in the right subderivation;
   this helps clarify the structure of derivations.  Accordingly,
   there may be additional formula occurrences $\Pi'$ and $\Theta'$ in
   the base sequent that are not copied up to the right subderivation.
   Finally, the right subderivation may address either the (textually)
   first disjunct or the second disjunct.  This leads to the two
   inference figures below.
\[
\begin{array}{c}
\derivii{$\vee\proveslabel^B_L$}
	{\Pi, \Pi' ; \Gamma, \m A \vee \m B^{\mu}, \m A^{\mu} 
	 \proves \Delta ; \Theta, \Theta'}	
	{\Pi, \m B^{\mu} ; \proves ; \Theta}	
	{\Pi, \Pi' ; \Gamma, \m A \vee \m B^{\mu}
	 \proves \Delta ; \Theta, \Theta'}	 \\
\derivii{$\vee\proveslabel^B_R$}
	{\Pi, \Pi' ; \Gamma, \m A \vee \m B^{\mu}, \m B^{\mu} 
	 \proves \Delta ; \Theta, \Theta'}	
	{\Pi, \m A^{\mu} ; \proves ; \Theta}	
	{\Pi, \Pi' ; \Gamma, \m A \vee \m B^{\mu}
	 \proves \Delta ; \Theta, \Theta'}	
\end{array}
\]	
	We call these inferences \emph{blocking} $(\vee\proveslabel)$
   inferences, or $(\vee\proveslabel^B)$ inferences.  We will appeal
   to two calculi in which these inferences appear.  The first, SCLU,
   permits both ordinary $(\vee\proveslabel)$ and
   $(\vee\proveslabel^B)$ inferences, without restriction.  SCLU is
   convenient for describing transformations between proofs.  The
   second, SCLB, permits $(\vee\proveslabel^B)$ inferences but not
   ordinary $(\vee\proveslabel)$ inferences.  

   	Blocks are more than just boundaries in the proof; they
   provide a locus for enforcing modularity.  We will ensure that a
   disjunct contributes inferences to the new block where it is
   introduced.  Thanks to this contribution, we can narrow down the
   choice of goals to restart in a modular way.  

	This result is made possible only by maintaining the right
   structure as we introduce $(\vee\proveslabel^B)$ inferences.  We
   use path prefixes to make explicit connections between program
   statements and any goals that they help establish.  The key notions
   are \emph{spanning}, \emph{simplicity} and \emph{balance} for
   sequents.  Spanned, simple and balanced sequents represent a
   consistent evolution of the state of proof search, which records a
   full set of restart goals and the corresponding assumptions, with
   no redundancy.
\begin{definition}[Carrier]
	The \emph{carrier} of a non-empty Herbrand prefix $\mu\eta$ is
   $B^{\mu\eta}_{X,\mu\eta}$ if $\eta$ is $\eta^u_{A\bimp_i B}(\mu,X)$
   and otherwise, when $\eta$ is $\eta^u_{\Box_i A}(\mu,X)$, is
   $A^{\mu\eta}_{X,\mu\eta}$.
\end{definition}
\begin{definition}[Spanned]
	Say one multiset of tracked prefixed formulas, $\Pi$, is {\em
   spanned} by another, $\Theta$, if for every expression occurrence
   $A^{\mu}_X$ in $\Pi$ and every non-empty prefix $\nu$ of $\mu$
   there is an occurrence of the carrier of $\nu$ in $\Theta$.  It is
   easy to see there is a minimal set $\Theta$ that spans $\Pi$ and
   that such $\Theta$ spans itself.  A sequent $\Pi; \Gamma \proves
   \Delta; \Theta$ is \emph{spanned} if $\Pi$ is spanned by $\Theta$,
   $\Gamma$ is spanned by $\Theta$, $\Delta$ is spanned by $\Theta$
   and $\Theta$ is spanned by $\Theta$.  A derivation or block is
   \emph{spanned} if every sequent in it is spanned.
\end{definition}
\begin{definition}[Simple]
	A multiset $\Psi$ is \emph{simple} if no expression occurs multiple
   times in $\Psi$; a sequent of the form $\Pi; \Gamma \proves \Delta;
   \Theta$ is \emph{simple} if $\Pi$ and $\Theta$ are simple.  A derivation
   or block is \emph{simple} iff every sequent in it is simple.
\end{definition}
\begin{definition}[Balanced]
	A pair of multisets of tracked, prefixed formulas $\Pi, \Theta$
   is \emph{balanced} if
\begin{itemize}
\item
	for any $\eta = \eta^u_{B\bimp_i C}(\mu,X)$, $\eta$ occurs in
   $\Theta$ exactly when the expression $B^{\mu\eta}_{X,\mu\eta}$
   occurs in $\Pi$ and exactly when the expression
   $C^{\mu\eta}_{X,\mu\eta}$ occurs in $\Theta$; and
\item
	for any $\eta = \eta^u_{\Box A}(\mu,X)$, $\eta$ occurs in
   $\Theta$ exactly when the expression $A^{\mu}_{X,\mu\eta}$ occurs
   in $\Theta$.
\end{itemize}
	A sequent $\Pi; \Gamma \proves \Delta; \Theta$ is
   \emph{balanced} if the pair $\Pi, \Theta$ is balanced.  A block or
   derivation is \emph{balanced} if every sequent in the block is
   balanced.
\end{definition}

	We use the notion of an \emph{isolated block} to obtain an
   even stronger characterization of proof search that proceeds in a
   well-regimented way.  In an isolated block, the only expressions
   preserved across a blocking inference are those that are in some
   sense intrinsic to the restart problem created by that inference.
   Specifically, each nested block must begin with the same
   end-sequent as the outer block, except for additional program
   statements that have to be added in order to introduce the
   newly-assumed disjunct, and the further goal and program statements
   required to obtain a balanced and spanned sequent.
\begin{definition}[Isolated]
	Let ${\cal D}$ be an SCLU derivation, and let ${\cal B}$ be a
   block of ${\cal D}$.  Write the end-sequent of ${\cal B}$ as $\Pi;
   \Gamma \proves \Delta; \Theta$ and consider the right subproof of
   some $(\vee\proveslabel^B)$ inference $L$ at the boundary of ${\cal
   B}$ has an end-sequent of the form $\Pi', E; \proves ; \Theta'$.
   The \emph{exported} expressions in $\Pi'$, $\Pi'_e$, consist of the
   occurrences of expressions $F$ in $\Pi'$ such that either is $F$
   based in an occurrence of $F$ in $\Pi$ or is based in an occurrence
   of $F$ as the side expression of an inference in which $E$ is also
   based. 

	${\cal B}$ is \emph{isolated} if the right subproof of each
   $(\vee\proveslabel^B)$ inference $L$ at the boundary of ${\cal B}$
   has an end-sequent of the form $\Pi', E; \proves ; \Theta'$ meeting
   the following conditions: $E$ is the side-expression of $L$;
   $\Theta'$ is the minimal multiset of expressions which spans
   $\Pi'_e, E, \Theta$ and includes $\Theta$; and $\Pi'$ is the
   smallest multiset including $\Pi'_e, E$ for which $\Pi', \Theta'$
   is balanced.  ${\cal D}$ is \emph{isolated} iff every block of
   ${\cal D}$ is isolated.
\end{definition}
	Isolation allows us to keep close tabs on the uses of formulas
   within blocks, which is important for establishing modularity later.  In
   particular, isolation provides a key notion in formalizing the obvious
   fact that an inference that makes no contribution to an SCLU derivation
   can be omitted.  

	Finally at this stage, we refine the form of proofs which we
   are willing to count as goal-directed.  Now it will often happen
   that, while each block of a derivation may be eager, the derivation
   as a whole will not be eager.  As observed in
   \cite{nadathur/loveland:disj}, derivations with blocks can
   nevertheless be seen as eager throughout by reconstructing the
   (restart) rule as backchaining against the negation of a subgoal.
   But we will simply consider \emph{blockwise eager} derivations from
   now on.
\begin{definition}[Blockwise delayed]
   	\m R is \emph{blockwise delayed} exactly when there is a tree
   of contiguous inferences ${\cal D}'$ within a single block of
   ${\cal D}$ where: ${\cal D}'$ contains R; ${\cal D}'$ has a left
   inference $L$ at the root; and the principal $E$ of $R$ is based in
   an occurrence of $E$ in the end-sequent of ${\cal D}'$.
\end{definition}
\begin{definition}[Blockwise eager]
	{\cal D} is {\em blockwise eager} exactly when it contains no
   blockwise delayed applications of right rules.
\end{definition}

	Obviously, we can use weakening to transform an SCLB or SCLU
   derivation into a SCLS derivation, so the blocking inference
   figures are sound.  The completeness of SCLB is a consequence of
   Lemma~\ref{block-lemma}.
\begin{lemma} \label{block-lemma}
	We are given a blockwise eager SCLS derivation ${\cal D}$ whose
   end-sequent is spanned and balanced and takes the form:
\[
	\Pi ; \proves ; \Theta
\]
	We transform ${\cal D}$ into a blockwise eager SCLB derivation
   in which every block is canceled, linked, isolated, simple,
   balanced and spanned.
\end{lemma}
	\textbf{Proof.}  We can transform individual blocks to achieve
   a streamlined form, which already implicitly reflects the logic
   programming search strategy of focused search on particular goals
   and program statements.  By pursuing a suitable ordering strategy
   as we inductively repeat this inductive transformation, we can
   create the desired SCLB proofs with an overall modular block
   structure.  See Appendix~\ref{block-proof-subsec}.  $\qed$

\subsection{Modularity}
\label{modularity-subsec}

	We now derive SCLP from SCLB.  SCLP proofs can be rewritten to
   SCLB rules by a weakening transformation.  Conversely, rewriting
   SCLB proofs to SCLP proofs is accomplished by induction on the
   structure of proofs.  The transformation is possible because
   multiple formulas in sequents are needed only for passing
   ambiguities and work done across branches in the search; this is
   ruled out by the use of $(\vee\proveslabel^{\s B}_{\s L})$,
   $(\vee\proveslabel^{\s B}_{\s R})$ and $(\imp\proveslabel^{\s S})$.
\begin{lemma} \label{lp-seq-lemma}
    	Given a blockwise eager SCLB derivation ${\cal D}$, with
   end-sequent
\[
	\Pi; \proves ;\Theta
\]
	in which every block is linked, simple and spanned, we
   can construct a corresponding SCLP derivation of the same
   end-sequent in which every block remains linked.
\end{lemma}
	\textbf{Proof.}  By induction on the structure of proofs.  See
   Appendix~\ref{modularity-proof-subsec}.  $\qed$

\section{Assessment and conclusions}
\label{assessment-sec}

   	To execute modal specifications requires leveraging both the
   flexibility of efficient classical theorem-proving and the
   distinctive modularity of modal logic.  This is a significant
   problem because the two are at odds.  On the one hand, flexible
   search strategies impose no constraints on the relationships among
   inferences.  By ignoring modularity, they can leave open
   inappropriate possibilities for search.  On the other hand,
   brute-force modular systems may place such strong constraints on
   the order in which search must proceed that it becomes impossible
   to guide that search in a predictable, goal-directed way.  In this
   paper, we have explored one strategy for balancing the flexibility
   of classical goal-directed search with the modularity of modal
   logic.  This strategy culminates in the development of a modular
   logic programming sequent calculus SCLP.

	\cite{my-thesis} describes a preliminary implementation of
   proof search in SCLP as a logic programming interpreter \dialup.  I
   close by summarizing \emph{how}
   (Section~\ref{implementation-subsec}) and no less importantly
   \emph{why} (Section~\ref{use-subsec}) I developed this
   implementation.

\subsection{Implementation}
\label{implementation-subsec}

   	An effective implementation of SCLP requires further
   treatments of \emph{unification} and \emph{search control}.

	In general, to implement first-order sequent calculus proof
   search, we must \emph{lift} the inference figures.  That is, we
   adapt the inferences that require instantiation to specific terms so
   that they introduce \emph{logic variables} instead.  As we
   construct the proof, we accumulate \emph{constraints} on the values
   of these variables---for example, we get constraints when an axiom
   link in the proof requires two formulas to be identical.  In the
   lifted system, each proof we find represents the set of ground
   proofs that you get by replacing the variables with values that
   satisfy the constraints.  Lifting is the essence of the resolution
   procedure \cite{robinson:resolution} but can be regarded as a
   general metatheoretical strategy.
   \cite{lincoln/shankar:search,voronkov:constraint} offer
   particularly general discussions of this strategy at its most
   sophisticated.

	For first-order modal inference in prefixed calculi, lifting
   introduces two kinds of logic variables, and two corresponding
   kinds of constraints.  First-order quantifiers introduce logic
   variables over individuals, subject to the familiar constraints
   that give rise to term unification problems.  Modal inferences,
   meanwhile, introduce logic variables over prefixes, subject to path
   equations.  This leads to specialized problems of equational
   unification; good solutions are known for the general setting of
   multi-modal logic; see for example
   \cite{auffray:equations,debart:modal/lp,otten/kreitz,schmidt:uni}.  

	The logical fragment of SCLP makes path equations particularly
   simple.  Inspection of the SCLP proof rules shows that, at any
   point in proof search, we have enough path constraints to determine
   \emph{ground} substitutions for all the path variables in the
   sequent except possibly for variables in the current program
   statement that are about to be unified with a goal.  In many cases,
   this makes path equations easy to solve---a compact representation
   of all possible solutions can be computed in polynomial time.  The
   details are beyond the scope of this paper, but see
   \cite{my-thesis}.

	Search control is the other issue.  An implementation has to
   make commitments about what statements to try and what rules to use
   to process those statements.  The fact that SCLP program and goal
   statements are labeled with ground prefixes means that we can
   easily test that a statement's label is a prefix of the goal label
   before attempting to match the statement and the goal.  We can also
   identify an atomic subformula of the statement nondeterministically
   as the \emph{head}, and commit to match that head with the goal.
   Before doing so, we can for example test that the head and the goal
   share the same predicate symbol.

	In the case of disjunction, we also want to make sure that we
   avoid reporting duplicate proofs, despite the duplicate rules for
   disjunction that we have.  Loveland considers a number of
   heuristics for this \cite{loveland:nhp}, and we expect that they
   apply in SCLP as well as in Near-Horn Prolog.  But here is another
   heuristic.  As motivated in Section~\ref{intro-example-sec},
   $(\vee\proveslabel_R)$ is required only for cancellation.  When we
   use it, we expect to cancel an assumption (like $B$ in
   Figure~\ref{explicit-modular-fig}) that could not be canceled
   otherwise.  We can make this precise: $(\vee\proveslabel_R)$ should
   only be used in a restart block, and the assumption that is
   canceled in that block ought not to be used in the subsequent
   restart block initiated by the $(\vee\proveslabel_R)$ inference.
   Otherwise, we will independently construct an alternative proof
   that uses $(\vee\proveslabel_L)$ instead.  Naturally, the kind of
   block analysis illustrated in the proof of
   Theorem~\ref{lp-sequent-thrm} can be used to show that this
   restriction is complete.

\subsection{Applications in modal representation}
\label{use-subsec}

   	In classical logic, indefinite information is a bit exotic.
   Rather than developing an indefinite specification, we much prefer
   to collect the additional information required to describe the
   world in a precise, definite way.  This is not true at all with
   modal specifications.  Modal specifications get much of their
   interest from their ability to contrast different perspectives or
   sources of information.  What one source of information represents
   with specific, definite information, another source represents with
   abstract, indefinite information.  Computation from modal
   specifications involves the coordinated exchange of information
   between these sources.

	In particular, problems of \emph{planning} \cite{aaai98} and
   problems of \emph{communication} \cite{jlac00} depend on indefinite
   modal specifications.  In planning, one agent, the
   \emph{scheduler}, has to allocate a task to another agent, the
   \emph{executive}.  (The executive may just be the scheduler at a
   later point in time!)  It is unrealistic to expect that the
   scheduler will know \emph{exactly} what the executive \emph{will}
   do; this almost certainly requires information that is not
   available to the scheduler.  Rather, the scheduler should merely
   know what the executive \emph{can} do.  This means that, to be
   useful, the scheduler must have an \emph{indefinite} modal
   specification that abstractly describes the information that will
   be available to the executive.  For examples, see
   \cite{moore:action,morgenstern:preconditions,scherl/levesque:knowledge,davis:preconditions}
   as well as \cite{aaai98}.

	In communication, the task of one agent, the \emph{speaker},
   is to formulate an utterance that allows another agent, the
   \emph{hearer}, to answer a question.  There are many cases where
   the speaker does not have enough information to answer the question
   directly.  However, the speaker can still design an utterance that
   allows the hearer to infer the right answer, because the hearer
   knows something the speaker does not.  Concretely, a user of a
   computer interface might want to know what action to take next.
   The right answer might be for the user to type \emph{jdoe} into a
   certain text box.  The speaker might know to say \emph{enter your
   user ID}, even if the speaker does not know what the user ID is.
   Again, the speaker can make such choices meaningfully only from an
   indefinite modal specification that says what the hearer knows
   abstractly but not definitely.  See \cite{jlac00} for a worked-out
   formal case study.

\appendix

\section{Proof of Theorem~\ref{eager-thrm}}
\label{eager-proof-sec}

\noindent
	\emph{ Any SCL(I) derivation ${\cal D}$ is equal to an eager
   derivation ${\cal D}'$ up to permutations of inferences.}

   	The proof depends on a generalization of delayed inferences,
   which we can term \emph{misplaced} inferences since we intend to
   eliminate them.  We assume an overall derivation ${\cal D}$, and
   consider a right inference \m R that applies to principal $E$ within
   some subderivation ${\cal D}'$ of ${\cal D}$. 
\begin{definition} \label{misplace-def}
   	 We say a right inference $R$ is \emph{right-based} on an
   inference $R'$ in ${\cal D}$ if $R = R'$ or $R$ is based on $R'$
   and every inference on which $R$ is based above and including $R'$
   is a right inference.  Then \m R is {\em misplaced} in ${\cal D}'$
   exactly when there are inferences $M$ and $R'$ in ${\cal D}'$ such
   that, in ${\cal D}$, $M$ is based on an inference $L$, $R$ is
   right-based on $R'$, and $R'$ is delayed with respect to $L$.
\end{definition}
	In this case we will also say $R$ is misplaced \emph{with
   respect to} $M$.  We can abstract a key case of misplaced
   inferences by the following schematic derivation:
\[
\begin{derive}{}
\derivitem{$R$}{}
\derivitem{$M$}{
\makebox[0in][r]{\parbox{2in}{\begin{singlespace}
Right inferences and inferences $R$ not based in
\end{singlespace}}}
\left\{
\makebox[1in][c]{$\vdots$}
\right.
}
\derivitem{$R'$}{
\makebox[0in][r]{\parbox{2in}{
\begin{singlespace}$R'$ delayed wrt $L$ \\ ($M$ based in $L$)
\end{singlespace}}}
\left\{
\makebox[1in][c]{$\begin{array}{c}\ldots E \ldots  \\
\downarrow
\end{array}$} \right.}
\derivitem{$L$}{
\ldots E \ldots
}
\end{derive}
\]
	This schematic derivation shows informally how \emph{misplaced
   inferences} help provide an inductive characterization of the
   inferences that stand in the way of obtaining an eager derivation.
   In an eager derivation, it will be impossible for $R$ to appear
   above $L$.  For $R'$ cannot be delayed with respect to $L$, but
   once $R'$ and $L$ are interchanged, we will obtain a new delayed
   inference that $R$ is based in, until finally we must interchange
   $L$ and $R$.  Of course, to do this, we must first interchange $R$
   with the \emph{misplaced} inferences, such as $M$, which stand
   between $R$ and $L$ and cannot themselves be interchanged with $L$
   because they are based in $L$.

	Observe that the relation $R$ is misplaced with respect to $M$
   is asymmetrical.  To see this, suppose $R$ is misplaced with
   respect to $M$.  By definition, $R$ is right-based on $R'$ which is
   delayed with respect to a left inference $L$ on which $M$ is based.
   Meanwhile, for $M$ to be misplaced with respect to $R$, by
   definition, we must have $M$ right-based on $M'$ and $R$ based in
   some left rule $L_R$.  Any such $M'$ would have to be based in $L$
   since no left inferences intervene between $M$ and $M'$; $M'$ must
   thus appear \emph{inside} a schematic like that above.  At the same
   time, since no left inferences intervene between $R$ and $R'$, $R'$
   would have to be based in any such $L_R$, which must thus appear
   \emph{outside} such a schematic, closer to the root of the overall
   derivation.  Accordingly, any such $L_R$ must occur closer to the
   root of ${\cal D}$ than $L$; meanwhile the principal of $M'$ is
   introduced further from the root than $L$.  So we will not have
   $M'$ delayed with respect to $L_R$.

	Call $R$ \emph{badly misplaced} in ${\cal D}'$ if $R$ is
   misplaced with respect to $M$ and $M$ occurs closer to the root
   than $R$.  A subderivation ${\cal D}'$ with no badly misplaced
   inferences will be called \emph{good}.  An overall good
   derivation is also eager, since any delayed inference is badly
   misplaced.  

	We can now present the proof in full using a lemma.
\begin{lemma}
\label{lp-local-eager-lemma}
	Consider a subderivation ${\cal D}'$ of an overall derivation
   ${\cal D}$, with the property that ${\cal D}'$ has good immediate
   subderivations and that ${\cal D}'$ ends in inference $M$.
   From ${\cal D}'$ we can construct a derivation with the same
   end-sequent that is good.
\end{lemma}
	{\bf Proof.}  The assumption that the immediate subderivations
   of ${\cal D}'$ are good is a very powerful one.  For suppose that
   some inference is badly misplaced with respect to some other in
   ${\cal D}'$.  Then we can only have some rule $R$ badly misplaced
   with respect to $M$---anything else would contradict that
   assumption.

	In fact, we can show that some such $R$ must be adjacent to
   $M$.  Consider an inference $S$ that intervenes between $R$ and
   $M$: we will show that $S$ must be badly misplaced with respect to
   $M$ too.  By the definition of misplaced, $M$ is based on some left
   rule $L$ in ${\cal D}$, $R$ is right-based on $R'$, and $R'$ is
   delayed with respect to $L$.  Now consider the inferences that $S$
   is based on above $L$.  If any of these is a left inference $L'$,
   or $S$ is itself a left inference, then $R$ is also misplaced with
   respect to $S$---indeed, badly misplaced.  This contradicts the
   assumption that the subderivations of ${\cal D}'$ are good.  So
   none of these inferences can be a left inference, which means $S$
   is a right inference that is right-based on some inference $S'$
   above $L$.  $S'$ must be delayed with respect to $L$.  Hence $S$ is
   badly misplaced with respect to $M$.

	Now we can proceed after \cite[Lemma 10]{kleene:permute}.
   Define the \emph{grade} of ${\cal D}'$ as the number of badly
   misplaced inferences in ${\cal D}'$.  We show by induction on the
   grade that ${\cal D}'$ can be transformed to a good one.

	The base case is a derivation of grade 0.  This case has ${\cal
   D}'$ itself good.  Thus, suppose the lemma holds for derivations of
   grade \m g, and consider ${\cal D}'$ of grade $\m g+1$.  By the
   argument just given, one immediate subderivation---call it ${\cal
   D}''$---must end with an inference $\m R$ which is badly misplaced
   with respect to $M$.  Such an $R$ of course cannot be based in $M$,
   so we interchange inferences $R$ and $M$.  In the result, the
   subderivation(s) ending in $M$ satisfy the condition of the lemma
   with grade \m g or less.  By applying the induction hypothesis, we
   can replace these subderivations with good ones.  By asymmetry, $M$
   is not now badly misplaced with respect to $R$, nor can any of the
   other inferences be badly misplaced with respect to $R$, since they
   were not so in the original derivation.  It follows that the result
   is a good derivation.  $\qed$

	Now, continuing the proof of Theorem~\ref{eager-thrm}, define
   the \emph{reluctance} of ${\cal D}$ to be the number of rule
   applications \m R such that the subderivation ${\cal D}_R$ of
   ${\cal D}$ rooted in $R$ is not good.  We proceed by induction on
   reluctance.  If reluctance is zero, ${\cal D}$ is itself good.

	Now suppose the theorem holds for derivations of reluctance \m
   d, and consider ${\cal D}$ of reluctance $\m d+1$.  Since ${\cal
   D}$ is finite, there must be a highest inference $R$ such that some
   inference is badly misplaced with respect to $R$ in the
   subderivation ${\cal D}_R$ rooted at $R$.  This ${\cal D}_{\s R}$
   satisfies the condition of Lemma~\ref{lp-local-eager-lemma}.
   Therefore this ${\cal D}_{\s R}$ can be replaced with a
   corresponding eager derivation, giving a new derivation of smaller
   reluctance.  The induction hypothesis then shows that the resulting
   derivation can be made eager.  $\qed$

\section{Proof of Theorem~\ref{lp-sequent-thrm}}
\label{lp-proof-sec}

	\emph{Let $\Gamma$ and $\Delta$ be multisets of tracked
   prefixed expressions in which each formula is tracked by the empty
   set and prefixed by the empty prefix.  There is a proof of $\Gamma
   \proves \Delta$ in SCL exactly when there is a proof of $\Gamma;
   \proves ;\Delta$ in SCLP in which every block is canceled.}

	\textbf{Proof.}  As observed already in
   Section~\ref{uniform-proof-subsec}, there is an SCL proof of
   $\Gamma \proves \Delta$ exactly when there is an SCLI proof of
   $\Gamma \proves \Delta$.  By Theorem~\ref{eager-thrm} of
   Section~\ref{uniform-proof-subsec}, there is an SCLI proof of
   $\Gamma \proves \Delta$ exactly when there is an \emph{eager} SCLI
   proof of $\Gamma \proves \Delta$.  By
   Lemma~\ref{articulation-lemma}, there is an eager SCLI proof of
   $\Gamma \proves \Delta$ exactly when there is an eager articulated
   SCLI proof of $\Gamma; \proves ;\Delta$.  And by
   Lemma~\ref{segment-lemma}, there is an eager articulated SCLI proof
   of $\Gamma; \proves ;\Delta$ exactly when there is an eager SCLS
   proof of $\Gamma; \proves ;\Delta$.

	Continuing through the argument, By the Contraction Lemma, we
   may assume without loss of generality that $\Gamma; \proves
   ;\Delta$ is a simple sequent.  We know from its lack of prefixes
   that the sequent $\Gamma; \proves; \Delta$ is also spanned and
   balanced.  By Lemma~\ref{block-lemma} of
   Section~\ref{block-proper-subsubsec}, then, there is an eager SCLS
   proof of $\Gamma; \proves ;\Delta$ exactly when there is a
   blockwise eager SCLB derivation of $\Gamma; \proves ;\Delta$ in
   which every block is canceled, linked, isolated, simple, balanced
   and spanned.  And by Lemma~\ref{lp-seq-lemma}, there is a blockwise
   eager SCLB derivation of $\Gamma; \proves ;\Delta$ in which every
   block is canceled, linked, isolated, simple, balanced and spanned
   exactly when there is an SCLP derivation of $\Gamma; \proves
   ;\Delta$ in which every inference is linked.  And if every
   inference is linked, every block is canceled.  $\qed$

\subsection{Proof of Lemma~\ref{segment-lemma}}
\label{segment-proof-subsec}

	We show in this section that an articulated SCLI proof with
   end-sequent $\Pi; \proves ;\Theta$ corresponds to an SCLS proof
   with end-sequent $\Pi; \proves ;\Theta$, and vice versa.  In fact,
   to transform SCLS to articulated SCLI we have a simple structural
   induction which replaces $(\imp\proveslabel^S)$ with
   $(\imp\proveslabel)$ using the weakening lemma; the soundness of
   SCLS over SCLI then follows by Lemma~\ref{articulation-lemma}.
   Thus, here we are primarily concerned with completeness of a new
   sequent inference figure.

	The use of $(\imp\proveslabel^S)$ in eager derivations ensures
   that the processing of each new goal refers directly to global
   program statements.  To formalize this idea, we introduce the
   notion of a \emph{fresh} inference.
\begin{definition}[Fresh]
	Let ${\cal D}$ be an SCLV derivation.  An inference $R$ in
   ${\cal D}$ is \emph{fresh} exactly when $R$ is a right inference
   and the path from $R$ to the root never follows the left spur of
   any $(\imp\proveslabel)$ inference.
\end{definition}
\begin{lemma}
\label{new-imp-ncxt-lemma}
	Let ${\cal D}$ be an eager SCLV derivation with an end-sequent of
   the form
\[
	\Pi ; \proveslabel \Delta ; \Theta
\]
	and consider a subderivation ${\cal D}'$ of ${\cal D}$ rooted
   in a fresh inference \m R.  Then the end-sequent of ${\cal
   D}'$ also has the form 
\[
	\Pi' ; \proveslabel \Delta' ; \Theta'
\]
	for some $\Pi'$, $\Delta'$ and $\Theta'$.
\end{lemma}
	{\bf Proof.}  Suppose otherwise, and consider a maximal ${\cal
   D}'$ whose end-sequent contains a non-empty multiset of local
   statements $\Gamma$.  We can describe ${\cal D}'$ equivalently as
   the subderivation of ${\cal D}$ that is rooted in a lowest fresh
   inference $R$ when the end-sequent of ${\cal D}$ contains some
   local statements.  $R$ cannot be the first inference of ${\cal D}$,
   so there must be an inference \m S in ${\cal D}$ immediately below
   \m R.  If \m S is a left rule, then the fact that ${\cal D}$ is
   eager leads to a contradiction.  $R$ must be based in $S$, or else
   $R$ will be delayed.  This means $S$ is an implication inference;
   but given that $R$ is fresh, \m R must appear along the branch of
   $(\imp\proveslabel^S)$ without local statements.  Meanwhile, if \m
   S is a right rule, it follows from the formulation of the rules
   that if the end-sequent of ${\cal D}_{\s R}$ has non-empty local
   statements then the end-sequent of ${\cal D}_{\s L}$ must also.
   This contradicts the assumption that \m R is first.  $\qed$

	Now we proceed with the proof of Lemma~\ref{segment-lemma}.
   We assume an eager SCLV derivation ${\cal D}$ with such an
   end-sequent; we show that we can transform it into an eager SCLS
   derivation ${\cal D}'$ with the same end-sequent.  The proof is by
   induction on the number of occurrences of $(\imp\proveslabel)$
   inferences in ${\cal D}$.

	In the base case, there are no $(\imp\proveslabel)$ inferences
   and ${\cal D}'$ is just ${\cal D}$.

	Suppose the claim holds for derivations where
   $(\imp\proveslabel)$ is used fewer than $n$ times, and suppose
   ${\cal D}$ is a derivation in which $(\imp\proveslabel)$ is used \m
   n times.  Choose an inference \m L of $(\imp\proveslabel)$ with no
   other $(\imp\proveslabel)$ inference closer to the root of ${\cal
   D}$; we must rewrite the left subderivation at $L$ to match the
   $(\imp\proveslabel^S)$ inference figure.  We distinguish a
   subderivation ${\cal D}'$ of ${\cal D}$ as a function of $L$ and
   draw on the inferences in ${\cal D}'$ to replace this
   subderivation---in particular, we identify ${\cal D}'$ as the
   largest subderivation of ${\cal D}$ containing $L$ but no right
   inferences or segment boundaries below $L$.

	Using Lemma~\ref{new-imp-ncxt-lemma}, we develop a schema of
   ${\cal D}'$ thus:
\[
\setlength{\localjunklength}{1.5em}
\deriviitheni{}
	     {\raisebox{\baselinestretch\localjunklength}{$L$}}
	     {\begin{array}{c}
		{\cal D}^A \\
		\Pi; \Gamma, A\imp B^{\mu}_X 
		\proveslabel A^{\mu}_X, \Delta; \Theta
	      \end{array}}
	     {\begin{array}{c}
		{\cal D}^B \\
		\Pi; \Gamma, A\imp B^{\mu}_X, B^{\mu}_X 
		\proveslabel \Delta; \Theta
	      \end{array}}
	     {{\cal D}^L
	      \left\{
	       \centerbox{$\begin{array}{c}
			   \Pi; \Gamma, A\imp B^{\mu}_X
			   \proves \Delta; \Theta \\
   			   \vdots \\
			   \Pi; \proves \Delta; \Theta
			  \end{array}$}
		\right.}
	     {\mbox{(Segment boundary or right rule)}}
\]
	We suppose \m L applies to an expression $\m A \imp \m
   B^{\mu}_X$; the left subderivation of $L$, ${\cal D}^A$ adds the
   goal $A$; the right, ${\cal D}^B$, uses the assumption $B$.  The
   subderivation of ${\cal D}'$ from the end-sequent of \m L abstracts
   the left inferences performed elsewhere in this segment (and any
   subgoals that these inferences trigger).  We notate this tree of
   inferences ${\cal D}^L$.  By Lemma~\ref{new-imp-ncxt-lemma}, ${\cal
   D}'$ ends with a sequent of the form $\Pi ; \proves \Delta ;
   \Theta$.  Because of the form of the intervening rules, we have the
   same succedent $\Delta; \Theta$ at $L$, as well as the same global
   statements $\Pi$.

	We use ${\cal D}^L$ to construct an eager SCLS derivation
   ${\cal A}$ corresponding to ${\cal D}^A$; we will substitute the
   result for the left subtree at $L$ to revise $L$ to fit the
   $(\imp\proveslabel^S)$ figure.  In outline, the derivation we aim
   for is an eager SCLS version of:
\[
\derivi{}
       {{\cal D}^A}
       {{\cal D}^L + A^{\mu}_X}
\]
	The problem is that if ${\cal D}^A$ is rooted in a right
   inference to $A$, we will not obtain an eager derivation when we
   reassemble $L$.  The SCLS derivation ${\cal A}$ we use is actually
   constructed by recursion on the structure of ${\cal D}^A$, applying
   this kind of transformation at appropriate junctures.  At each
   stage, we call the subderivation of ${\cal D}^A$ we are considering
   ${\cal D}'^A$.

	For the base case, this subderivation is an axiom, and we
   construct this subderivation as a result.  If ${\cal D}'^A$ ends in
   a right rule, the construction proceeds inductively by constructing
   corresponding subderivations and recombining them by the same right
   rule.  With a right inference here, the resulting derivation must
   be eager since the subderivations are eager.

	If ${\cal D}'^A$ ends in a left inference, the construction is
   not inductive.  We observe that ${\cal D}'^A$ has an end-sequent of
   the form
\[
	\Pi, \Pi' ; \proves \Delta, \Delta'; \Theta, \Theta'
\]
	(The inventory of expressions can only be expanded, and that
   only in certain places, as we follow right inferences to reach
   ${\cal D}'^A$.)  So we first weaken ${\cal D}^L$ by the needed
   additional expressions---$\Pi'$ on the left and $\Delta'$ (locally)
   and $\Theta'$ (globally) on the right; then we identify the open
   leaf in ${\cal D}^{\s L}$ with ${\cal D}'^A$, obtaining a larger
   derivation ${\cal D}_I$ defined as:
\[
\derivi{}
       {{\cal D}'^A}
       {\Pi' + {\cal D}'^L + A^{\mu}_X + \Delta'; \Theta'}
\]
	Any delayed inference in ${\cal D}_I$ would in fact be delayed
   in ${\cal D}'^A$, so this is an eager derivation.  The result has,
   moreover, fewer than $n$ $(\imp\proveslabel)$ inferences, since it
   omits at least $L$ from ${\cal D}'$.  Then the induction hypothesis
   applies to give the needed SCLS derivation ${\cal A}$.

	Given the derivation ${\cal A}$ so constructed, we substitute
   ${\cal A}$ for ${\cal D}^A$ in ${\cal D}$.  The result ${\cal D}^*$ is
   an eager derivation; ${\cal D}^*$ contains an $(\imp\proveslabel^S)$
   inference corresponding to $L$ and therefore contains fewer than \m n
   uses of $(\imp\proveslabel)$.  The induction hypothesis applies to
   transform ${\cal D}^{*}$ to the needed overall derivation.  $\qed$

\subsection{Proof of Lemma~\ref{block-lemma}}
\label{block-proof-subsec}

\subsubsection{Replacing Herbrand terms}
\label{replace-subsubsec}

	To begin, it is convenient to observe that the use of indexed
   Herbrand terms allows us to rename Herbrand terms in a proof under
   certain conditions.  
\begin{lemma}[Substitution] \label{renaming-lemma}
	Let ${\cal D}$ be an SCLU derivation with end-sequent
\[
	\Pi; \proves ; \Theta
\]
	in which no Herbrand terms or Herbrand prefixes appear;
   consider a spanned simple subderivation ${\cal D}'$ in which a
   modal Herbrand function $\eta^u_A$ occurs in some sequent, but does
   not occur in the end-sequent.  Let $\eta^v_A$ be a Herbrand
   function that does not occur in ${\cal D}$.  Then we can construct
   a proof ${\cal D}^*$ containing corresponding inferences in a
   corresponding order to ${\cal D}$ but in which Herbrand terms and
   Herbrand prefixes are adjusted so that $\eta^v_A$ is used in place
   of $\eta^u_A$ precisely in the subderivation corresponding to
   ${\cal D}'$.
\end{lemma}
	The \textbf{proof} is by induction on the structure of
   derivations.  A complex substitution may be required, because the
   Herbrand calculus may require not only the replacement of
   $\eta^u_A$ itself but also the replacement of Herbrand terms that
   depend indirectly on $\eta^u_A$.  It is convenient to begin by
   replacing any first-order Herbrand term not introduced by a
   $(\exists\proveslabel)$ or $(\proveslabel\forall)$ inference by a
   distinguished constant $c_0$---starting with leaves of the
   derivation and working downward.  This replacement is to ensure
   that each first-order and modal Herbrand term in ${\cal D}$ is
   determined from an expression in the end-sequent of ${\cal D}$ by a
   finite number of steps of inference.  We continue with the
   systematic replacement of $\eta^u_A$ and its dependents.  In both
   cases, the form of ${\cal D}$ ensures that a finite substitution
   can systematically rename all these Herbrand terms as required.  We
   use the fact that each sequent is simple and spanned to extend this
   substitution inductively upward.  Because each sequent is spanned
   the substitution does not need to be extended at
   $(\Box\proveslabel)$ inferences; because each sequent is simple the
   substitution can be extended freshly at $(\proveslabel\Box)$ and
   $(\proveslabel\bimp)$ inferences.  Finally, the form of first-order
   Herbrand terms ensures that a finite extension of the substitution
   suffices for $(\proveslabel\exists)$ and $(\forall\proveslabel)$
   inferences.  $\qed$.

\subsubsection{Rectifying blocks}
\label{rect-subsubsec}

	The transformation of individual blocks appeals to the
   following definition of \emph{required} elements of proofs.
\begin{definition}[Required]
	Given a derivation ${\cal D}$ with end-sequent 
\[
	\Pi; \Gamma \proves \Delta; \Theta
\]
	we say that an expression occurrence $E$ in $\Theta$ or $\Pi$
   is \emph{required} iff either it is linked or some block in ${\cal
   D}$ is adjacent to the root block and has an end-sequent
\[
	\Pi'; \proves ; \Theta'
\]
	in which $\Pi'$ or $\Theta'$ contains an expression occurrence
   based in $E$.
\end{definition}
\begin{lemma}[Rectification] \label{linking-lemma}
	We are given a blockwise eager SCLU derivation ${\cal D}$ such
   that: every block in ${\cal D}$ is canceled and isolated; every block in
   ${\cal D}$ other than the root is spanned, linked, balanced and simple;
   and the end-sequent of ${\cal D}$ is balanced.  We transform ${\cal D}$
   to an SCLU derivation ${\cal D}'$ in which every block is canceled,
   linked, isolated, balanced and simple and every block other than the
   root is spanned.  Every block in ${\cal D'}$ other than the root block
   is identical to a block of ${\cal D}$; and the inferences in the root
   block of ${\cal D}$ correspond to inferences in the same order in ${\cal
   D}$ (and so ${\cal D}'$ is blockwise eager).  If the end-sequent of
   ${\cal D}$ is spanned then ${\cal D}'$ is spanned and isolated.  
\end{lemma} 
	\textbf{Proof.}  We describe a transformation that establishes the
   following inductive property given ${\cal D}$.  There are simple
   multisets $\Pi_M \subseteq \Pi$ and $\Theta_M \subseteq\Theta$, together
   with multisets $\Gamma' \subseteq \Gamma$ and $\Delta' \subseteq \Delta$
   such that: any $\Theta'$ that spans $\Pi_M$ includes $\Theta_M$; and for
   any simple $\Pi'$ with $\Pi_M \subseteq \Pi' \subseteq \Pi$ and any
   simple $\Theta'$ with $\Theta' \subseteq \Theta$ such that $\Pi'$ and
   $\Theta'$ are spanned by $\Theta'$ and the pair $\Pi', \Theta'$ is
   balanced, there is a ${\cal D}'$ in which every block is canceled,
   linked, balanced, balanced and simple, with end-sequent:
\[
	\Pi'; \Gamma' \proves \Delta'; \Theta'
\]
	In this ${\cal D}'$, each expression in $\Gamma'$ is linked; each
   expression in $\Delta'$ is linked; each $\Pi_M$ expression that occurs
   in $\Pi'$ is required and each $\Theta_M$ expression that occurs in
   $\Theta'$ is linked.  Every block in ${\cal D'}$ other than the root
   block is identical to a block of ${\cal D}$; and the inferences in the
   root block of ${\cal D}$ correspond to inferences in the same order in
   ${\cal D}$.  Finally, if $\Gamma'$ and $\Delta'$ are spanned by
   $\Theta'$ then ${\cal D}'$ is spanned; if ${\cal D}$ is linked then
   ${\cal D}'$ contains all the axioms of ${\cal D}$.

	At axioms, for ${\cal D}$ of
\[
	\Pi; \Gamma, A^{\mu}_X \proves A^{\mu}_Y, \Delta; \Theta
\]
	$\Pi_M$ and $\Theta_M$ are empty, while $\Gamma' = A^{\mu}_X$ and
   $\Delta' = A^{\mu}_X$.  Assume we are given simple $\Pi'$ from $\Pi$ and
   simple $\Theta'$ from $\Theta$ with $\Pi'$ and $\Theta'$ spanned by
   $\Theta'$.  We construct ${\cal D}'$ of
\[
	\Pi'; A^{\mu}_X \proves A^{\mu}_Y; \Theta'
\]
	If $A^{\mu}_X$ is spanned by $\Theta'$, this axiom is spanned too;
   the remaining conditions are immediate.

	At inferences, consider as a representative case
   $(\vee\proveslabel)$.  ${\cal D}$ ends:
\[
\derivii{}
	{\begin{array}{c}
	 {\cal D}_1 \\
	 \Pi; \Gamma, A\vee B^{\mu}_X, A^{\mu}_X \proves
	 \Delta; \Theta
	 \end{array}}
	{\begin{array}{c}
	 {\cal D}_2 \\
	 \Pi; \Gamma, A\vee B^{\mu}_X, B^{\mu}_X \proves
	 \Delta; \Theta
	 \end{array}}
	{\Pi; \Gamma, A\vee B^{\mu}_X \proves
	 \Delta; \Theta}
\]
	The blocks of ${\cal D}_1$ and ${\cal D}_2$ either contain the root
   or are blocks from ${\cal D}$; the Herbrand prefixes in the end-sequents
   of ${\cal D}_1$ and ${\cal D}_2$ occur with the same distribution as in
   ${\cal D}$.  Therefore we can apply the induction hypothesis to get
   $\Pi_{M1}$, $\Theta_{M1}$, $\Gamma_1'$ and $\Delta_1'$ for ${\cal D}_1$;
   we can apply it to get $\Pi_{M2}$, $\Theta_{M2}$, $\Gamma_2'$ and
   $\Delta_2'$ for ${\cal D}_2$.  To transform ${\cal D}$ itself, we
   perform case analysis on $\Gamma_1'$ and $\Gamma_2'$.

   	If $\Gamma_1'$ does not contain an occurrence of $A^{\mu}_X$, then
   $\Pi_M = \Pi_{M1}$, $\Theta_M = \Theta_{M1}$, $\Gamma' = \Gamma_1'$ and
   $\Delta' = \Delta_1'$; ${\cal D}_1'$ suffices to carry through the
   induction hypothesis.

	Similarly, if $\Gamma_2'$ does not contain an occurrence of
   $B^{\mu}_X$, then $\Pi_M = \Pi_{M2}$, $\Theta_M = \Theta_{M2}$, $\Gamma'
   = \Gamma_2'$ and $\Delta' = \Delta_2'$; ${\cal D}_2'$ suffices to carry
   through the induction hypothesis.

	Otherwise, we will set up $\Pi_M = \Pi_{M1} \cup \Pi_{M2}$ and
   $\Theta_M = \Theta_{M1} \cup \Theta_{M2}$ (as sets); by the inductive
   characterization of $\Pi_{M1}$, $\Pi_{M2}$, $\Theta_{M1}$ and
   $\Theta_{M2}$, any $\Theta'$ that spans both $\Pi_{M2}$ and $\Pi_{M2}$
   includes both $\Theta_{M1}$ and $\Theta_{M2}$.  We also set up $\Gamma'$
   as the multiset containing at least one occurrence of $A\vee B^{\mu}_X$
   and as many expression occurrences of any expression as either are found
   in $\Gamma_1' \backslash A^{\mu}_X$ or are found in $\Gamma_2'
   \backslash B^{\mu}_X$; we set up $\Delta'$ as the multiset containing as
   many expression occurrences of any expression as are found in either
   $\Delta_1'$ or $\Delta_2'$.

	To continue, we now consider simple $\Pi'$ from $\Pi$ and simple
   $\Theta'$ from $\Theta$ such that $\Pi_{M1} \subseteq \Pi'$, $\Pi_{M2}
   \subseteq \Pi'$, $\Pi'$ and $\Theta'$ are spanned by $\Theta'$, and the
   pair $\Pi', \Theta'$ is balanced.  We know that $\Theta'$ includes
   $\Theta_M$.  We can apply the inductive property to transform ${\cal
   D}_1$ and ${\cal D}_2$ into derivations with the inductive property:
\[
\begin{array}{c}
{\cal D}_1' \\
\Pi'; \Gamma_1' \proves \Delta_1'; \Theta'
\end{array}
\;\;\;\;\;\;\;
\begin{array}{c}
{\cal D}_2' \\
\Pi'; \Gamma_2' \proves \Delta_2'; \Theta'
\end{array}
\]
	We weaken \emph{the lowest block} of ${\cal D}_1'$ on the left
   by the expressions in $\Gamma^+$ and not already in $\Gamma'$ and
   on the right by the expressions in $\Delta^+$ and not already in
   $\Delta'$, giving ${\cal D}_1^+$.  We similarly weaken the lowest
   block of ${\cal D}_2'$ on the left by the expressions in $\Gamma^+$
   and not already in $\Gamma'_2$ and on the right by the expressions
   in $\Delta^+$ and not already in $\Delta'_2$, giving ${\cal
   D}_2^+$.  Only the lowest blocks are affected by the weakening
   transformations, so other blocks remain canceled, linked, spanned,
   isolated and simple; the lowest block in each case remains
   canceled.  The lowest blocks also remain linked since no inferences
   are added; and they remain simple (and balanced) because no
   weakening occurs in the global areas.  Construct ${\cal D}'$ as
\[
\derivii{}
	{\begin{array}{c}
	 {\cal D}_1^+ \\
	 \Pi'; \Gamma^+, A^{\mu}_X \proves \Delta^+; \Theta'
	 \end{array}}
	{\begin{array}{c}
	 {\cal D}_2^+ \\
	 \Pi'; \Gamma^+, B^{\mu}_X \proves
	 \Delta^+; \Theta'
	 \end{array}}
	{\Pi'; \Gamma^+ \proves
	 \Delta^+; \Theta'}
\]
	The end-sequent is simple and balanced so the root block is simple
   and balanced; the inference is linked since $A^{\mu}_X$ and $B^{\mu}_X$
   are linked in the subderivations, so the root block is linked.  The root
   block remains canceled as always.

	Any $\Pi_M$ expression is required here because it is required
   either in ${\cal D}_1^+$ in virtue of its membership in $\Pi_{M1}$ or in
   ${\cal D}_2^+$ in virtue of its membership in $\Pi_{M2}$; likewise any
   $\Theta_M$ expression is linked here because it is linked either in
   ${\cal D}_1^+$ in virtue of its membership in $\Theta_{M1}$ or in ${\cal
   D}_2^+$ in virtue of its membership in $\Theta_{M2}$.  Thus, except for
   the spanning conditional, we have shown everything we need
   of this ${\cal D}'$.

	Finally, then, if $\Gamma'$ and $\Delta'$ is spanned by $\Theta'$,
   $\Delta_1'$ and $\Delta_2'$ are spanned by $\Theta'$ and $\Gamma_1'$ and
   $\Gamma_2'$ are spanned by $\Theta'$ in the resulting (spanned)
   subderivations ${\cal D}_1'$ and ${\cal D}_2'$.  This shows that the
   end-sequent of ${\cal D}'$ is also spanned, so ${\cal D}'$ itself is
   spanned. 

	This reasoning is representative of the construction required also
   for $(\wedge\proveslabel)$, $(\exists\proveslabel)$,
   $(\forall\proveslabel)$, $(\proveslabel\wedge)$, $(\proveslabel\vee)$,
   $(\proveslabel\exists)$, $(\proveslabel\forall)$, (decide) and
   (restart).  It applies also for $(\imp\proveslabel^S)$, with the obvious
   caveat that we do not weaken the left subderivation to match local left
   expressions, since the form of the $(\imp\proveslabel^S)$ inference
   requires there to be none. 

	Next we have $(\vee\proveslabel^B)$; we consider the
   representative case of $(\vee\proveslabel^B_L)$.  ${\cal D}$ ends:
\[
\derivii{}
	{\begin{array}{c}
	 {\cal D}_1 \\
	 \Pi_0, \Pi; \Gamma, A\vee B^{\mu}_X, A^{\mu}_X \proves
	 \Delta; \Theta_0, \Theta
	 \end{array}}
	{\begin{array}{c}
	 {\cal D}_2 \\
	 \Pi_0, B^{\mu}_X; \proves
	 \Theta_0
	 \end{array}}
	{\Pi_0, \Pi; \Gamma, A\vee B^{\mu}_X \proves
	 \Delta; \Theta_0, \Theta}
\]
	We treat this specially to respect the block boundary before ${\cal
   D}_2$.  In particular, we apply the induction hypothesis to ${\cal D}_1$
   (as we may since its end-sequent has the same distribution of Herbrand
   prefixes as does that of ${\cal D}$), to get $\Pi_{M1}$, $\Theta_{M1}$,
   $\Gamma_1'$ and $\Delta_1'$.  If $A^{\mu}_X$ does not occur in
   $\Gamma_1'$, we let $\Pi_M = \Pi_{M1}$, $\Theta_M = \Theta_{M1}$,
   $\Gamma' = \Gamma_1'$ and $\Delta' = \Delta_1'$; any derivation ${\cal
   D}_1'$ constructed from appropriate $\Pi'$ and $\Theta'$ suffices to
   carry through the induction hypothesis.

	Otherwise, we get $\Pi_M = \Pi_{M1} \cup \Pi_{e0}$ (as a set),
   $\Theta_M = \Theta_{M1}$; any $\Theta'$ that spans $\Pi_M$ also
   spans $\Pi_{M1}$ and so includes $\Theta_M$.  $\Delta' = \Delta_1'$
   and $\Gamma'$ contains $\Gamma_1'$ with the occurrence of
   $A^{\mu}_X$ removed, together with an occurrence of $A\vee
   B^{\mu}_X$ if $\Gamma_1'$ does not already contain such an
   expression.

	Assume simple $\Pi'$ with $\Pi_M \subseteq \Pi' \subseteq \Pi$ and
   simple $\Theta'$ with $\Theta' \subseteq \Theta$ with $\Pi'$ and
   $\Theta'$ spanned by $\Theta'$ and the pair $\Pi', \Theta'$ balanced.
   As before, we must have $\Theta_M$ included in $\Theta'$.  We therefore
   obtain ${\cal D}_1'$ by the inductive property; we then weaken ${\cal
   D}_1'$ locally within the lowest block by $A\vee B^{\mu}_X$ on the left
   if necessary, to obtain a good derivation ${\cal D}_1^*$.

	The needed ${\cal D}'$ is now constructed as:
\[
\derivii{}
	{\begin{array}{c}
	 {\cal D}_1^* \\
	 \Pi'; \Gamma', A^{\mu}_X \proves
	 \Delta' ; \Theta'
	 \end{array}}
	{\begin{array}{c}
	 {\cal D}_2 \\
	 \Pi_0, B^{\mu}_X ; \proves
	 \Theta_0
	 \end{array}}
	{\Pi'; \Gamma' \proves \Delta' ; \Theta'}
\]
	We first argue that the construction instantiates the
   $(\vee\proveslabel^B_L)$ inference rule.  Every Herbrand prefix in
   $\Pi_{0e}$ and $B^{\mu}_X$ occurs in $\Pi'$ or $\Gamma'$, so
   $\Pi_{0e}$ and $B^{\mu}_X$ are spanned by $\Theta'$.  But because
   the root block in ${\cal D}$ is isolated, $\Pi_{0e}$ and
   $B^{\mu}_X$ are spanned minimally by $\Theta_0$.  Thus $\Theta_0
   \subseteq \Theta'$.  $\Pi_{0e} \subseteq \Pi_M$ by construction; by
   isolation $\Pi_0$ is the smallest set such that the pair of $\Pi_0,
   \Theta_0$ is balanced.  But since $\Pi', \Theta'$ is balanced,
   $\Pi_0 \subseteq \Pi'$.

	Now we show that ${\cal D}'$ so constructed has the needed
   properties.  The end-sequent is simple and balanced so the root block is
   simple and balanced.  The inference is linked: $A^{\mu}_X$ is linked in
   ${\cal D}_1'$ by the induction hypothesis; $B^{\mu}_X$ is linked in
   ${\cal D}_2$ because ${\cal D}_2$ begins a new block which by assumption
   is canceled.  The root block remains canceled as always.  Any $\Pi_M$
   expression is required here because either a corresponding expression
   $\Pi_{0e}$ in the new block at the left subderivation is based on it, or
   because it is required in ${\cal D}_1'$.  Every $\Theta_M$ is linked
   because it is linked in ${\cal D}_1^*$.

	Finally, if $\Gamma'$ and $\Delta'$ are spanned by $\Theta'$, then
   $\Delta_1'$ and $\Gamma_1'$ are spanned by $\Theta_1'$.  The new
   subderivation ${\cal D}_1'$ is therefore spanned by the inductive
   property; this ensures that the overall derivation is spanned.

	Next consider $(\Box\proveslabel)$.  ${\cal D}$ ends:
\[
\derivi{}
       {\begin{array}{c}
	{\cal D}_1 \\
	\Pi; \Gamma, \Box_i A^{\mu}_X, A^{\mu\nu}_{X,\mu\nu}
	 \proves \Delta; \Theta
	\end{array}}
	{\Pi; \Gamma, \Box_i A^{\mu}_X
	 \proves \Delta; \Theta}
\]
	As always, we apply the induction hypothesis to ${\cal D}_1$ (as we
   may since the Herbrand prefixes on $\Pi$ and $\Theta$ formulas remain
   the same) to obtain $\Pi_{M1}$, $\Theta_{M1}$, $\Gamma_1'$ and
   $\Delta_1'$.  If $A^{\mu\nu}_{X,\mu\nu}$ does not occur in $\Gamma_1'$,
   we let $\Pi_M = \Pi_{M1}$, $\Theta_M = \Theta_{M1}$, $\Gamma' =
   \Gamma_1'$ and $\Delta' = \Delta_1'$; any subderivation ${\cal D}_1'$
   obtained by the inductive property suffices to witness the inductive
   property for ${\cal D}$.

	Otherwise we obtain $\Gamma'$ by extending $\Gamma_1'$ by the
   principal expression $\Box_i A^{\mu}_X$ if necessary and
   eliminating the side expression $A^{\mu\nu}_{X,\mu\nu}$; $\Pi_M =
   \Pi_{M1}$, $\Theta_M = \Theta_{M1}$ and $\Delta' = \Delta_1'$.
   (Since these are common to the subderivation, any $\Pi'$ that spans
   $\Pi_M$ includes $\Theta_M$.)  Now we consider $\Pi'$ with $\Pi_M
   \subseteq \Pi' \subseteq \Pi$ and $\Theta'$ with $\Theta' \subseteq
   \Theta$, $\Pi'$ and $\Theta'$ spanned by $\Theta'$ and the pair
   $\Pi', \Theta'$ balanced.  As always, we have $\Theta_M \subseteq
   \Theta'$.  We obtain ${\cal D}_1'$ using $\Pi'$ and $\Theta'$, and
   weaken the lowest block by local formulas; calling the result
   ${\cal D}_1^+$, we can produce ${\cal D}'$ by the following
   construction:
\[
\derivi{}
       {\begin{array}{c}
	{\cal D}_1^+ \\
	\Pi'; \Gamma', A^{\mu\nu}_{X,\mu\nu}
	 \proves \Delta'; \Theta'
	\end{array}}
	{\Pi'; \Gamma'
	 \proves \Delta'; \Theta'}
\]
	Everything is largely as before.  The key new reasoning comes
   when we assume that $\Gamma'$ and $\Delta'$ are spanned by
   $\Theta'$.  We must argue that $\Gamma', A^{\mu\nu}_{X,\mu\nu}$ is
   in fact spanned by $\Theta'$.  Since $A^{\mu\nu}_{X,\mu\nu}$ is
   linked in ${\cal D}^+_1$, there must be an axiom in this block
   which is based in $A^{\mu\nu}_{X,\mu\nu}$; indeed, since the
   expression occurs as a local antecedent, this axiom must occur
   within the segment.  This axiom must pair expressions prefixed by a
   path $\mu'$ where $\mu\nu$ is a prefix of $\mu'$.  But because
   ${\cal D}'$ remains blockwise eager, no inferences apply to
   $\Delta'$ or $\Theta'$ formulas within the segment (nor can they in
   this fragment augment the $\Delta'$ or $\Theta'$ formulas within
   the segment); therefore some $\Delta'$ expression is associated
   with Herbrand prefix $\mu'$.  But since $\Delta'$ is spanned by
   $\Theta'$, we have that every prefix of $\mu'$ is associated with
   some $\Theta'$ expression; so every prefix of $\mu\nu$ is
   associated with some $\Theta'$ expression.  Thus ${\cal D}_1^+$ is
   spanned and in turn ${\cal D}'$ is spanned.

	We have one last representative class of inferences in ${\cal D}$:
   $(\proveslabel\Box)$ and $(\proveslabel\bimp)$.  We illustrate with the
   case where ${\cal D}$ ends in $(\proveslabel\bimp)$:
\[
\derivi{}
       {\begin{array}{c}
	{\cal D}_1 \\
	\Pi, A^{\mu\eta}_{X,\mu\eta}; \Gamma \proves
	\Delta, A\bimp_i B^{\mu}_X ;
	\Theta, B^{\mu\eta}_{X,\mu\eta}
	\end{array}}
       {\Pi; \Gamma \proves
	\Delta, A\bimp_i B^{\mu}_X;
	\Theta}
\]	
	We begin by applying the induction hypothesis to ${\cal D}_1$ (as
   we can, given the symmetric extension of $\Pi$ and $\Theta$ by labeled
   expressions).  We obtain $\Theta_{M1}$, $\Pi_{M1}$, $\Gamma_1'$ and
   $\Delta_1'$; we consider alternative cases in response to $\Theta$ and
   $\Theta_{M1}$.  First we suppose $B^{\mu\eta}_{X,\mu\eta} \not\in
   \Theta$.  It follows by our assumption about ${\cal D}$ that
   $A^{\mu\eta}_{X,\mu\eta} \not\in \Pi$ either, nor does $\eta$ occur in
   $\Theta$.  For this case, we start by defining an overall $\Pi_M$ and
   $\Theta_M$: $\Theta_M$ is $\Theta_{M1}$ with any occurrence of
   $B^{\mu\eta}_{X,\mu\eta}$ eliminated; $\Pi_M$ is $\Pi_{M1}$ with any
   occurrence of $A^{\mu\eta}_{X,\mu\eta}$ eliminated.  $\Pi_M$ contains no
   occurrences of $\mu\eta$, since $\Pi$ does not; thus given the inductive
   property of $\Theta_{M1}$ and $\Pi_{M1}$, any $\Theta'$ that spans
   $\Pi_M$ spans $\Theta_M$.  We define $\Gamma'$ and $\Delta'$ so that
   $\Gamma' = \Gamma_1'$ and $\Delta'$ contains $\Delta_1'$ together with
   an occurrence of $A\bimp_i B^{\mu}_X$, provided $\Delta_1'$ does not
   already contain one and $B^{\mu\eta}_{X,\mu\eta} \in \Theta_{M1}$ or
   $A^{\mu\eta}_{X,\mu\eta} \in \Pi_{M1}$. So, assume we are given simple
   $\Pi'$ with $\Pi_M \subseteq \Pi' \subseteq \Pi$ and simple $\Theta'$
   with $\Theta' \subseteq \Theta$ (and so $\Theta_M \subseteq \Theta'$)
   such that $\Pi'$ and $\Theta'$ are spanned by $\Theta'$ and the pair
   $\Pi', \Theta'$ is balanced.

	We consider whether $B^{\mu\eta}_{X,\mu\eta} \in \Theta_{M1}$ or
   $A^{\mu\eta}_{X,\mu\eta} \in \Pi_{M1}$.  If neither, we apply the
   induction hypothesis to ${\cal D}_1$ for the case that $\Theta_1'$ is
   $\Theta'$ and $\Pi_1'$ is $\Pi'$.  The resulting derivation ${\cal
   D}_1'$ serves as ${\cal D}'$.

	Otherwise, $B^{\mu\eta}_{X,\mu\eta} \in \Theta_{M1}$ or
   $A^{\mu\eta}_{X,\mu\eta} \in \Pi_{M1}$; we apply the inductive property
   of ${\cal D}_1$ for the case that $\Theta_1'$ is $\Theta',
   B^{\mu\eta}_{X,\mu\eta}$ and $\Pi_1'$ is $\Pi', A^{\mu\eta}_{X,\mu\eta}$
   (clearly $\Pi_1'$ and $\Theta_1'$ are spanned by $\Theta_1'$ assuming
   $\Pi'$ and $\Theta'$ are spanned by $\Theta'$; the pair $\Pi_1',
   \Theta_1'$ is also balanced given its symmetric extension).  If
   $B^{\mu\eta}_{X,\mu\eta} \in \Theta_{M1}$, by the inductive property it
   is linked.  If $A^{\mu\eta}_{X,\mu\eta} \in \Pi_{M1}$, it is required,
   but we shall show that it is in fact linked.  By the definition of being
   required, the other possibility is that there is a block adjacent to the
   root block of ${\cal D}_1'$ with end-sequent
\[
	\Pi'', E; \proves \Theta''
\]
	in which the $(\vee\proveslabel^B)$ inference $R$ that bounds the
   block is based in $E$ and $\Pi'', E$ or $\Theta''$ contains an
   expression occurrence based in $A^{\mu\eta}_{X,\mu\eta}$.  But since the
   original block is isolated in the original ${\cal D}$, it is $E$ that
   must be based in $A^{\mu\eta}_{X,\mu\eta}$.  But then $R$ is based in
   $A^{\mu\eta}_{X,\mu\eta}$ and $R$ is linked: in particular its side
   expression in the left spur) must be linked; so
   $A^{\mu\eta}_{X,\mu\eta}$ is linked too.

	Thus we can weaken ${\cal D}_1'$ in its lowest block if necessary
   by $A\bimp_i B^{\mu}_{X}$ as a local right formula (in $\Gamma$),
   producing ${\cal D}_1^+$; ${\cal D}_1^+$ remains good by the same
   argument as the earlier cases.  Thus we can construct ${\cal D}'$ as:
\[
\derivi{}
	      {\begin{array}{c}
		{\cal D}_1^+ \\
		\Pi', A^{\mu\eta}_{X,\mu\eta} ; 
		\Gamma' \proves \Delta', 
		A\bimp_i B^{\mu}_{X};
		\Theta', B^{\mu\eta}_{X,\mu\eta}
		\end{array}}
	      {\Pi'; \Gamma' \proves \Delta'; \Theta'}
\]
	The end-sequent here is simple and balanced, so the whole root
   block is simple and balanced.  The new inference is linked (in virtue of
   the linked occurrence of one side expression---$A^{\mu\eta}_{X,\mu\eta}$
   or $B^{\mu\eta}_{X,\mu\eta}$) so the whole root block is linked.  The
   root block is of course canceled.  Each element of $\Pi_M$ is required
   because it is an element of $\Pi_{M1}$ and required in the immediate
   subderivation; each element of $\Theta_M$ is linked, because it is an
   element of $\Theta_{M1}$ and therefore linked in the immediate
   subderivation.
	
	To conclude the case, suppose the end-sequent of ${\cal D}$ is
   spanned and that $\Gamma'$ and $\Delta'$ are spanned by $\Theta'$; it
   follows that same property applies to ${\cal D}_1$ so the subderivation
   is spanned.  Then the end-sequent must also be spanned.

	The alternative case has $B^{\mu\eta}_{X,\mu\eta} \in \Theta$.  By
   assumption, it also has $A^{\mu\eta}_{X,\mu\eta} \in \Pi$.  We therefore
   define an overall $\Pi_M$ and $\Theta_M$ directly as $\Pi_{M1}$ and
   $\Theta_{M1}$, respectively; similarly $\Gamma' = \Gamma_1'$ and
   $\Delta' = \Delta_1'$.  To construct the needed ${\cal D}'$ for
   appropriate $\Pi'$ and $\Theta'$, we simply apply the induction
   hypothesis to ${\cal D}_1$ for the case that $\Theta_1'$ is $\Theta'$
   and $\Pi_1'$ is $\Pi'$.  The resulting derivation ${\cal D}_1'$
   suffices.  

	Having completed the induction, we argue that we can obtain an
   overall ${\cal D}'$ that is isolated, assuming the original ${\cal D}$
   is not only isolated but spanned.  Apply the inductive result to ${\cal
   D}$ for the case $\Pi' = \Pi$ and $\Theta' = \Theta$; since $\Gamma'
   \subseteq \Gamma$ and $\Delta' \subseteq \Delta$ we obtain a spanned
   derivation ${\cal D}'$ ending
\[
	\Pi; \Gamma' \proves \Delta'; \Theta
\]
	Consider the end-sequent of any block other than the root in ${\cal
   D}'$; it is
\[
	\Pi_0, E ; \proves ; \Theta_0
\]
 	where a corresponding block occurs in ${\cal D}$.  I argue by
   contradiction that for any $F \in \Pi_0$ either $F \in \Pi$ or $F$
   is based in an occurrence of $F$ as the side expression of an
   inference in ${\cal D}'$ in which $E$ is also based.  (This will
   show that ${\cal D}'$ is isolated.)  So consider an exceptional
   $F$.  Since ${\cal D}$ is isolated, if $F \not\in \Pi$, $F$ is
   based in an occurrence of $F$ as the side expression of an
   inference in ${\cal D}$ in which $E$ is also based; this inference
   introduces some path symbol $\eta$ which occurs in the label of $F$
   and $E$.  In ${\cal D}'$, $E$ can not be based in such an
   inference; otherwise $F$ would also be based in that inference,
   since ${\cal D}'$ is simple.  (We have assumed that $F$ is not
   based in such an inference.)  But in this case the expression in
   the end-sequent of ${\cal D}'$ on which $E$ is based must contain
   $\eta$.  Because the end-sequent of ${\cal D}'$ is spanned the form
   of $\Pi$ and $\Theta$ is constrained in ${\cal D}$, $F$ must occur
   in $\Pi$.  This is absurd.  $\qed$

	We conclude Section~\ref{rect-subsubsec} by observing some
   facts about this construction.  First, let ${\cal D}'$ be a
   derivation obtained by the construction of
   Lemma~\ref{linking-lemma}, and suppose ${\cal D}'$ is weakened (in
   a spanned and balanced way) to ${\cal D}''$ by adding occurrences
   of global expressions that either already occur in the end-sequent
   of ${\cal D}'$ or never occur as global expressions in ${\cal D}'$.
   Then a straightforward induction shows that ${\cal D}'$ is obtained
   again from ${\cal D}''$ by the construction of
   Lemma~\ref{linking-lemma}.

	Second, observe that if ${\cal D}'$ is a derivation obtained
   by the construction of Lemma~\ref{linking-lemma}, and ${\cal D}''$
   is obtained from ${\cal D}''$ by the renaming of Herbrand prefixes
   (as in Lemma~\ref{renaming-lemma}), then straightforward induction
   shows that ${\cal D}''$ is obtained again from ${\cal D}''$ by the
   construction of Lemma~\ref{linking-lemma}.

	Third, let ${\cal D}'$ be a derivation for which the
   construction of Lemma~\ref{linking-lemma} yields itself.  Let $\nu$
   be a prefix and let the $\Pi; \Theta$ be the smallest balanced pair
   where $\Theta$ contains all the carriers of prefixes of $\nu$
   introduced in ${\cal D}'$.  Suppose each expression in $\Pi$ and
   $\Theta$ has the property that at most one inference of ${\cal D}'$
   has an occurrence of that expression as a side expression.
   Consider a derivation ${\cal D}''$ obtained from ${\cal D}'$ by
   weakening globally by $\Pi$ (on the left) and by $\Theta$ (on the
   right).  Let ${\cal D}^*$ be the result of applying the
   construction of Lemma~\ref{linking-lemma} to ${\cal D}''$.  Then
   ${\cal D}^*$ contains any subderivation of ${\cal D}'$ whose
   end-sequent contains $\Pi$ and $\Theta$ as global formulas.  Again
   this is a straightforward induction; the base case considers a
   subderivation of ${\cal D}'$ whose end-sequent contains $\Pi$ and
   $\Theta$ as global formulas; in this case we apply the first
   observation.  Unary inferences extend the claim immediately.  At
   binary inferences, one subderivation must be unchanged, by the
   first observation: since $\Pi$ and $\Theta$ are introduced on a
   unique path, each $\Pi$ and $\Theta$ formula never occurs or
   already occurs in the end-sequent in that subderivation.  Thus the
   other subderivation necessarily appears in the derivation obtained
   by the construction of Lemma~\ref{linking-lemma}.

\subsubsection{Block conversion}
\label{block-proper-subsubsec}

	We now have the background required to perform the conversion to
   block structure, and complete the proof of Lemma~\ref{block-lemma}.

	\emph{We are given a blockwise eager SCLS derivation ${\cal
   D}$ whose end-sequent is spanned and balanced and takes the form:
\[
	\Pi ; \proves ; \Theta
\]
	We can transform ${\cal D}$ into a blockwise eager SCLB
   derivation in which every block is canceled, linked, isolated,
   simple, balanced and spanned.}

	\textbf{Proof.} Our induction hypothesis is stronger than the
   lemma.  We assume a blockwise eager SCLU derivation ${\cal D}$ with
   end-sequent of the form
\[
	\Pi ; \proves ; \Theta
\]
	in which every block is canceled, linked, isolated, simple,
   balanced and spanned, such that that the subproof rooted at any
   $(\vee\proveslabel)$ inference in ${\cal D}$ is an SCLS derivation.  And
   we identify a distinguished expression occurrence $E$ in the end-sequent
   of ${\cal D}$ which is linked.  By Lemma~\ref{linking-lemma}, it is
   straightforward to obtain such a derivation from the SCLS derivation
   (containing only a single block) that we have assumed.  We transform
   ${\cal D}$ into a blockwise eager SCLB derivation in which every block
   is canceled, linked, isolated, simple, balanced and spanned and in which
   $E$ is also linked; we perform induction on the number of
   $(\vee\proveslabel)$ inferences in ${\cal D}$.

	In the base case there are no $(\vee\proveslabel)$ inferences,
   so ${\cal D}$ itself is an SCLB derivation.

	In the inductive case, we assume ${\cal D}$ with \m n
   $(\vee\proveslabel)$ inferences, and assume the hypothesis true for
   derivations with fewer.  We find an application $L$ of
   $(\vee\proveslabel)$ with no other closer to the root of ${\cal
   D}$.  We will transform ${\cal D}$ to eliminate $L$.

	Let ${\cal D}'$ denote the smallest subderivation of ${\cal D}$
   containing the full block of ${\cal D}$ in which $L$ occurs.
   Explicitly, ${\cal D}'$ may be ${\cal D}$ itself; otherwise, ${\cal D}'$
   is rooted at the right subderivation of the highest
   $(\vee\proveslabel^B)$ inference below $L$---an inference we will refer
   to as $H$.  In either case, our assumptions allow us to identify a
   distinguished linked expression $F$ in the end-sequent of ${\cal D}'$:
   either the assumed $E$ from ${\cal D}$, or the side expression of the
   inference $H$ (assumed canceled).  Suppose $A\vee B^{\nu}_Y$ is the
   principal of $L$.  We can apply Lemma~\ref{renaming-lemma} to rename
   $A\vee B^{\nu}_Y$ to $A\vee B^{\mu}_X$ in such a way that each symbol in
   $\mu$ that is introduced in ${\cal D}'$ is introduced by a unique
   inference there.  Now we can infer the following schema for ${\cal D}'$:
\[
\begin{array}{c}
\left[
\centerbox{\derivii{$L$}
	{\begin{array}{c}
   	{\cal D}^A \\
	\Pi_0, F, \Pi; \Gamma, A\vee B^{\mu}_X, A^{\mu}_X \proves 
	\Delta ; \Theta_0, \Theta 
	 \end{array}}
	{\begin{array}{c}
	{\cal D}^B \\
	\Pi_0, F, \Pi; \Gamma, A\vee B^{\mu}_X, B^{\mu}_X \proves 
	\Delta ; \Theta_0, \Theta 
	\end{array}}
	{\Pi_0, F, \Pi; \Gamma, A\vee B^{\mu}_X \proves
	 \Delta; \Theta_0, \Theta}~~~}
\right] \\
{\cal D}^L \\
\Pi_0,  F; \proves ; \Theta_0
\end{array}
\]
	That is, the subderivation of ${\cal D}'$ below \m L is${\cal
   D}^{\s L}$; the right subderivation above \m L (in which \m B is
   assumed) is ${\cal D}^{\s B}$; the left is ${\cal D}^A$.  

	We will use the inferences from ${\cal D}^L$ to construct
   alternative smaller derivations in place of ${\cal D}^A$ and ${\cal
   D}^B$.  By $\Theta'$, indicate the minimal set of formulas required in
   addition to $\Theta_0$ to span $A^{\mu}_X$; by $\Pi'$ indicate the
   minimal set of formulas required in addition to $\Pi_0, F$ and
   $A^{\mu}_X$ to ensure that the pair given by $\Pi_0, \Pi', F, A^{\mu}_X$
   and $\Theta_0, \Theta'$ is balanced.  (This is well-defined because the
   sequent $\Pi_0, F \proves \Theta_0$ is already spanned and balanced.)
   Now we can construct two new subderivations ${\cal D}'^A$ and ${\cal
   D}'^B$ given respectively as follows:
\[
\begin{array}{c}
\left[
\centerbox{\derivi{decide}{
	\begin{array}{c}
	{} \Pi' + A^{\mu}_X + {\cal D}^A + \Theta' \\
	\Pi_0, F, \Pi, \Pi', A^{\mu}_X; 
	\Gamma, A\vee B^{\mu}_X, A^{\mu}_X \proves 
	\Delta ; \Theta_0, \Theta, \Theta'
	\end{array}}
	{\Pi_0, F, \Pi, \Pi', A^{\mu}_X; 
	\Gamma, A\vee B^{\mu}_X \proves 
	\Delta ; \Theta_0, \Theta, \Theta'}~~~~~~~~~~~~~~~~~}
\right] \\
\Pi' + A^{\mu}_X + {\cal D}^L + \Theta' \\
\Pi_0, F, \Pi', A^{\mu}_X; \proves ;\Theta_0, \Theta'
\end{array}
\]
\[
\begin{array}{c}
\left[
\centerbox{\derivi{decide}{
	\begin{array}{c}
	{} [ \Pi' + B^{\mu}_X + {\cal D}^B + \Theta' ] \\
	\Pi_0, F, \Pi, \Pi', B^{\mu}_X; 
	\Gamma, B\vee B^{\mu}_X, B^{\mu}_X \proves 
	\Delta ; \Theta_0, \Theta, \Theta'
	\end{array}}
	{\Pi_0, F, \Pi, \Pi', B^{\mu}_X; 
	\Gamma, B\vee B^{\mu}_X \proves 
	\Delta ; \Theta_0, \Theta, \Theta'}~~~~~~~~~~~~~~~~~}
\right] \\
\Pi' + B^{\mu}_X + {\cal D}^L + \Theta' \\
\Pi_0, F, \Pi', B^{\mu}_X; \proves ;\Theta_0, \Theta'
\end{array}
\]
	That is, we weaken ${\cal D}^A$ and ${\cal D}^B$ by global versions
   of the side expression of inference $L$ throughout their \emph{lowest
   blocks}; we apply a (decide) inference to obtain a new subderivation to
   substitute for the subderivation rooted at $L$ in ${\cal D}^L$.  We
   weaken by sufficient additional formulas globally in the \emph{lowest
   blocks} to ensure that the end-sequents of these derivations are
   balanced and spanned.

	Since we have changed only the lowest block here, and have
   ensured that this block remains isolated and canceled, we can now
   apply Lemma~\ref{linking-lemma} to obtain corresponding derivations
   ${\cal D}_I^A$ and ${\cal D}_I^B$ in which every block is canceled,
   linked, isolated, simple, balanced and spanned.  In light of our
   first observation about the construction of
   Lemma~\ref{linking-lemma}, we can see that the inferences of ${\cal
   D}^A$ are preserved up to the new (decide) inference.  And in light
   of our third observation about the construction of
   Lemma~\ref{linking-lemma}, given the unique inferences introducing
   $\Theta_0$ and $\Pi_0$, this (decide) inference must be preserved
   in ${\cal D}_I^A$.  Thus $A^{\mu}_X$ is linked in ${\cal D}_I^A$
   and for analogous reasons $B^{\mu}_X$ is linked in ${\cal D}_I^B$.
   These derivations satisfy the induction hypothesis as deductions
   with fewer than $n$ $(\vee\proveslabel)$ inferences; we can apply
   the induction hypothesis with $A^{\mu}_X$ and $B^{\mu}_X$ as the
   distinguished linked formulas to preserve.  This results in SCLB
   derivations ${\cal A}$ and ${\cal B}$ with the same end-sequents as
   ${\cal D}'^A$ and ${\cal D}'^B$, in which every block is canceled,
   linked, isolated, simple and spanned, and in which respectively
   $A^{\mu}_X$ and $B^{\mu}_X$ are linked.

	We need only one of ${\cal A}$ and ${\cal B}$ to reconstruct
   ${\cal D}'$ using blocking inferences.  For example, we obtain a
   proof using $(\vee\proveslabel^B_L)$ by using ${\cal B}$ in place
   of ${\cal D}^B$ as schematized below:
\[
\begin{array}{@{}c@{}}
\left[
\centerbox{\derivii{$\vee\proveslabel^B_L$}
	{\begin{array}{c}
   	{\cal D}^A \\
	\Pi_0, F, \Pi; \Gamma, A\vee B^{\mu}_X, A^{\mu}_X \proves 
	\Delta ; \Theta_0, \Theta 
	 \end{array}}
	{\begin{array}{c}
	{\cal B} \\
	\Pi_0, F, \Pi', B^{\mu}_X \proves \Theta_0, \Theta'
	\end{array}}
	{\Pi_0, F, \Pi; \Gamma, A\vee B^{\mu}_X \proves 
	\Delta ; \Theta_0, \Theta }~~~~~~~~~~}
\right] \\
{\cal D}^L \\
\Pi_0, F; \proves ; \Theta_0
\end{array}
\]
	In a complementary way, we obtain a proof using
   $(\vee\proveslabel^B_R)$ by using ${\cal A}$ in place of ${\cal
   D}^A$ as schematized below:
\[
\begin{array}{@{}c@{}}
\left[
\centerbox{\derivii{$\vee\proveslabel^B_L$}
	{\begin{array}{c}
   	{\cal D}^B \\
	\Pi_0, F, \Pi; \Gamma, A\vee B^{\mu}_X, B^{\mu}_X \proves 
	\Delta ; \Theta_0, \Theta 
	 \end{array}}
	{\begin{array}{c}
	{\cal A} \\
	\Pi_0, F, \Pi', A^{\mu}_X \proves \Theta_0, \Theta'
	\end{array}}
	{\Pi_0, F, \Pi; \Gamma, A\vee B^{\mu}_X \proves 
	\Delta ; \Theta_0, \Theta}~~~~~~~~~~}
\right] \\
{\cal D}^L \\
\Pi_0, F; \proves ;\Theta_0
\end{array}
\]
	Note that the root block is isolated in both cases, because we have
   added only as many formulas to $\Pi'$ and $\Theta'$ as are necessary to
   obtain a balanced, spanned sequent; the remaining expressions originate
   in the end-sequent of the previous block, which we know was isolated.
   Thus, in both cases, we have blockwise eager derivations in which every
   block is canceled, isolated, simple, balanced and spanned, in which
   fewer than $n$ $(\vee\proveslabel)$ inferences are used, and in which
   only the root block may fail to be linked.  We thus need to apply the
   construction of Lemma~\ref{linking-lemma} again to ensure that the root
   block is linked.  It is possible for the distinguished occurrence of $F$
   not to be linked in one of the resulting derivations, but not both.  To
   see this, consider applying the construction of
   Lemma~\ref{linking-lemma} to ${\cal D}'$ itself, as a test: the result
   will be ${\cal D}'$ since ${\cal D}'$ is linked.  Starting from ${\cal
   D}^A$ and ${\cal D}^B$ and axioms elsewhere, each inference in ${\cal
   D}'$ corresponds to an inference in the alternative derivations
   schematized above.  We can argue by straightforward induction that no
   formula is linked in the reconstructed ${\cal D'}$ unless it is also
   linked in the one of the corresponding reconstructed alternative
   derivations.  And $F$ is linked in ${\cal D}'$.
	
	Call the derivation in which $F$ is linked ${\cal D}''$; we
   substitute ${\cal D}''$ for ${\cal D}'$ in ${\cal D}$.  Since $F$
   remains linked in ${\cal D}''$, when we do so, we obtain a blockwise
   eager SCLU derivation with an appropriate end-sequent, with fewer
   original $(\vee\proveslabel)$ inferences, and in which every block
   remains canceled, linked, isolated, simple, balanced and spanned, and in
   which $(\vee\proveslabel)$ inferences lie at the root of SCLS
   derivations.  Applying the induction hypothesis to the result gives the
   required SCLB derivation.  $\qed$

\subsection{Proof of Lemma~\ref{lp-seq-lemma}}
\label{modularity-proof-subsec}

	\emph{We are given a blockwise eager SCLB derivation ${\cal
   D}$, with end-sequent
\[
	\Pi;\Gamma \proves \Delta;\Theta
\]
	in which every block is linked, simple and spanned.  We
   construct an SCLP derivation ${\cal D}'$ of which four additional
   properties hold:
\begin{itemize}
\item
	the end-sequent of ${\cal D}'$ takes the form
\[
	\Pi;\Gamma' \proves \Delta';\Theta
\]
	with $\Gamma' \subseteq \Gamma$ and $\Delta' \subseteq
   \Delta$;
\item
	${\cal D}'$ contains in each segment or block all and only the
   axioms of the corresponding segment or block of ${\cal D}$;
\item
	whenever ${\cal D}'$ contains a sequent of the form
\[
	\Pi^*; \Gamma^* \proveslabel \m F ; \Theta^*
\]
	$F$ is the only right formula on which an axiom in that block
   is based; and
\item
	whenever ${\cal D}'$ contains a sequent of the form 
\[
	\Pi^* ; \m F \proveslabel \Delta^*; \Theta^*
\]
	then \m F is the only left formula on which an axiom in that
   segment is based.
\end{itemize}}

	In the base case, ${\cal D}$ is 
\[
	{\Pi; \Gamma, \m A^{\mu}_{\s X} \proves 
	\m B^{\nu}_{\s X}, \Delta; \Theta}
\]
	and ${\cal D}'$ is
\[
	{\Pi; \m A^{\mu}_{\s X} \proves  \m B^{\nu} ; \Theta}
\]
	Supposing the claim true for proofs of height \m h, consider a
   proof ${\cal D}$ with height \m h+1.  We consider cases for the
   different rules with which ${\cal D}$ could end.

	The treatment of $(\proveslabel\wedge)$ is representative of
   the case analysis for the right rules other than
   $(\proveslabel\bimp)$.  ${\cal D}$ ends
\[
\derivii{$\proveslabel\wedge$}
	{\Pi; \proves A^{\mu}_X, A\wedge B^{\mu}_X, \Delta; \Theta}
	{\Pi; \proves B^{\mu}_X, A\wedge B^{\mu}_X, \Delta; \Theta}
	{\Pi; \proves A\wedge B^{\mu}_X, \Delta; \Theta}
\]
	(It is a consequence of Lemma~\ref{new-imp-ncxt-lemma} that in
   the initial derivation there is an empty local area.)  We simply
   apply the induction hypotheses to the immediate subderivations.  If
   the resulting derivations end with (restart), consider the
   immediate subderivation of the results, otherwise consider the
   results themselves.  These derivations end
\[
\begin{array}{c}
\Pi; \proves C ; \Theta \\
\Pi; \proves D ; \Theta
\end{array}
\]
	We must have \m C = \m A; we know from the structure of ${\cal
   D}$ that \m A is linked, and \m A could not be linked in ${\cal D}$
   unless $\m C = \m A$ since ${\cal D}'$ shows that all of the axioms
   in ${\cal D}$ derive from \m C.  For the same reason \m D = \m B.
   So we can combine the resulting proofs by an $(\proveslabel\wedge)$
   inference to give the needed ${\cal D}'$.

   	The case of $(\proveslabel\bimp)$ proceeds similarly, but relies
   on an additional observation.  ${\cal D}$ ends
\[
\derivi{$\proveslabel\bimp$}
       {\begin{array}{c}
	{\cal D}_1 \\
	\Pi, A^{\mu\eta}_{X,\mu\eta}; \proves 
	\Delta, A\bimp_i B^{\mu}_X; B^{\mu\eta}_{X,\mu\eta}, \Theta
	\end{array}}
       {\Pi; \proves 
	\Delta, A\bimp_i B^{\mu}_X; \Theta}
\]
	We apply the induction hypothesis to ${\cal D}_1$ and
   eliminate any final (restart) inference.  This gives us a
   derivation ${\cal D}_1'$ of
\[
	\Pi, A^{\mu\eta}_{X,\mu\eta} ; \proves 
	E ; B^{\mu\eta}_{X,\mu\eta}, \Theta
\]
	If we know that the $B$-side expression of this inference is
   linked in this block, then we can conclude, as before, that $E$ is
   an occurrence of the expression $B^{\mu\eta}_{X,\mu\eta}$.  We show
   this as follows.  We know from the structure of ${\cal D}$ only
   that \emph{one} of the $A$-expression and the $B$-expression must
   be linked.  However, it is straightforward to show that no left
   expression $A^{\mu\eta}_{X,\mu\eta}$ is linked in an SCLP
   derivation with a local goal $C^{\nu}_Y$ unless $\mu\eta$ is a
   prefix of $\nu$.  (The argument is a straightforward variant for
   example of \cite[Lemma 2]{permute-paper}.)  Since ${\cal D}$ is
   simple and spanned, $\eta$ must be new;
   $B^{\mu\eta}_{X,\mu\eta}$ is the only expression whose associated
   path term has $\mu\eta$ as a prefix.

	Thus, we construct ${\cal D}'$ using an SCLP inference as
\[
\derivi{$\proveslabel\bimp$}
       {\begin{array}{c}
	{\cal D}_1' \\
	\Pi, A^{\mu\eta}_{X,\mu\eta} ; \proves 
	B^{\mu\eta}_{X,\mu\eta} ; B^{\mu\eta}_{X,\mu\eta}, \Theta
	\end{array}}
	{\Pi; \proves 
	A\bimp_i B^{\mu}_{X} ; \Theta}
\]

	Now suppose ${\cal D}$ ends in a left rule other than
   $(\imp\proveslabel^S)$ or $(\vee\proveslabel^B)$.  We take
   $(\wedge\proveslabel)$ as a representative case; then ${\cal D}$ is:
\[
\derivi{$\wedge\proveslabel$}
       {\begin{array}{c}
	{\cal D}_1 \\
	\Pi; \Gamma, A \wedge B^{\mu}_X, A^{\mu}_X, B^{\mu}_X	
	\proves \Delta; \Theta
	\end{array}}
       {\Pi; \Gamma, A \wedge B^{\mu}_X
	\proves \Delta; \Theta}
\]
	Apply the induction hypothesis to ${\cal D}_1$.  If the result
   ends in a (decide) inference, let ${\cal D}_1'$ be the immediate
   subderivation of the result; otherwise let ${\cal D}_1'$ be the
   result itself.  ${\cal D}_1'$ is an SCLP derivation with an
   end-sequent of the form:
\[
	\Pi ; \m E \proves \m F; \Theta
\]
	$E$ must be a side expression of the inference in question,
   here $(\wedge\proveslabel)$; otherwise the corresponding inference
   could not have been linked in ${\cal D}$.  One of the inference
   figures $(\wedge\proveslabel_L)$ and $(\wedge\proveslabel_R)$ must
   apply depending on which side expression $E$ is.  For concrete
   illustration, we suppose $E$ is $A^{\mu}_X$; then we construct
   ${\cal D}'$ as:
\[
\derivi{$\wedge\proveslabel_L$}
       {\begin{array}{c}
	{\cal D}_1' \\
	\Pi ; A^{\mu}_X \proves F; \Theta 
	\end{array}}
       {\Pi; A\wedge B^{\mu}_X \proves F; \Theta}
\]

	Next, we suppose ${\cal D}$ ends in $(\imp\proveslabel^S)$, as
   follows:
\[
\derivii{$\imp\proveslabel^S$}
	{\begin{array}{c}
	 {\cal D}_1 \\
	 \Pi; \proves A^{\mu}_X, \Delta; \Theta
	\end{array}}
	{\begin{array}{c}
	 {\cal D}_2 \\
	 \Pi; \Gamma, A\imp B^{\mu}_X, B^{\mu}_X \proves \Delta; \Theta
	\end{array}}
	{\Pi; \Gamma, A\imp B^{\mu}_X \proves \Delta; \Theta}
\]
   	We begin by applying the induction hypothesis to the
   subderivation ${\cal D}_1$.  After stripping off any (restart), we
   obtain an SCLP derivation ${\cal D}_1$ with end-sequent
\[
	\Pi ; \proves C; \Theta
\]
	By the usual linking argument, the expression $C$ must be
   identical to $A^{\mu}_X$.  We then apply the induction hypothesis
   also to the right subderivation.  Again, after stripping off any
   (decide), we get an SCLP derivation ${\cal D}_2$ with end-sequent
\[
	\Pi; D \proves E; \Theta
\]
	By the usual linking argument, $D$ must in fact be identical
   to $B^{\mu}_X$.  Thus we obtain the needed ${\cal D}'$ by combining
   the two derivations by the SCLP $(\imp\proveslabel)$ rule:
\[
\derivii{$\imp\proveslabel$}
	{\begin{array}{c}
	 {\cal D}_1' \\
	 \Pi; \proves A^{\mu}_X ; \Theta
	 \end{array}}
	{\begin{array}{c}
	 {\cal D}_2' \\
	 \Pi; B^{\mu}_X \proves E; \Theta
	 \end{array}}	
	{ \Pi; A\imp B^{\mu}_X \proves E; \Theta }
\]

	Finally, for $(\vee\proveslabel^B)$, we consider the
   representative case of ${\cal D}$ as schematized below:
\[
\derivii{$\vee\proveslabel^B_L$}
	{\begin{array}{c}
	 {\cal D}_1 \\
	 \Pi; \Gamma, A^{\mu}_X \proves \Delta; \Theta
	 \end{array}}
	{\begin{array}{c}
	 {\cal D}_2 \\
	 \Pi', B^{\mu}_X ; \proves ; \Theta'
	 \end{array}}
	{\Pi; \Gamma, A\vee B^{\mu}_X \proves \Delta; \Theta}
\]
	We begin by applying the induction hypothesis to ${\cal D}_1$,
   the subderivation in the current block; if necessary, we strip off
   any initial (decide) inference, obtaining ${\cal D}_1'$ with an
   end-sequent that by linking takes the form:
\[
	\Pi; A^{\mu}_X \proves E ; \Theta
\]
	Next, we apply the induction hypothesis to the other
   subderivation.  Since both local areas are empty in the input
   subderivation, they remain empty in the result subderivation: this
   gives ${\cal D}_2'$ with end-sequent:
\[
	\Pi', B^{\mu}_X ; \proves ; \Theta'
\]
	The two subderivations can be recombined by the SCLP
   $(\vee\proveslabel_L)$ inference to obtain the needed ${\cal D}'$:
\[
\derivii{$\vee\proveslabel_L$}
	{\begin{array}{c}
	 {\cal D}_1' \\
	 \Pi; A^{\mu}_X \proves E ; \Theta
	\end{array}}
	{\begin{array}{c}
	 {\cal D}_2' \\
	 \Pi', B^{\mu}_X ; \proves ; \Theta'
	\end{array}}
	{\Pi; A\vee B^{\mu}_X ; \proves E; \Theta}
\]
	$\qed$